\documentclass[11pt, a4paper]{article}

\pdfoutput=1

\usepackage{fullpage}
\usepackage{setspace}
\usepackage{amsmath}
\usepackage{amsfonts}
\usepackage{amssymb}
\usepackage{fix-cm}
\usepackage[toc]{appendix}
\usepackage{xcolor}
\usepackage{hyperref}
\hypersetup{colorlinks, allcolors=blue}
\usepackage[numbers]{natbib}
\usepackage{graphicx}
\usepackage{booktabs}
\usepackage[UKenglish]{babel}

\DeclareMathOperator{\prob}{Pr}
\DeclareMathOperator{\sinc}{sinc}
\DeclareMathOperator{\Tr}{Tr}
\newcommand{\Tq}{\ensuremath{T_\text{q}}}
\newcommand{\Tb}{\ensuremath{T_\text{b}}}
\newcommand{\Tbkg}{\ensuremath{T_\text{bkg}}}
\newcommand{\vp}{\ensuremath{\nu_\text{p}}}
\newcommand\smallO{
	\mathchoice
	{{\scriptstyle\mathcal{O}}}
	{{\scriptstyle\mathcal{O}}}
	{{\scriptscriptstyle\mathcal{O}}}
	{\scalebox{.7}{$\scriptscriptstyle\mathcal{O}$}}
}

\begin{document}

\begin{titlepage}

\centering

{\huge\bfseries{Superconducting Microwave Detector Technology for Ultra-Light Dark Matter Haloscopes and other Fundamental Physics Experiments: Background Theory (Part I)}\par}

\vspace{0.5cm}
{Christopher N. Thomas$^1$, Stafford Withington$^2$ and David J. Goldie$^1$ \par}

\vspace{0.5cm}
{$^1$\emph{Cavendish Laboratory, University of Cambridge, JJ Thomson Avenue, Cambridge, CB3 0HE, UK}\par}
{$^2$\emph{Clarendon Laboratory, University of Oxford, Parks Road, Oxford, OX1 3PU, UK}\par}

\vspace{0.5cm}
\today

\begin{abstract}
We consider how superconducting microwave detector technology might be applied to the readout of cavity-axion haloscopes and similar fundamental physics experiments.
Expressions for the sensitivity of two detection schemes are derived: 1) a dispersive spectrometer, and 2) a direct-conversion/homodyne receiver using detectors as mixing elements.
In both cases the semi-classical/Poisson-mixture approach is used to account for quantum effects.
Preliminary sensitivity calculations are performed to guide future development work.
These suggest the homodyne scheme offers a near-term solution for realising near-quantum-noise limited receivers with improved usability compared with parametric amplifiers.
Similarly, they show that the dispersive spectrometer offers a potential way to beat the quantum noise limit, but that significant technological development work is needed to do so.
\end{abstract}

\end{titlepage}

\newpage

\onehalfspacing

\tableofcontents

\newpage

\section{Introduction}\label{sec:introduction}

One of the most technologically challenging applications of microwave radiometry is the readout of low frequency axion haloscopes.
Here the aim is to detect thermal line emission due to the axion against a background of thermal radiation from a microwave resonant cavity at frequencies up to 10\,GHz.
This line is theorised to be $\sim$1 kHz wide and to contain $\sim$10$^{-22}$\,W of power, and must be detected, for example, against a background power loading of $\sim$10$^{-20}$\,W in the case of a typical 1\,GHz cavity cooled to $\sim$1\,K\,\cite{daw1998search}.
Expressed in terms of noise temperatures, the challenge is to measure 10 mK temperature excess against a 1\,K background and any contribution from instrument noise, given only a linewidth of bandwidth.
The corresponding background photon arrival rate is only fifteen per second, so operation of the radiometer is further complicated by quantum effects.
For comparison, measurements of the cosmic microwave background (CMB) have achieved $\mu$K resolution against a 4\,K background, but they do so thanks to much larger photon arrival rates. 

Haloscopes have traditionally used coherent radiometers for read out.
In a typical system the signal from the cavity is band-filtered, amplified and then down-converted for digitisation.
Digital signal processing techniques are then used to form multiple radiometer channels simultaneously across the down-converted bandwidth.
The receiver noise temperature of a well engineered system is determined by the initial low noise amplifier, with the current state of the art for transistor based designs being of order 1.3\,K below 3 \,GHz\,\cite{lnf}.

Reductions in receiver noise temperatures are required for deeper searches and to fully exploit lower temperature cavities.
Unfortunately, the minimum receiver noise temperature of a coherent system is limited to $T_q \sim h \nu_0 / (2 k_b)$ by quantum effects, where $h$ is Planck's constant, $\nu_0$ is the channel frequency and $k_b$ is Boltzmann's constants ($T_q \approx$  24\,mK at 1\,GHz)\,\cite{kerr1997receiver}.
New parametric amplifier designs are starting to approach this limit, leaving little room for improvement.
Further, they are typically more difficult to operate and narrower in bandwidth than typical transistor-based designs.
As such, there is increasing interest in alternate receiver technologies that are easier to operate and/or have growth potential in terms of sensitivity and with increasing tuning range up to mm-wavelengths.

One such technology is direct detection radiometers.
In such systems the input signal is band-filtered and then then the power in the filtered signal is measured directly using a power detector such as a bolometer.
Since they do not use amplifiers and or mixers prior to detection, direct detection radiometers are not subject to the quantum noise limit and can, in principle, achieve better sensitivities than coherent systems in a low background power limit\,\cite{zmuidzinas2003thermal,obrient2010log}.
However, to build multi-channel systems requires either a dispersive element, such as a filter-bank or grating, or a Fourier transform spectrometer.

Direct detection radiometers based on superconducting detectors including transition edge sensors (TESs) and Kinetic Inductance Detectors (KIDs) have been extensively developed for astronomical observations in the sub-mm range (above 40\,GHz for TESs and above 100\,GHz for KIDs).
Here the observation problem is similar to that of the axion case; the background photon count can become low, in which case the quantum noise limit associated with coherent receivers becomes a significant sensitivity penalty\,\cite{obrient2010log}.
Noise equivalent powers (NEPs) below 10$^{-18}$\,W/$\sqrt{\text{Hz}}$ are now readily achieved and multi-channel, system-on-chip, filter-bank spectrometers have been demonstrated.
Additionally, the technology to support these detectors, such as cryogenic coolers and readout electronics, is mature and commercially available.
Work is ongoing to extend these technologies to longer wavelengths ($<$ 20\,GHz).

In this report we will consider the application of superconducting detector technology developed in the astronomical context to fundamental physics measurements, using readout of the cavity haloscope as an example application.
We will consider two different measurement schemes for measuring the power spectral density of a microwave input signal.
The first is a dispersive spectrometer in which the input signal is first filtered into different frequency channels at its native frequency, then the output of each channel is measured by a superconducting microwave power detector.
We will refer to this scheme generically as a filter-bank spectrometer.
The second is a homodyne (or direct-conversion) receiver implemented using superconducting microwave detectors as the mixing element.
In the case of the homodyne scheme, the overall behaviour of the system is that of a coherent system and as such it is subject to a quantum noise limit.
However, we believe it will be much easier to operate then equivalent parametric amplifier designs; the pump level will not need precise adjustment and it will be tuneable over a wide input frequency range.
Homodyne readout of cavity haloscopes using power detectors has already been promoted by Omarov\,\cite{omarov2023speeding}; here will consider an alternate scheme using a balanced receiver architecture and present a sensitivity analysis using concepts more familiar to microwave engineers.

In Section \ref{sec:basic_overview} of the report we will describe the basic operation of the two schemes and highlight their key advantages.
In Section \ref{sec:signal_analysis} we then present a noise and responsivity analysis of both configurations.
This is based on a Poisson-mixture model of the photon detection problem\,\cite{saklatvala2007coupled,saklatvala,zmuidzinas2015use} and will allow us to explore the behaviour in both the quantum (few-photon) and classical (many-photon) regimes, as well as the cross-over region between the two.
This is, to our knowledge, the first time the Poisson-mixture model has been used to analyse a homodyne receiver.
Finally, in Sections \ref{sec:ultimate_sensitivity} we consider modes of operation of the two receivers and the corresponding sensitivities.
Detector technologies for realising the receivers are discussed in a companion report\,\cite{goldie2024low}.

\section{Detection schemes considered}\label{sec:basic_overview}

In this section we will describe the detection schemes that will be considered.
It will be useful to consider some basic aspects of power detector theory prior to doing so, as this will help motivate the two designs.

\subsection{Microwave power detector theory}\label{sec:bo_microwave_power_detector_theory}

In this section we will consider how power detectors are normally operated and the corresponding measures of their sensitivity.
The canonical example of a power detector is the bolometer.

A power detector responds to the total incident power $p(t)$ in the signal as a function of time $t$, as opposed to the amplitude as in the case of an amplifier or mixer.
For a linear device, the readout output $y(t)$ can be modelled by
\begin{equation}\label{eqn:det_theory_det_output}
	y(t) = \int_{-\infty}^{\infty} g(t - t') p(t') \, dt' + n(t).
\end{equation}
Here $g(t)$ is the response function of the detector and $n(t)$ is the noise at the output.
This noise may come from a variety of sources, including the readout electronics, the detector and the input signal itself.
The response function $g(t)$ encodes both the responsivity of the device and the fact its response rate to changes in the input power is limited; response times for superconducting microwave power detectors range from microseconds for tunnel junctions to milliseconds for transition edge sensors.

The sensitivity of a power detector is usually measured using a figure of merit called the Noise Equivalent Power (NEP).
This is canonically defined as the steady-state input power level that gives a signal to noise ratio of unity in 1\,Hz of output bandwidth.
A reference output bandwidth is given because it is always possible to reduce the output bandwidth by filtering and thereby increase the sensitivity.
NEP is strictly an overall system quantity and depends not only on the detector, but the readout electronics, detector, background input signal and particular modulation strategy.

Modulation is used here to mean converting the steady-state power level of interest into a time-varying power level.
This is common practice as it allows the signal-of-interest to be moved to parts of the output bandwidth of the detector where the noise is lower, thereby improving sensitivity.
An example of modulation strategy is `chopping', whereby the signal of interest is turned on an off cyclically to encode the input level as the amplitude of several harmonics at the detector output (e.g. using a chopper wheel or by pointing the detector on and off target).
The disadvantage of modulation is that input signal power may be lost, e.g. when the source is off in a chopped measurement, or if power is modulated into frequency components that are not subsequently measured.
We will define the efficiency $\eta_m^2$ of a modulation as the ratio of the time-averaged signal power flow for the modulated signal to that for the unmodulated signal.
Poor modulation efficiency degrades sensitivity.

A more general definition of NEP that accounts for modulation effects is as the constant of proportionality in the relationship between the noise in a power measurement and the length for which the output signal is recorded.
Let $\Delta P$ be the RMS error in a power measurement and $\tau$ the time the signal is recorded for, then we can define the $\text{NEP} (f_0)$ when the steady-state signal is modulated onto a tone with frequency $f_0$ by
\begin{equation}\label{eqn:dp_from_nep}
	\Delta P = \frac{\text{NEP} (f_0)}{\eta_m \sqrt{2 \tau}}
\end{equation}
where $\eta_m$ is the modulation efficiency assuming only that tone is detected.
This definition reduces to the canonical definition when the single is unmodulated; $\eta_m = 1$ and the output bandwidth for time averaging over $\tau$ is $\Delta \nu = 1 / \sqrt{2 \tau}$.
However, it extends the definition to modulated signals in a manner that distinguishes the intrinsic sensitivity of the detector system and reductions in that sensitivity due to imperfect modulation.
(\ref{eqn:dp_from_nep}) is derived in Appendix \ref{sec:nep}.

The NEP can be decomposed into two components according to
\begin{equation}
	\text{NEP}^2 = \text{NEP}_i^2 + \text{NEP}_\text{photon}^2
\end{equation}
where $\text{NEP}_i$ is the NEP that would be measured in the absence of an input signal and $\text{NEP}_\text{photon}$ is, therefore, the noise due to the input signal.
$\text{NEP}_i$ arises from noises sources internal to the detector and readout electronics and we will refer to it as the \emph{intrinsic} NEP of the detector system.
Similarly, we will refer to $\text{NEP}_\text{photon}$ as the \emph{photon} NEP in the particular application.
The latter terminology follows historical convention and should not be interpreted as being restricted to noise associated with the photonic nature of the light.
Instead it should be read as meaning \emph{any} noise associated with the input photons, be it shot noise in the quantum regime or wave-noise in the classical regime.
In most applications of interest the signal is relatively weak and so the photon noise will be dominated by any background radiation.

\begin{figure}
\centering
\includegraphics{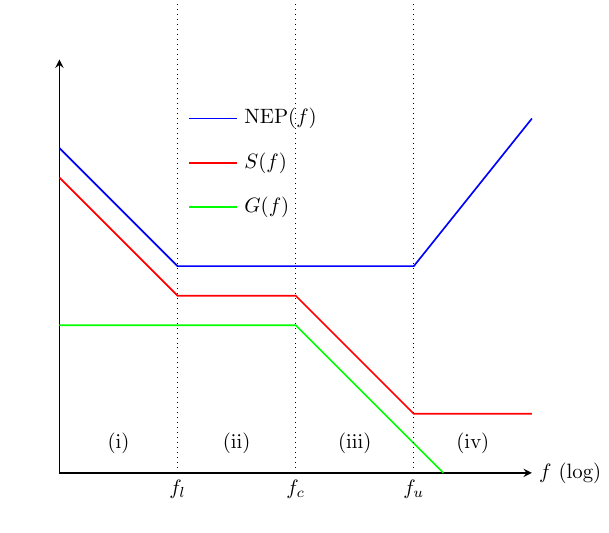}
\caption{
\label{fig:nep_cartoon}
Behaviour of $S(f)$, $G(f)$ and $\text{NEP}$ for a typical power detector.
}
\end{figure}

Figure \ref{fig:nep_cartoon} shows the behaviour of $\text{NEP}_i (f_0)$ for a typical superconducting power detector.
Three quantities are plotted: $\text{NEP}_i (f_0)$, the noise power spectral density $S(f)$ at the detector output and the frequency domain responsivity $R(f) = |g(f)|^2$.
We can normally distinguish four regions, (i)--(iv), in the output noise power spectral density $S(f)$.
Below a certain shoulder frequency, region (i), $S(f)$ is dominated by $1/f$ noise resulting from long timescale drift in system parameters.
This is followed by a region, (ii) and (iii) together, where the noise from the detector itself dominates.
This usually rolls off with the detector responsivity $G(f)$, as seen in region (iii).
Finally there is a region, (iv), where the detector responsivity has rolled off sufficiently that noise from the readout electronics now dominates.
The result is the NEP is lowest over some range of non-zero frequencies.

The valley in Figure \ref{fig:nep_cartoon} is seen generally and motivates modulation as a strategy to improve detector sensitivity.
This is most simply illustrated by considering chopping, which converts the steady-state input power level into an input square-wave.
This generates a strong fundamental tone at the detector output at frequency $f_0$, the amplitude and phase of which can be measured using lock-in techniques and used to infer the original power level.
The noise in the measurement depends on the input-referred noise power spectral density at the fundamental frequency and hence we can improve sensitivity by choosing $f_l < f_0 < f_u$.
If instead we want to think about the problem in the time-domain, the rising NEP at low frequencies is due to $1/f$ noise, which results from drifts in the output signal over long timescales caused by system instability.
These fluctuations prevent direct time-averaging, but chopping between the unknown signal and a known source (i.e. zero) allows for calibration of the drift and its subsequent removal.
Overcoming detector instability is critical to achieving optimum performance and motivates the homodyne approach.

Finally, it is useful later to be able to convert NEP into the equivalent system noise temperature $T_\text{sys}$ of a coherent radiometer, so as to allow a sensitivity comparison.
In a radiometric application the aim is to measure the noise temperature $T_s$ as defined by $P_s = k_b T_s \Delta \nu$, where $\Delta \nu$ is the input radiation bandwidth of the power measurement.
Therefore, whereas a power detector measures total power incident over some input bandwidth, a coherent radiometer measures the power spectral density in that bandwidth.
It follows from (\ref{eqn:dp_from_nep}) that the RMS temperature error achievable with a power detector is
\begin{equation}\label{eqn:rms_temp_noise_power_detector}
	\Delta T_\text{rms} = \frac{\text{NEP}(f_0)}{k_b \eta_m \Delta \nu \sqrt{2 \tau}}.
\end{equation}
The equivalent result for a coherent radiometer is simply the radiometer equation,
\begin{equation}\label{eqn:rms_temp_noise_coherent_receiver}
	\Delta T_\text{rms} = \frac{T_\text{sys}}{\sqrt{ \tau \Delta \nu}},
\end{equation}
hence we see that the equivalent system noise temperature of a power detector system is
\begin{equation}\label{eqn:sys_temp_power_detector}
	T_\text{sys} = \frac{\text{NEP}(f_0)}{k_b \eta_m \sqrt{2 \Delta \nu}}.
\end{equation}
Equivalently, the effective NEP of a coherent radiometer system is
\begin{equation}
	\text{NEP} = k_b T_\text{sys} \sqrt{2 \Delta \nu}.
\end{equation}

\subsection{Filter-bank spectrometer}\label{sec:bo_filterbank}

\begin{figure}
\centering
\includegraphics{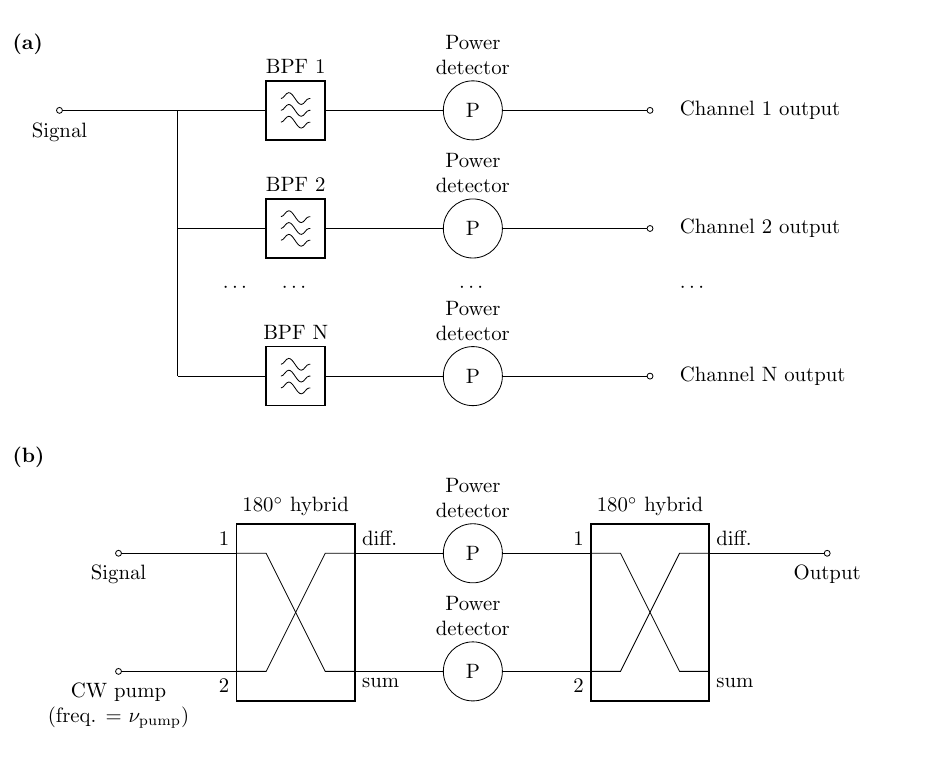}
\caption{\label{fig:receiver_architecture}
The two types of radiometer considered.
Panel a) shows the concept of a filterbank spectrometer and b) a direct conversion receiver implemented with power detectors.
}
\end{figure}

The first of the two detection schemes that we will consider in this report is the filter-bank spectrometer, as illustrated in Figure \ref{fig:receiver_architecture}a.
This is conceptually the simplest way of realising a multichannel radiometer with power detectors.
A filter-bank circuit is used to disperse the signal into the different spectral channels, then the power in each channel is measured by a separate detector.
Here we will assume the interest is in implementing multiple channels to give spectral sensitivity and mimic the behaviour of a coherent receiver.
However the analysis applies equally well to the case where a single detector is used to measure the power in a bandpass signal, e.g. for a whole power measurement on a cavity haloscope.

A recent development in radio astronomy is cyrogenic filterbank spectrometers that integrate the filterbank and detectors together as a single microwave monolithic integrated circuits (MMICs).
These MMICs utilise superconducting, rather than semiconduting, electronics and this enables the fabrication of both ultrasensitive detectors and low-loss, narrow-bandwidth, filter channels and ultrasensitive detectors.
Imaging arrays with few-colour pixels at mm-wavelengths ($>$40 GHz) have already been deployed\,\cite{sobrin2022design,westbrook2018polarbear,dahal2020class} and true spectrometer designs with are in development\,\cite{goldie2020first,endo2019wideband,redford2022superspec}.
It should be possible to scale this technology to axion search frequencies ($<$40 GHz) by developing lumped element filter designs.
Alternatively, the cavity haloscope itself could be used as a single, tuneable, filter for a single detector.

The main advantages of this design is that it can, in principle, achieve better sensitivity than a coherent radiometer in the case of low-background loading, as expected for an axion haloscope.
This will demonstrated in Section \ref{sec:sensitivity_filterbank_spectrometer}.
However, this advantage is also widely discussed in the literature, e.g. \cite{zmuidzinas2003thermal}.

The main disadvantages are operational.
Firstly, unlike coherent radiometers they are not retuneable; location of the channels are fixed at production.
Secondly, a stabilisation method such as Dicke-switching is needed to overcome $1/f$ noise and achieve ultimate sensitivity.
The microwave switch needed to do the latter, indicated in the figure, is technologically challenging to realise, as it must operate at cryogenic temperatures and add the minimal amount of structure possible to the receiver's spectral response.

\subsection{Homodyne scheme}\label{sec:bo_homodyne_scheme}

The second detection scheme we will consider is illustrated in Figure \ref{fig:receiver_architecture}b.
It comprises a pair of identical power detectors that are connected to the sum and difference outputs of a 180$^\circ$ microwave hybrid.
The noise signal of interest is applied to one input of the hybrid and a continuous wave (CW) to the other, with the frequency of the latter chosen to be similar to that of the signal of interest.
The output signal is the difference of the two detector outputs.

This arrangement functions as a double-balanced mixer in the classical limit.
To see this, consider the case where continuous waves are applied to both ports.
If the hybrid is ideal, the inputs to the two detectors are
\begin{equation}\label{eqn:bo_a1_and_a2}
\begin{aligned}
	a_1 (t) &= \Re[ b_1 e^{2 \pi i \nu_1 t} + b_2 e^{2 \pi i \nu_2 t} ] \\
	a_2 (t) &= \Re[ b_1 e^{2 \pi i \nu_1 t} - b_2 e^{2 \pi i \nu_2 t} ]
\end{aligned}
\end{equation}
where the $b_n$ and $\nu_n$ are the amplitude and frequency of the tone at input $n$.
The power detectors respond to the square of these signals, so the outputs are
\begin{equation}\label{eqn:bo_p1_and_p2}
\begin{aligned}
	p_1 (t) &= \frac{\kappa}{2} \left\{
		|b_1^{\vphantom *} |^2 + |b_2^{\vphantom *} |^2
		+ 2 \Re[b_1^{\vphantom *}  b_2^* e^{2 \pi i (\nu_1 - \nu_2) t} ]
	\right\} \\
	p_2 (t) &= \frac{\kappa}{2} \left\{
		|b_1^{\vphantom *} |^2 + |b_2^{\vphantom *} |^2
		- 2 \Re[b_1^{\vphantom *}  b_2^* e^{2 \pi i (\nu_1 - \nu_2) t} ]
	\right\} \\
\end{aligned}
\end{equation}
for some responsivity $\kappa$.
The difference signal $d(t)$ is then
\begin{equation}
	d(t) = p_1(t) - p_2(t) = 2 \Re \bigl[
		b_1^{\vphantom *}  b_2^* e^{2 \pi i (\nu_1 - \nu_2) t} \bigr],
\end{equation}
which is the product of the input signals.
When a more general signal is applied at one port, it therefore appears in the difference signal shifted down in frequency by an amount equal to the frequency of the continuous wave on the second port.

\begin{figure}
\centering
\includegraphics{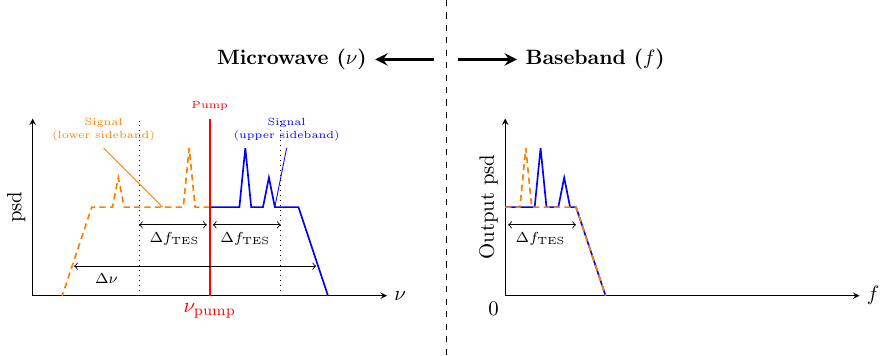}
\caption{\label{fig:direct_downconversion}
Operation of the receiver arrangement in Figure \ref{fig:receiver_architecture}b as a direct conversion receiver.
}
\end{figure}

For axion searches we propose operating the arrangement as a direct downconversion receiver,
as illustrated in Figure \ref{fig:direct_downconversion}.
The frequency of the CW tone would be chosen to down-convert some bandwidth of interest into the output bandwidth of the detectors, i.e. close to the centre of the bandwidth of the signal.
Ideally the frequency should be chosen to place the down-converted signal in the range of the NEP valley discussed in Section \ref{sec:bo_microwave_power_detector_theory}, so as to maximise sensitivity.
The down converted signal will then appear as noise in the difference signal and its power spectral density can be recovered using standard methods for measuring the NEP of a detector.
If the difference signal were filtered to match the line width and the total power then measured, the receiver could be used as a scanned matched filter.

As discussed in the introduction, such a system can in principle function as a quantum noise limited radiometer when the CW power is greater than the input noise signal power.
A detailed argument will be presented in Section \ref{sec:sensitivity_homodyne_receiver}, but we provide a qualitative argument here.
If the internal noise is made sufficiently low, the detectors' NEPs will be dominated by photon noise from background power loading.
This comprises two parts: a shot noise term of quantum origin and a wave noise term of classical origin.
The wave noise term is correlated between detectors that see the same signal and cancels in the difference signal.
The photon noise remains and scales as the product of the photon energy with the energy in the signal.
Three noise terms are therefore seen in the difference signal: the down-converted signal, the photon noise generated by the CW signal and the photon noise generated by the input noise signal.
The first two contributions scale with the CW power, while the third scales with the signal power.
When the difference signal is rescaled by the CW power to recover the signal temperature, the photon noise term from the input noise signal becomes negligible.
However, the photon noise from the CW remains and is of order $h \nu / k_b$, i.e. twice the quantum noise limit.

If the radiometric sensitivity limit is the same as more conventional coherent receivers based on amplifiers, one might ask why this architecture is attractive?
The answer is that we believe it will be easier to operate.
Current quantum noise limited receivers use parametric amplifiers, which typically require the amplitude of the CW pump signal be finely adjusted to operate and must be recalibrated on retuning.
The proposed architecture should, by contrast, be able to operate with a range of CW signal amplitudes and at any frequency where both the hybrid functions and microwave power can be coupled efficiently into the detector.
High coupling efficiencies have been demonstrated for microstrip coupled TESs over instantaneous frequency ranges of 0--700  GHz\,\cite{rostem2009chip}, so the tuning range can in principle be made very large and only be limited by hybrid design.

The main disadvantage of this arrangement is that the instantaneous bandwidth is much smaller than a traditional coherent receiver.
This is because the instantaneous bandwidth is set by the bandwidth of the detector, which is of order 1\,kHz for typical low noise detectors (see the companion note).
This compares with a bandwidth of a few MHz for parametric amplifiers.
However, since the bandwidth of the output of cavity haloscope is only on the few tens of kHz this is not a critical issue.

\section{Signal analysis}\label{sec:signal_analysis}

\begin{figure}
\centering
\includegraphics{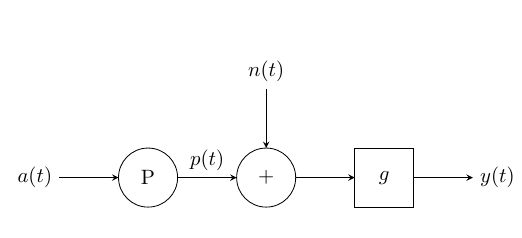}
\caption{\label{fig:single_detector_noise_model}
Detector noise model.
}
\end{figure}

In this section we will develop a model for the output signals of the receivers in the two schemes, which will form the basis of the sensitivity analysis in Section \ref{sec:ultimate_sensitivity}.
The starting point is to analyse the input-output behaviour of single detector, which models a single channel of the filter-bank spectrometer.
We may then analyse the output of the homodyne receiver by combining the outputs of two such detectors for appropriate input signals.

Let us begin by defining what is meant by the input of the detector in this context.
We will assume all the detectors are fed from transmission lines and that input is the component of the wavefield on the line that is travelling towards the detector.
We also assume that this input wave is quasi-monochromatic with centre frequency $\nu_0$ and bandwidth $\Delta \nu \ll 2 \nu_0$, which is appropriate for the intended application and significantly simplifies the subsequent analysis.
A convenient choice of input signal is then the complex analytic signal representation of the instantaneous amplitude of the input wave at the detector port, which we denote $a(t)$.
We assume the normalisation of the amplitude has been chosen such that the time-averaged power flow into the detector is given by $|a(t)|^2 /2$.
In this case, $a(t)$ is simply the time-domain form of the Kurokawa power wave amplitudes used in frequency domain in microwave analysis.
More details of $a(t)$ are given in Appendix \ref{sec:tl_signals}.

Next we must specify the basic detector model.
We will assume each detector responds linearly to the power absorbed at its input port whilst also containing noise in its output generated by sources internal to the detector.
Figure \ref{fig:single_detector_noise_model} shows a model of such a detector, comprising an ideal power detector, additive noise source and output filter in series.
The ideal power detector converts $a(t)$ to a power signal $p(t)$, to which is added a noise signal $n(t)$ and then the result is filtered to give the final output signal $y(t)$:
\begin{equation}\label{eqn:sa_output_equation}
	y(t) = \int_{-\infty}^t g(t, t') \{ p(t') + n(t') \} \, dt'.
\end{equation}
The noise signal $n(t)$ is the effective input referred noise, i.e. it is the equivalent input signal that would give the total noise measured at the detector output for $p(t) = 0$.
The filter function $g(t)$ is assumed to contain the effects of any detector response time, but not of further filtering operations like time-averaging; the latter will be dealt with as part of the sensitivity analysis as it is not intrinsic to the detector.

There is a subtly in the behaviour of the ideal power detector and it does not simply produce a signal proportional to $|a(t)|^2 / 2$.
In reality, the input signal is a stream of photons of energy $h \nu_0$ which are absorbed randomly, depositing their energy instantaneously, according to some average rate.
Further, this rate may itself vary randomly over time.
Typically the average of this rate is the quantity of interest, as it the `time-averaged' power flow into the detector.
However, we see its measurement will be complicated either in the limit where there are very few photons in the measurement period, or if the noise in the rate is significant over the measurement period.
Calculating the statistics of $p(t)$ will be a significant part of the analysis and we defer discussion to Section \ref{sec:sa_pow_mgf}.

With the input signal and detector specified, we can proceed with the signal analysis.
Our approach will be to work in terms of the moment generating functionals (MGFs) of all the signals, which will allow us to mechanistically generate any order of correlation function for the output signals.
We will start by reviewing MGFs (Section \ref{sec:sa_mgf}), then proceed to calculate the MGF for the output of the ideal power detector (Section \ref{sec:sa_pow_mgf}), the output of a noisy single power detector (Figure, \ref{fig:single_detector_noise_model}, Section \ref{sec:mgf_sd}) and the output of the homodyne receiver (Section \ref{sec:mgf_homo}).
These will then be used to calculate the correlation functions of the output signals in the two schemes to the orders necessary for the sensitivity analysis.

\subsection{Moment generating functionals}\label{sec:sa_mgf}

In this section we review the basics of MGFs as necessary for the analysis.

To begin with, let us consider the moment generating \emph{function} of a real continuous random variable $x$.
If the probability density of $x$ is $\prob (x)$, then the moment generating function $M_x$ of x is defined by
\begin{equation}\label{eqn:def_moment_generating_function}
	M_x (s) = \int_{-\infty}^\infty e^{s x} \prob (x) \, dx
\end{equation}
for dummy variable $s$.
As the name suggests, the moment generating function can be used to calculate the moments of the variable.
This is is achieved by differentiation and it is straightforward to show that
\begin{equation}\label{eqn:scalar_moments_from_moment_generating_functions}
	\langle x^n \rangle = \left( \frac{d^n M_x}{ds^n} \right)_{s=0}.
\end{equation}
Examples of moment generating functions include
\begin{equation}\label{eqn:gaussian_moment_generating_function}
	M_x (s) = e^{\mu s + \frac{1}{2} \sigma^2 s^2}
\end{equation}
if $x$ is Gaussian distributed with mean $\mu$ and variance $\sigma^2$ and
\begin{equation}\label{eqn:poisson_moment_generating_function}
	M_x (s) = e^{(e^s - 1) \lambda}
\end{equation}
for a Poisson distributed variable with rate $\lambda$.

This definition is straightforwardly extended to a set of jointly distributed random variables, $\{ x_n | n = 1, 2, \dots N\}$.
Representing the values of the variables as a vector $\boldsymbol{x}$ and denoting the joint probability distribution by $\prob (\boldsymbol{x})$, we may define the multivariate moment generating function by
\begin{equation}\label{eqn:def_vec_moment_generating_function}
	M_{\boldsymbol{x}} (\boldsymbol{s}) = \int
		e^{\boldsymbol{s}^T \cdot \boldsymbol{x}}
		\prob (\boldsymbol{x})
	\, d^N\boldsymbol{x}.
\end{equation}
The different joint moments of the variables can then be generated by differentiating $M_{\boldsymbol{x}}$ appropriately:
\begin{equation}\label{eqn:joint_moments_from_moment_generating_function}
	\langle x_i x_j \dots x_k \rangle
	= \left( \frac{\partial}{\partial s_i} \frac{\partial}{\partial s_j}
	\dots \frac{\partial M_{\boldsymbol{x}}}{\partial s_k} \right)_{\boldsymbol{s}=\boldsymbol{0}}
\end{equation}
As an example, the moment generation function for a multivariate Gaussian distribution is
\begin{equation}\label{eqn:multivariate_gaussian_moment_generating_function}
	M_x (s) = e^{
		\boldsymbol{\mu}^T \cdot \boldsymbol{s}
		+ \frac{1}{2} \boldsymbol{s}^T \cdot \boldsymbol{\Sigma} \cdot \boldsymbol{s}
	}
\end{equation}
where $\boldsymbol{\mu}$ is the vector of means and $\boldsymbol{\Sigma}$ is the covariance matrix.

It follows from (\ref{eqn:def_moment_generating_function}) that if the variables are statistically independent then
\begin{equation}\label{eqn:multivariate_moment_generating_function_independent_variables}
	M_{\boldsymbol{x}} (\boldsymbol{s})
	= \prod_{i=1}^N M_{x_i} (s_i),
\end{equation}
where the $M_{x_i} (s_i)$ are the moment generating functions of the independent variables.
Similarly, if $\boldsymbol{y} = \boldsymbol{A} \cdot \boldsymbol{x} + \boldsymbol{b}$ for matrix $\boldsymbol{A}$ and vector $\mathbf{b}$, then
\begin{equation}\label{eqn:moment_generating_function_linear_transform}
	M_{\boldsymbol{y}} (\boldsymbol{s})
	= e^{\boldsymbol{b}^T \cdot \mathbf{s}}
	M_{\boldsymbol{x}} (\boldsymbol{A}^T \cdot \boldsymbol{s}).
\end{equation}
We will make use of both of these results later.

A generalisation of the multivariate results can be used to treat sets of multivariate complex variables.
To do so it is convenient to use Wirtinger calculus, whereby we may treat $z_i$ and $z_i^*$ as independent variables.
Then we may define
\begin{equation}\label{eqn:def_complex_multivariate_moment_generating_function}
	M_{\boldsymbol{z}} (\mathbf{w}, \mathbf{w}^*) 
	= \int \int e^{
		\boldsymbol{w}^\dagger \cdot \boldsymbol{z}
		+ \boldsymbol{z}^\dagger \cdot \boldsymbol{w}
	}
	\prob (\boldsymbol{z}, \boldsymbol{z}^*)
	\, d^N \boldsymbol{z} d^N \boldsymbol{z}^*.
\end{equation}
The joint moments of the variables can then be calculated using
\begin{equation}
	\langle z^{\vphantom{*}}_i z^{\vphantom{*}}_j \dots z^{\vphantom{*}}_k
		z^*_l z^*_m \dots z^*_n \rangle
	= \left( \frac{\partial}{\partial w^*_i} \frac{\partial}{\partial w^*_j}
		\dots \frac{\partial}{\partial w^*_k}
		\frac{\partial}{\partial w_l} \frac{\partial}{\partial w_m}
		\dots \frac{\partial M_{\boldsymbol{z}}}{\partial w_n}
	\right)_{\boldsymbol{w}, \boldsymbol{w}^* =\boldsymbol{0}}
\end{equation}
where the Wirtinger derivatives are assumed.

We are now in a position to introduce moment generating functionals.
Let $x(t)$ be a real valued random process.
Assume $x(t)$ is drawn from the Hilbert space of finite, real-valued, functions on $-\infty < t < \infty$ with inner product
\begin{equation}\label{eqn:function_space_ip}
	f(t) \circ g(t) = \int_{-\infty}^{\infty} f(t) g(t) \, dt.
\end{equation}
If $\{h_n (t) | n = 1, 2, \dots\}$ is a set of orthonormal basis functions for the space, then $x(t)$ can be decomposed as
\begin{equation}\label{eqn:function_space_decomposition}
\begin{aligned}
	x(t) &= \sum_{n=1}^\infty x_n h_n(t) \\
	x_n &= h_n (t) \circ x(t)
\end{aligned}
\end{equation}
whereby `finite function' we mean $\sum_{n=1}^\infty |x_n| < \infty$.
The moment generating function is defined by assuming the decomposition coefficients $\{x_n | n = 1, 2, \dots \}$ are a set of jointly distributed random variables.
Substituting (\ref{eqn:function_space_decomposition}) into (\ref{eqn:def_moment_generating_function}), we obtain the moment generating functional
\begin{equation}\label{eqn:moment_generating_functional}
	\mathcal{M}_x [ s(t) ]
	= \int e^{s(t) \circ x(t)} \mathcal{P} [x(t)] \mathcal{D}[x(t)] 
\end{equation}
where $\mathcal{P}$ is the probability density functional of $x(t)$ and $\mathcal{D}$ is the functional integration measure.
The functional equivalent of the joint moments is the many time correlation functions, which are found by the operation
\begin{equation}\label{eqn:correlation_functions_from_mgf}
	\langle x(t_i) x(t_j) \dots x (t_k) \rangle = \left(
		\frac{\delta}{\delta s(t_i)}
		\frac{\delta}{\delta s(t_j)}
		\dots \frac{\delta \mathcal{M}_x}{\delta s(t_k)}
	\right)_{s(t_i), s(t_j), \dots, s(t_k) = 0}
\end{equation}
where $\delta / (\delta s(t_n))$ denotes the functional derivative with respect to $s(t_n)$.

To illustrate this procedure, let us consider a real-valued Gaussian process $x(t)$ with zero mean.
We start with the probability density functional.
Substituting $x_n$ into the probability density function of a zero-mean multivariate Gaussian, we obtain
\begin{equation}\label{eqn:pdf_gaussian}
	\mathcal{P}[ x(t) ]
	= \frac{1}{\sqrt{\det[\Gamma (t_1, t_2)]}}
	\exp \left( -\frac{1}{2} x(t_1) \circ \Gamma^{-1} (t_1, t_2) \circ x(t_2) \right),
\end{equation}
where
\begin{equation}\label{eqn:correlation_function}
	\Gamma (t_1, t_2) = \langle x(t_1) x(t_2) \rangle
	= \sum_{m=1}^\infty \sum_{n=1}^\infty \Sigma_{mn} x(t_1) x(t_2),
\end{equation}
is the autocorrelation function of $x(t)$, $\Gamma^{-1} (t_1, t_2)$ is the functional inverse defined by
\begin{equation}\label{eqn:inv_correlation_function}
	\int_{-\infty}^\infty \Gamma^{-1} (t_1, t_2) \Gamma(t_2, t_3) \, dt_2
	= \delta (t_1 - t_3)
\end{equation}
and we assume the factors of $1/\sqrt{2 \pi}$ associated with each dimension are absorbed into the integration measure.
(\ref{eqn:inv_correlation_function}) follows from the definition of the inverse covariance matrix as used in the probability density function: $\boldsymbol{\Sigma}^{-1} \cdot \boldsymbol{\Sigma} \cdot \boldsymbol{x} = \boldsymbol{x}$.
Similarly, the determinant of the correlation function can be defined by
\begin{equation}
	\det[\Gamma (t_1, t_2)] = \det(\boldsymbol{A})
\end{equation}
for matrix $\boldsymbol{A}$ with components $A_{mn} = h_m(t_1) \circ \Gamma(t_1, t_2) \circ h_n(t_2)$ in the limit the number of dimensions tends to infinity.
The moment generating functional is much easier to calculate and substituting $x_n$ as defined by (\ref{eqn:function_space_decomposition}) into (\ref{eqn:multivariate_gaussian_moment_generating_function}) for $\boldsymbol{\mu} = \boldsymbol{0}$ yields
\begin{equation}\label{eqn:mgf_gaussian}
	\mathcal{M}_x [s(t)]
	= e^{\frac{1}{2} s(t_1) \circ \Gamma(t_1, t_2) \circ s(t_2)}.
\end{equation}

As an aside, although we we will use moment generating functionals throughout, it is perfectly possible to carry out the calculations that follow using the discretised forms, then take the continuum limits once the moments have been calculated.
To do so the following set of substitutions can be be used.
Let
\begin{equation}\label{eqn:fvec_def}
	a_i = a(t_i)
\end{equation}
\begin{equation}\label{eqn:fmat_def}
	F_{ij} = f(t_i, t_j)
\end{equation}
and $t_{i+1} - t_{i} = \delta t$ for all $i$.
Then in the continuum limit we make the substitutions
\begin{equation}
	f_i \rightarrow f(t),
\end{equation}
\begin{equation}
	F_{ij}\rightarrow f(t_i, t_j),
\end{equation}
\begin{equation}
	\delta t \sum_{j} F_{ij} a_j \rightarrow \int f(t_i, t) a(t) \, dt,
\end{equation}
\begin{equation}
	\delta_{ij} / \delta t \rightarrow \delta (t_1, t_2),
\end{equation}
\begin{equation}
	F_{ij}^{-1} / (\delta t)^2 \rightarrow f^{-1} (t_1, t_2)
\end{equation}
and
\begin{equation}
	\det(\delta t \boldsymbol{F}) \rightarrow \det(f(t_1, t_2)),
\end{equation}
where $f^{-1}$ is defined by the relationship
\begin{equation}
	\iint f^{-1} (t_1, t_2) f(t_2, t_3) g(t_3) \, dt_2 dt_3 = g(t_1)
\end{equation}
for all possible $g(t)$ in the problem considered.

\subsection{Moment generating functional for the absorbed power}\label{sec:sa_pow_mgf}

The power detector, labelled $P$ in Figure \ref{fig:single_detector_noise_model}, converts the input electromagnetic signal, $a(t)$, into a $p(t)$, where the expected value of $p(t)$ is proportional to the total incident power.
Detectors of this type respond to absorbed photons; for example a TES responds to the change in temperature due to the energy deposited by the photon, while in an STJ the absorption of a photon allows charge to tunnel across the junction.
Detection is therefore, fundamentally, a quantum process and $p(t)$ is an intrinsically noisy.

We are interested primarily of radiation of thermal origin and a full quantum calculation shows the noise in the power of such a signal comprises two terms.
The first term, usually referred to as photon shot noise, results from the randomness of the arrival times of the photons.
The second, usually referred to as wave- or bunching-noise, is normally interpreted as arising from thermal fluctuations in the emission rate of the source.
Shot noise dominates for low arrival rates and typically determines the sensitivity of optical detectors.
Wave noise dominates in the classical limit when the photon arrival rate is high and typically determines the sensitivity of microwave radiometers.
Because both low- and high-rate signals (signal and pump) are present simultaneously in the homodyne scheme, we must account for both types of noise.
Note that it is not the signal of interest that determines the arrival rate, but the \emph{total} signal; the regime a detector operates in is therefore usually determined by the background.

As an alternative to the full quantum mechanical analysis, $p(t)$ can instead be calculated approximately using a semi-classical approach\,\cite{zmuidzinas2015use,saklatvala}.
In this approach the absorption of photons is treated as a \emph{nonhomogenous} Poisson process for which the instantaneous arrival rate is proportional to the instantaneous classical incident power.
A nonhomogenous Poisson process is a generalisation of a Poisson process in which the arrival rate is allowed to vary in time.
If the classical incident power is itself a stochastic signal, then strictly the ultimate signal has a mixed distribution; this is the origin of the approach's other name, Poisson-mixture model\,\cite{saklatvala}.
In such a mixed model, the photon shot noise arises from the Poisson statistics and the wave noise from the statistics of the rate.
The semi-classical approach has been shown to be completely equivalent to the full calculation in certain limits, such as the case of black body radiation\,\cite{sudarshan1963equivalence} and more generally where the classical statistics of the source are Gaussian, and can be viewed as a complementary physical picture\,\cite{zmuidzinas2015use}. 
However, the semi-classical approach results is algebraically simpler and we will adopt it here.

Let us formalise the preceding discussion.
Let $N(t)$ denote the total number of photons absorbed in the detector since some start time $t = t_s$ for which $N(t_s) = 0$.
Since we are only considering quasimonochromatic sources, we can approximate each photon as depositing energy $h \nu_0$, so the total energy $U$ absorbed in the detector up to time $t$ is
\begin{equation}\label{eqn:total_abs_energy_from_n}
	U = h \nu_0 N(t).
\end{equation}
Classically we would expect
\begin{equation}\label{eqn:total_abs_energy_from_p}
	U = \int_{t_s}^t p(t) \, dt
\end{equation}
and therefore we may make the identification
\begin{equation}\label{eqn:p_from_n}
	p(t) = h \nu_0 \frac{dN}{dt}.
\end{equation}
According to the semiclassical approach $N(t)$ is a nonhomogenous process with time variable rate $\lambda(t)$.
We expect
\begin{equation}\label{eqn:lambda_as_exp_dNdt}
	\lambda (t) = \left< \frac{dN}{dt} \right> = \frac{\langle p(t) \rangle}{h \nu_0}
\end{equation}
and therefore for the detector we can write
\begin{equation}\label{eqn:lambda_from_a}
	\lambda (t) = \frac{1}{2 h \nu_0} |a(t)|^2,
\end{equation}
which links the photon arrival rate to the input signal.

Given our knowledge of the statistics of $N(t)$, we can now calculate the statistics of $p(t)$.
The MGF of $p(t)$ on $t_s < t < t_f$ is the expected value of the exponent of
\begin{equation}\label{eqn:ip_s_p}
	\int_{t_s}^{t_f} s(t) p(t) dt.
\end{equation}
Later on, we will let the domain of integration tend to all time, but for now it is simplest to consider a finite domain.
(\ref{eqn:ip_s_p}) may be approximated as a discrete sum
\begin{equation}\label{eqn:ip_s_p_discretised}
	\int_{t_s}^{t_f} s(t) p(t) dt
	\approx \sum_{n=1}^N s(t_n) \int_{t_n}^{t_n+\delta t} p(t) dt.
\end{equation}
Using (\ref{eqn:p_from_n}) to substitute for the power and evaluating the integrals, we can simplify
\begin{equation}\label{eqn:ip_s_p_discretised_in_n}
	\int_{t_s}^{t_f} s(t) p(t) dt
	\approx h \nu_0 \sum_{m=1}^N s(t_m) n_m
\end{equation}
where the
\begin{equation}\label{eqn:def_increment}
	n_m = N(t_m + \delta) - N(t_m)
\end{equation}
constitute what is known as a set of increments of the Poisson process.
It follows that the MGF of $p(t)$ can be found as the continuum limit of the moment generating function
\begin{equation}\label{eqn:discrete_p_mgf}
	M_{\boldsymbol{p}} (\boldsymbol{s}) = M_{\mathbf{n}} (h \nu_0 \boldsymbol{s}),
\end{equation}
where $M_{\boldsymbol{n}} (\boldsymbol{s})$ is the moment generating function of the vector of increments $\boldsymbol{n}$.

From the earlier discussion, we expect $\boldsymbol{n}$ to depend on the values $\{a(t_n)\}$ of the signal over the increments, which we represent with the vector $\boldsymbol{a}$.
In the general case where $a(t)$ is a stochastic signal, we therefore have
\begin{equation}
	M_{\boldsymbol{n}} (\boldsymbol{s})
	= \frac{1}{i} \iiint e^{\boldsymbol{s} \cdot \boldsymbol{n}}
		\Pr (\boldsymbol{n} | \boldsymbol{a}, \boldsymbol{a}^*)
		\Pr (\boldsymbol{a}, \boldsymbol{a}^*)
	\, d^N \boldsymbol{n}\,  d^N \boldsymbol{a} \, d^N \boldsymbol{a}^*,
\end{equation}
where $\Pr (\boldsymbol{a}, \boldsymbol{a}^*)$ is the probability density of $\boldsymbol{a}$ (remembering it is complex).
We may rewrite this as
\begin{equation}\label{eqn:mgf_n_full}
	M_{\boldsymbol{n}} (\boldsymbol{s})
	= \frac{1}{i} \iint e^{\boldsymbol{s} \cdot \boldsymbol{n}}
		M_{\boldsymbol{p}} (\boldsymbol{s} | \boldsymbol{a}, \boldsymbol{a}^*)
		\Pr (\boldsymbol{a}, \boldsymbol{a}^*)
	\, d^N \boldsymbol{a} \, d^N \boldsymbol{a}^*.
\end{equation}
where 
\begin{equation}\label{eqn:mgf_n_given_a}
	M_{\boldsymbol{n}} (\boldsymbol{s} | \boldsymbol{a}, \boldsymbol{a}^*)
	= \int e^{\boldsymbol{s} \cdot \boldsymbol{n}}
		\Pr (\boldsymbol{n} | \boldsymbol{a}, \boldsymbol{a}^*)
	\, d^N \boldsymbol{n}
\end{equation}
is the moment generating function of $\boldsymbol{n}$ given $\boldsymbol{a}$.
We see that calculating $M_{\boldsymbol{n}} (\boldsymbol{n})$ comprises two steps:
1) Calculating $M_{\boldsymbol{p}} (\boldsymbol{s} | \boldsymbol{a}, \boldsymbol{a}^*)$, which is signal independent.
2) Integrating this with respect to the probability density of the signal of interest.

The moment generating function of the increments given $\boldsymbol{a}$ can be calculating using the defining characteristics of a nonhomogenous Poisson process.
For $N(t)$ to be a non-homogenous poisson process with rate $\lambda (t)$, the following conditions must be satisfied\,\cite{nhp}:
\begin{enumerate}
\item $N(t_s) = 0$.
\item For all $t_s \leq t_1 \leq t_2 \leq t_3 \dots \leq t_n$, the random variables $N(t_2) - N(t_1)$, $N(t_3 ) - N(t_2)$, $\dots$, $N(t_n) - N(t_{n-1})$, called increments, are independent.
\item For any $t$ satisfying $t_s < t < \infty$, then
\begin{equation}
\begin{aligned}
	\Pr(N(t + \delta t) - N(t) = 0) = 0) &= 1 - \lambda(t) \delta t + o(\delta t) \\
	\Pr(N(t + \delta t) - N(t) = 0) = 1) &= \lambda(t) \delta t + o(\delta t) \\
	\Pr(N(t + \delta t) - N(t) = 0) = 2) &=  o(\delta t),
\end{aligned}
\end{equation}
using little-o notation.
\end{enumerate}
Condition 2) implies the overall moment generating function is simply the product of the moment generating functions of the individual increments.
Condition 3) in turn implies the moment generating function of a single increment is
\begin{equation}\label{eqn:mgf_increment}
\begin{aligned}
	M_{n_i} (s_i) &= 1 + (e^{s_i} - 1) \lambda(t) \delta t + o(\delta t) \\
	&= e^{ (e^{s_i} - 1) \lambda(t) \delta t} + o(\delta t),
\end{aligned} 
\end{equation}
Comparing (\ref{eqn:mgf_increment}) and (\ref{eqn:poisson_moment_generating_function}), we see sufficiently small $\delta t$ each increment is Poisson distributed with rate $\lambda(t) \delta t$.
We could also argue this on the basis that provided $\delta t$ is sufficiently small that the rate does not vary significantly over an increment then the arrivals over that increment should be Poisson distributed.
Hence we obtain
\begin{equation}\label{eqn:mgf_all_increments}
	M_\mathbf{n} (\mathbf{s}) = \prod_{i=1}^N e^{ (e^{s_i} - 1) \lambda(t_i) \delta t}
\end{equation}
Using (\ref{eqn:lambda_from_a}) to substitute for the rate, we therefore find
\begin{equation}\label{eqn:discrete_p_mgf_given_a}
	M_{\boldsymbol{n}} (\boldsymbol{s} | \boldsymbol{a}) =
	\exp \Bigl( 
		\tfrac{1}{2 h \nu_0} \boldsymbol{a}^\dagger \cdot
		(e^{\boldsymbol{S}} - \boldsymbol{I}) \cdot
		\boldsymbol{a} \, \delta t
	\Bigr)
\end{equation}
where
\begin{equation}\label{eqn:def_s_matrix}
	S_{ij} = s(t_i) \delta_{ij}
\end{equation}
and the exponential of a matrix is interpreted in the normal sense.

If $a(t)$ is deterministic, (\ref{eqn:discrete_p_mgf}) is all that is needed.
Taking the continuum limit, we find
\begin{equation}\label{eqn:mgf_p_deterministic}
	\mathcal{M}_{p} [s] =
	\exp \biggl( 
		\frac{1}{2 h \nu_0} \int_{t_f}^{t_s}
		(e^{h \nu_0 s(t) } - 1)
		|a(t)|^2 \, dt
	\biggr).
\end{equation}
Although the classical power limit is deterministic, noise still arises from the photon absorption process.

To illustrate the full calculation when $a(t)$ is a stochastic signal, we will consider the case where $a(t)$ is a zero mean stationary Gaussian process.
This is relevant to the analysis of the filterbank spectrometer and is representative of the real world noise signals of interest.
In this case
\begin{equation}\label{eqn:pr_a_gaussian}
	\text{Pr} \bigl[ \boldsymbol{a}, \boldsymbol{a}^* \bigr]
	= \frac{1}{\det(2 \pi \boldsymbol{\Sigma})}
	\exp \biggl( -\boldsymbol{a}^\dagger
		\cdot \boldsymbol{\Sigma}^{-1} \cdot
		\boldsymbol{a}
	\biggr)
\end{equation}
where 
\begin{equation}\label{eqn:covar_matrix_from_correlation_function}
	\Sigma_{ij} = \Gamma (t_i - t_j) = \langle a(t_i) a^*(t_j) \rangle
\end{equation}
and $\Gamma(t_1 - t_2)$ is the autocorrelation function of the process.
A zero mean Gaussian process is fully characterised by its correlation function.
If the input signal is generated by several independent sources of this nature, then $\boldsymbol{\Sigma} = \sum_n \boldsymbol{\Sigma}_n$, where $\boldsymbol{\Sigma}_n$ is the covariance matrix of the $n^\text{th}$ source.

The full moment generating function is calculated by substituting (\ref{eqn:discrete_p_mgf_given_a}) and (\ref{eqn:pr_a_gaussian}) into (\ref{eqn:mgf_n_full}) and evaluating the integral.
The latter step can be carried out conveniently using the identity
\begin{equation}\label{eqm:gaussian_integral_identity}
\begin{aligned}
	&\frac{1}{\det(\boldsymbol{\Sigma})} \int
		\exp \left(
			\boldsymbol{z}^\dagger \cdot \boldsymbol{A} \cdot \boldsymbol{z}
			- \boldsymbol{b}^\dagger \cdot \boldsymbol{z}
			- \boldsymbol{z}^\dagger \cdot \boldsymbol{b}
			- (\boldsymbol{z} - \boldsymbol{\mu})^\dagger
			\cdot \boldsymbol{\Sigma}^{-1} \cdot (\boldsymbol{z} - \boldsymbol{\mu})
		\right)
	\, \prod_n \frac{dz^{\vphantom *}_n dz^*_n}{2 \pi i} \\
	&= \frac{1}{\det(\boldsymbol{I} - \boldsymbol{\Sigma} \cdot \boldsymbol{A})}
	\exp \left(
		- \boldsymbol{\mu}^\dagger \cdot \boldsymbol{\Sigma}^{-1} \cdot \boldsymbol{\mu}
		+ (\boldsymbol{\mu} - \boldsymbol{\Sigma} \cdot \boldsymbol{b})^\dagger
		\cdot \boldsymbol{\Sigma}^{-1}
		\cdot (\boldsymbol{I} - \boldsymbol{\Sigma} \cdot \boldsymbol{A})^{-1}
		\cdot (\boldsymbol{\mu} - \boldsymbol{\Sigma} \cdot \boldsymbol{b})
	\right). 
\end{aligned}
\end{equation}
which is proved in Appendix \ref{sec:gaussian_integrals}.
Doing so yields
\begin{equation}
\begin{aligned}
	M_{\boldsymbol{n}} (\boldsymbol{s}) 
	&= \frac{1}{\det(\boldsymbol{I} - \delta t \boldsymbol{\Sigma} \cdot
	\{ \exp(\boldsymbol{S}) - \boldsymbol{I} \} / (2 h \nu_0))}.
\end{aligned}
\end{equation}
Taking the continuum limit, we then find
\begin{equation}\label{eqn:mgf_p_zero_mean_gaussian_noise}
	\mathcal{M}_{p} [s(t)]
	= \frac{1}{\det[
		\delta (t_1 - t_2) - 
		\Gamma (t_1, t_2) \{ e^{h \nu_0 s(t_2)} - 1 \} / (2 h \nu_0)
	]}.
\end{equation}

\subsection{Accounting for emission noise}\label{sec:emission_noise}

Strictly, (\ref{eqn:p_from_n}) should be
\begin{equation}\label{eqn:p_from_n_det_emission}
	p(t) = h \nu_0 \left\{ \frac{dN}{dt} - \frac{dN_e}{dt} \right\}
\end{equation}
where $N$ is as before and $N_e$ is cumulative number of photons emitted from the detector.
This assumes the detector is also limited to monochromatic emission, which is the case if the input is filtered.
In most cases $N \gg N_e$ and $N_e$ can be ignored.
However, this is not true if the detector and source are at nearly the same temperature, as might be the case for an axion cavity haloscope.
As such, we should consider how the results obtained so far are modified in the presence of emission.

Assume the output signal produced by the detector input is $b(t)$.
We expect $a(t)$ and $b(t)$ to be uncorrelated.
Repeating the analysis of Section \ref{sec:sa_pow_mgf}, we then find (\ref{eqn:mgf_p_deterministic}) and (\ref{eqn:mgf_p_zero_mean_gaussian_noise_a_and_b}) are modified to
\begin{equation}\label{eqn:mgf_p_deterministic_a_and_b}
\begin{aligned}
	\mathcal{M}_{p} [s] &=
	\exp \biggl( 
		\frac{1}{2 h \nu_0} \int_{t_f}^{t_s}
		(e^{h \nu_0 s(t) } - 1)
		|a(t)|^2 \, dt
	\biggr) \\
	&\times \exp \biggl( 
		\frac{1}{2 h \nu_0} \int_{t_f}^{t_s}
		(e^{-h \nu_0 s(t) } - 1)
		|b(t)|^2 \, dt
	\biggr)
\end{aligned}
\end{equation}
and
\begin{equation}\label{eqn:mgf_p_zero_mean_gaussian_noise_a_and_b}
\begin{aligned}
	\mathcal{M}_{p} [s(t)]
	&= \frac{1}{\det[
		\delta (t_1 - t_2) - 
		\Gamma_a (t_1, t_2) \{ e^{h \nu_0 s(t_2)} - 1 \} / (2 h \nu_0)
	]} \\
	&\times \frac{1}{\det[
		\delta (t_1 - t_2) - 
		\Gamma_b (t_1, t_2) \{ e^{-h \nu_0 s(t_2)} - 1 \} / (2 h \nu_0)
	]} 
\end{aligned}
\end{equation}
where $\Gamma_a (t_1, t_2)$ is the correlation function of the input signal and $\Gamma_b (t_1, t_2)$ is the correlation function of the output signal emitted by the detector.
We will not use these modified expressions in general, however we will revisit them in Section \ref{sec:psd_fb} when we consider the output power spectral noise density of a single filter-bank channel.

\subsection{Moment generating functional for the output of a single detector}\label{sec:mgf_sd}

We now have all the terms we need to calculate the moment generating functional of the output of the detector, $y(t)$, in Figure \ref{fig:single_detector_noise_model}.
The power signal $p(t)$ and noise signal $n(t)$ are statistically independent, so using the continuum limit of (\ref{eqn:moment_generating_function_linear_transform}) we have
\begin{equation}\label{eqn:mgf_y_from_mgf_n_and_mgf_p}
	\mathcal{M}_y [s(t)] = 
	\mathcal{M}_n \biggl[ \int_{-\infty}^\infty g(t', t) s(t') \, dt' \biggr]
	\mathcal{M}_p \biggl[ \int_{-\infty}^\infty g(t', t) s(t') \, dt' \biggr].
\end{equation}
The moment generating of $p(t)$ is given by (\ref{eqn:mgf_p_zero_mean_gaussian_noise}).
We will assume $n(t)$ is a zero mean Gaussian process, in which case
\begin{equation}\label{eqn:mgf_n}
	\mathcal{M}_n [s(t)] = 
	\exp \biggl( \frac{1}{2} \int_{-\infty}^\infty
		\Gamma_n (t_1, t_2) s(t_1) s(t_2)
	\, dt_1 dt_2 \biggr)
\end{equation}
using (\ref{eqn:mgf_gaussian}), where $\Gamma_n (t_1, t_2)$ is the time domain correlation function.
Combining these results we have
\begin{equation}\label{eqn:mgf_y_single_detector}
\begin{aligned}
	\mathcal{M}_y [s(t)] &= 
	\frac{1}{\det[
		\delta (t_1 - t_2) - 
		\Gamma (t_1, t_2) \{ e^{h \nu_0 \int_{-\infty}^{\infty}
			g(t_3, t_2) s(t_3)
			\, dt_3} - 1 \} / (2 h \nu_0)
	]} \\
	& \exp \biggl( \frac{1}{2} \iiint_{-\infty}^\infty
		\Gamma_n (t_1, t_2)  g(t_3, t_1) g(t_4, t_2) s(t_3) s(t_4)
	\, dt_1 dt_2 dt_3 dt_4 \biggr).
\end{aligned}
\end{equation}
This is the appropriate moment generating functional in the case of the filter bank spectrometer.

\subsection{Moment generating functional for the homodyne output}\label{sec:mgf_homo}

\begin{figure}
\centering
\includegraphics{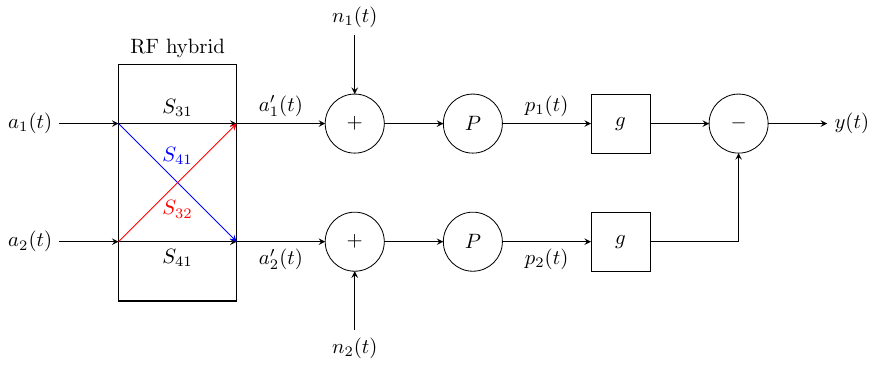}
\caption{\label{fig:homodyne_system_diagram}
Homodyne system noise model.
}
\end{figure}

The calculation of the moment generating function for the homodyne scheme uses many of the tools we have already developed, but they need to be applied in a slightly different way.
For simplicity we assume the ideal system, but will discuss how the results can be extended to the non-ideal case throughout.

Consider Figure \ref{fig:homodyne_system_diagram}, which shows the homodyne scheme in the same notation as Figure \ref{fig:single_detector_noise_model}.
We assume the output hybrid is ideal and the detector response functions are identical, in which case
\begin{equation}\label{eqn:output_homodyne_basic}
	y(t) = \int_{-\infty}^{\infty} g(t, t') \{ d(t') + n_1 (t') - n_2 (t') \}.
\end{equation}
where we call
\begin{equation}\label{eqn:def_difference_signal}
	d(t) = p_1 (t) - p_2(t).
\end{equation}
the difference signal.
We could, for example, guarantee this identical behaviour by logging the detector outputs directly and then synthesizing the response functions and hybrid in software during post processing.
The difference signal and noise signals from the two detectors must all be statistically independent as they arise from different physical sources, so the MGF of $y(t)$ is simply the product of those of the three functions.
Since we know the MGF of $n_1 (t)$ and $n_2(t)$ from previously, the problem is one of calculating the MGF of the difference signal.

We will assume the detection process (i.e. absorption of photons from the signal) in the two detectors is independent.
On this basis, we might expect the MGF of the difference signal to simply be the appropriate linear combination of the MGFs of $p_1 (t)$ and $p_1(t)$.
However, the inputs of detector 1 and detector 2 both depend on both $a_1(t)$ and $a_2(t)$, which means the fluctuations in arrival rates at each detector are not independent and therefore neither are $p_1 (t)$ and $p_2(t)$.
Instead we must calculate the moment generating function of $d(t)$ given $a_1 (t)$ and $a_2 (t)$ assuming detection is independent, then integrate over the probable values of the signals.
Explicitly, we have
\begin{equation}\label{eqn:mgf_d_from_conditional_mgf_and_p}
	\mathcal{M}_d [ s(t) ]
	= \iint \mathcal{M}_d [ s(t) | a_1(t), a_2 (t) ] \Pr [a_1 (t), a_2(t)]
	\, \mathcal{D}[a_1(t)] \, \mathcal{D}[a_2(t)]. 
\end{equation}
where we will call $\mathcal{M}_d [ s(t) | a_1(t), a_2 (t) ]$ the conditional MGF.

The conditional MGF can be calculated using (\ref{eqn:mgf_p_deterministic}) in combination with knowledge of the input signals to the detectors, which we have denoted by primed $a'_1(t)$ and $a'_2(t)$ in Figure \ref{fig:homodyne_system_diagram}.
Under the assumption of independent detection processes and using the continuum limits of (\ref{eqn:moment_generating_function_linear_transform}), we have
\begin{equation}\label{eqn:cond_mgf_d_det_inputs_full}
\begin{aligned}
	\mathcal{M}_d [ s(t) | a_1(t), a_2 (t) ]
	&= \exp \biggl( 
		\frac{1}{2 h \nu_0} \int_{-\infty}^{\infty}
		(e^{h \nu_0 s(t) } - 1) |a'_1 (t)|^2
		+ (e^{-h \nu_0 s(t) } - 1) |a'_2 (t)|^2 \, dt
	\biggr) \\
	&= \exp \biggl( 
		\frac{1}{2 h \nu_0} \int_{-\infty}^{\infty}
		\{ \cosh(h \nu_0 s(t))  - 1 \} \{ |a'_1 (t)|^2 + |a'_2 (t)|^2 \} \, dt
	\biggr) \\
	&\times \exp \biggl( 
		\frac{1}{2 h \nu_0} \int_{-\infty}^{\infty}
		\sinh(h \nu_0 s(t) \{ |a'_1 (t)|^2 - |a'_2 (t)|^2 \} \, dt
	\biggr) .
\end{aligned}
\end{equation}
In general,
\begin{equation}\label{eqn:hybrid_outputs_non_ideal}
\begin{aligned}
	a'_1 (t) &= S_{31} a_1 (t) + S_{32} a_2 (t) \\
	a'_2 (t) &= S_{41} a_1 (t) + S_{42} a_2 (t)
\end{aligned}
\end{equation}
where $S_{mn}$ is the scattering parameter between ports $m$ and $n$ of the hybrid at $\nu_0$ and these results can be carried through the analysis to model the non-ideal case.
However, for an ideal hybrid we may simplify
\begin{equation}\label{eqn:hybrid_outputs_ideal}
\begin{aligned}
	a'_1 (t) &= \frac{1}{\sqrt{2}} \{ a_1 (t) + a_2 (t) \} \\
	a'_2 (t) &= \frac{1}{\sqrt{2}} \{ a_1 (t) - a_2 (t) \}.
\end{aligned}
\end{equation}
Hence
\begin{equation}\label{eqn:cond_mgf_d_det_inputs}
\begin{aligned}
	\mathcal{M}_d [ s(t) | a_1(t), a_2 (t) ]
	&= \exp \biggl( 
		\frac{1}{2 h \nu_0} \int_{-\infty}^{\infty}
		\{ \cosh(h \nu_0 s(t))  - 1 \} \{ |a_1 (t)|^2 + |a_2 (t)|^2 \} \, dt
	\biggr) \\
	&\times \exp \biggl( 
		\frac{1}{ h \nu_0} \int_{-\infty}^{\infty}
		\sinh(h \nu_0 s(t) \Re[ a^{\vphantom *}_1 (t) a^*_2 (t) ] \, dt
	\biggr) .
\end{aligned}
\end{equation}

Now we carry out the integral over the probable input signals.
We will assume the signal at the first input is zero mean Gaussian noise and that at the second is a deterministic pump signal.
Denoting the pump signal as
\begin{equation}\label{eqn:pump_signal}
	a_2 (t) = a_p e^{2 \pi i \vp t},
\end{equation}
where $a_p$ is the pump amplitude and $\vp$ is the pump frequency, we can then simplify the conditional MGF to
\begin{equation}\label{eqn:cond_mgf_d_det_inputs_a2_known}
\begin{aligned}
	&\mathcal{M}_d [ s(t) | a_1(t), a_2 (t) ]
	= \mathcal{M}_d [ s(t) | a_1(t) ] \\
	&= \exp \biggl( 
		\frac{1}{2 h \nu_0} \int_{-\infty}^{\infty}
		\{ \cosh(h \nu_0 s(t))  - 1 \} \{ |a_1 (t)|^2 + |a_p|^2 \} \, dt
	\biggr) \\
	&\times \exp \biggl( 
		\frac{1}{ h \nu_0} \int_{-\infty}^{\infty}
			\sinh( h \nu_0 s(t) )
			\Re \Bigl[ a^*_p \, a^{\vphantom *}_1 (t) e^{-2 \pi i \vp t} \Bigr]
	\, dt \biggr) .
\end{aligned}
\end{equation}
Inserting (\ref{eqn:cond_mgf_d_det_inputs_a2_known}) into (\ref{eqn:mgf_d_from_conditional_mgf_and_p}), assuming $\Pr[a_1]$ is that for a Gaussian process, and then integrating using the continuum limit of (\ref{eqm:gaussian_integral_identity}), we find the MGF functional can be written as the product of three functionals
\begin{equation}\label{eqn:mgf_d_as_a_product}
	\mathcal{M}_d [s(t)] =
	\mathcal{M}_1 [s(t)] \mathcal{M}_2 [s(t)] \mathcal{M}_3 [s(t)],
\end{equation}
where
\begin{align}\label{eqn:def_m1_m2_m3}
&\mathcal{M}_1 [s(t)] = \exp \biggl( 
		\frac{1}{2 h \nu_0} |a_p|^2 \int_{-\infty}^{\infty}
		\{ \cosh(h \nu_0 s(t))  - 1 \} \, dt
	\biggr), \\
&\mathcal{M}_2 [s(t)] = \frac{1}{\det \bigl[k(t_1, t_2) \bigr]}, \\
&\mathcal{M}_3 [s(t)] \\
	&= \exp \biggl( \frac{|a_p|^2}{(2 h \nu_0)^2}
		\iint_{-\infty}^{\infty} k^{-1} (t_1, t_2) \Gamma (t_2, t_3)
		\sinh (h \nu_0 s(t_1)) \sinh (h \nu_0 s(t_3))
		e^{- 2 \pi i \vp (t_1 - t_3)}
		\, dt_1 dt_2 dt_3
	\biggr),
\end{align}
and
\begin{equation}\label{eqn:def_k_operator}
	k(t_1, t_2) = \delta(t_1 - t_2) - \Gamma(t_1, t_2) \{ \cosh(h \nu_0 s(t_2))  - 1 \} / (2 h \nu_0).
\end{equation}
$\mathcal{M}_1$ is associated with the shot noise in the detectors due to the pump, $\mathcal{M}_2$ with the wave and shot noise of the input signal and $\mathcal{M}_3$ with the mixing action of the detector.

The MGF of the output signal follows by combining the results for $d(t)$ with those for the internal noise in the detectors.
The final result is
\begin{equation}\label{eqn:mgf_final_homodyne}
	\mathcal{M}_y [s(t)]
	= \prod_{n=1}^5 \mathcal{M}_{n} [u(t)] 
\end{equation}
where
\begin{equation}\label{eqn:def_u}
	u(t_1) = \int_{-\infty}^{\infty} g(t_2, t_1) s(t_2) \, dt_2,
\end{equation}
\begin{equation}\label{eqn:def_m4}
	\mathcal{M}_4 [s(t)] 
	= \exp \biggl( \frac{1}{2} \int_{-\infty}^\infty
		\Gamma_{n_1} (t_1, t_2) s(t_1) s(t_2),
	\, dt_1 dt_2 \biggr)
\end{equation}
\begin{equation}\label{eqn:def_m5}
	\mathcal{M}_5 [s(t)] 
	= \exp \biggl( \frac{1}{2} \int_{-\infty}^\infty
		\Gamma_{n_2} (t_1, t_2) s(t_1) s(t_2)
	\, dt_1 dt_2 \biggr)
\end{equation}
and $\Gamma_{n_1} (t_1, t_2)$ and $\Gamma_{n_2} (t_1, t_2)$ are the autocorrelation functions of the internal noise in detector's 1 and 2.

\subsection{Time domain correlation functions}\label{sec:mgf_td}

In this section we will use the MGFs just derived to calculate the correlation functions of the output signal for each receiver architecture.
To do so we will make extensive use of the following functional derivatives, which are derived in Appendix \ref{sec:useful_functional_derivatives}:
\begin{equation}\label{eqn:deriv_functional_determinant}
	\frac{\delta \det \bigl[ f(t_2, t_3) \bigr]}{\delta s(t_1)}
	= \det \bigl[ f(t_2, t_3) \bigr] \iint
		f^{-1} (t_3, t_2)
		\frac{\delta  f(t_2, t_3)}{\delta s(t_1)}
	\, dt_2 dt_3
\end{equation}
\begin{equation}\label{eqn:deriv_functional_inverse}
	\frac{\delta f^{-1} (t_2, t_3)}{\delta s(t_1)}
	= -\iint f^{-1} (t_2, t_4)
		\frac{\delta f (t_4, t_5)}{\delta s(t_1)}
		 f^{-1} (t_5, t_2)
	\, dt_4 dt_5.
\end{equation}

\subsubsection{Filterbank spectrometer output correlation functions}\label{sec:mgf_td_fb}

In the case of the filterbank spectrometer, to calculate the sensitivity we will need the expected value $\langle y(t) \rangle$ of the output signal and the two-time correlation function of the fluctuations $\Delta y (t) = y(t) - \langle y(t) \rangle$ in the output around this value.
The latter can be calculated from moments of $y(t)$ using the relation
\begin{equation}\label{eqn:fb_fluctuation_correlation_function}
	\langle \Delta y(t_1) \Delta y(t_2) \rangle
	= \langle y(t_1) y(t_2) \rangle - \langle y(t_1) \rangle \langle y(t_2) \rangle.
\end{equation}

It follows from (\ref{eqn:correlation_functions_from_mgf}) that we can find the expected value of the signal by taking the functional derivative of the moment generating functional (\ref{eqn:mgf_y_single_detector}) and then setting $s(t) = 0$.
Using (\ref{eqn:deriv_functional_determinant}), the derivative evaluates to
\begin{equation}\label{eqn:first_deriv_mgf_fb}
\begin{aligned}
	\frac{\delta \mathcal{M} [s(t)]}{\delta s(t_1)} &= \biggl\{ 
	\frac{1}{2} \iint_{-\infty}^\infty h^{-1} (t'_2, t'_1)
		\Gamma (t'_1, t'_2) g(t_1, t'_2)
		e^{h \nu_0 \int_{-\infty}^{\infty} g(t'_3, t'_2) s(t'_3) \, dt'_3}
	\, dt'_1 dt'_2 \\
	&+ \frac{1}{2} \iint_{-\infty}^\infty
		\bigl\{ \Gamma_n (t'_1, t'_2) + \Gamma_n (t'_2, t'_1) \bigr\}
		g(t_1, t'_1) g(t'_3, t'_2) s(t'_3)
	\, dt'_1 dt'_2 dt'_3
	\biggr\} \mathcal{M} [s(t)],
\end{aligned}
\end{equation}
where
\begin{equation}\label{eqn:mgf_fb_def_h}
	h(t_1, t_2) = 
	\delta (t_1 - t_2) -  \Gamma (t_1, t_2) \{ e^{h \nu_0 \int_{-\infty}^{\infty}
	g(t_3, t_2) s(t_3) \, dt_3} - 1 \} / (2 h \nu_0)
\end{equation}
and we have relabelled some integration variables to obtain the result as presented.
The detector noise signal $n(t)$ is real, so $\Gamma_n (t_1, t_2) = \Gamma_n (t_2, t_1)$ by definition.
Hence we may simplify
\begin{equation}\label{eqn:first_deriv_mgf_fb_v2}
\begin{aligned}
	\frac{\delta \mathcal{M} [s(t)]}{\delta s(t_1)} &= \biggl\{ 
	\frac{1}{2} \iint_{-\infty}^\infty h^{-1} (t'_2, t'_1)
		\Gamma (t'_1, t'_2) g(t_1, t'_2)
		e^{h \nu_0 \int_{-\infty}^{\infty} g(t'_3, t'_2) s(t'_3) \, dt'_3}
	\, dt'_1 dt'_2 \\
	&+ \iint_{-\infty}^\infty
		\Gamma_n (t'_1, t'_2) g(t_1, t'_1) g(t'_3, t'_2) s(t'_3)
	\, dt'_1 dt'_2 dt'_3
	\biggr\} \mathcal{M} [s(t)]
\end{aligned}
\end{equation}
To obtain $\langle y(t_1) \rangle$ we set $s(t) = 0$ in (\ref{eqn:first_deriv_mgf_fb_v2}).
$h^{-1} (t_1, t_2) = \delta(t_1 - t_2)$ when $s(t) = 0$, so we obtain
\begin{equation}\label{eqn:expect_y_fb}
	\langle y(t) \rangle =
	\frac{1}{2 } \int_{-\infty}^\infty
		g(t, t') \Gamma (t', t')
	\, dt'.
\end{equation}

To find the fluctuation correlation function, we start by calculating $\langle y(t_1) y(t_2) \rangle$ and to do so we need the second functional derivative of (\ref{eqn:mgf_y_single_detector}).
This follows by using differentiating (\ref{eqn:first_deriv_mgf_fb_v2}), where (\ref{eqn:deriv_functional_inverse}) can be used to differentiate $h^{-1} (t_1, t_2)$.
The result is
\begin{equation}\label{eqn:second_deriv_mgf_fb}
\begin{aligned}
	&\frac{\delta^2 \mathcal{M} [s(t)]}{\delta s(t_1) \delta s(t_2)} = \biggl\{
	\frac{1}{2} h \nu_0 \iint_{-\infty}^\infty h^{-1} (t'_2, t'_1)
		\Gamma (t'_1, t'_2) g(t_1, t'_2) g(t_2, t'_2)
		e^{h \nu_0 \int_{-\infty}^{\infty} g(t'_3, t'_2) s(t'_3) \, dt'_3}
	\, dt'_1 dt'_2 \\
	&+ \frac{1}{4} \iiiint_{-\infty}^\infty
		h^{-1} (t'_2, t'_4) h^{-1} (t'_5, t'_1) g(t_1, t'_5) g(t_1, t'_2)
		\Gamma (t'_4, t'_5) \Gamma (t'_1, t'_2)  \\
	& \quad \times e^{h \nu_0 \int_{-\infty}^{\infty} g(t'_3, t'_2) s(t'_3) \, dt'_3}
		e^{h \nu_0 \int_{-\infty}^{\infty} g(t'_6, t'_5) s(t'_6) \, dt'_6}
	\, dt'_1 dt'_2 dt'_4 dt'_5 \\
	&+ \iint_{-\infty}^\infty
		g(t_1, t'_1) g(t_2, t'_2) \Gamma_n (t'_1, t'_2) 
	\, dt'_1 dt'_2
	\biggr\} \mathcal{M} [s(t)]
	+ \frac{1}{\mathcal{M}_y [s(t)]} \frac{\delta \mathcal{M} [s(t)]}{\delta s(t_1)}
	\frac{\delta \mathcal{M} [s(t)]}{\delta s(t_2)},
\end{aligned}
\end{equation}
using (\ref{eqn:deriv_functional_inverse}) to differentiate $h^{-1} (t_1, t_2)$.
Setting $s(t) = 0$ and using the fact
\begin{equation}\label{eqn:gamma_hermitian_symmetry}
	\Gamma (t_1, t_2) = \Gamma^* (t_2, t_1)
\end{equation}
by definition, gives
\begin{equation}\label{eqn:fb_y_correlation}
\begin{aligned}
	&\langle y(t_1) y(t_2) \rangle= \\
	&\iint_{-\infty}^{\infty} g(t_1, t'_1) g(t_2, t'_2) \biggr\{
		\frac{1}{2} h \nu_0 \Gamma (t'_1, t'_2) \delta (t'_1 - t'_2)
		+ \frac{1}{4} | \Gamma (t'_1, t'_2) |^2
		+ \Gamma_n (t'_1, t'_2)
	\biggr\} \, dt'_1 dt'_2 \\
	&+ \langle p(t_1) \rangle \langle p(t_2) \rangle.
\end{aligned}
\end{equation}
Hence
\begin{equation}\label{eqn:fb_dy_correlation}
\begin{aligned}
	&\langle \Delta y(t_1) \Delta y(t_2) \rangle= \\
	&\iint_{-\infty}^{\infty} g(t_1, t'_1) g(t_2, t'_2) \biggr\{
		\frac{1}{2} h \nu_0 \Gamma (t'_1, t'_2) \delta (t'_1 - t'_2)
		+ \frac{1}{4} | \Gamma (t'_1, t'_2) |^2
		+ \Gamma_n (t'_1, t'_2)
	\biggr\} \, dt'_1 dt'_2.
\end{aligned}
\end{equation}

\subsubsection{Homodyne scheme output correlation functions}\label{sec:mgf_td_homo}

The calculation of the output correlation functions in the homodyne scheme follows exactly the same procedure as for the filter-bank spectrometer.
However, the number of terms increases rapidly with successive derivatives.
The result is the calculations, while straightforward, are algebraically involved and the results are lengthy to state.
Full expressions for the correlation functions up to fourth-order are given in Appendix \ref{sec:full_td_calc_homo}.

For what follows we will focus on what we will call the \emph{strong-pump} regime, where the correlation functions simplify.
In this regime the amplitude of the pump signal is assumed to be much stronger than the input signal, such that the $n^\text{th}$ order correlation function of $d(t)$ is dominated by the terms that scale as order $|a_p|^n$.
Define a new function
\begin{equation}\label{eqn:def_q}
	q(t) = d(t) + n_1 (t) - n_2 (t),
\end{equation}
such that we can express the correlation functions of $y(t)$ as
\begin{equation}\label{eqn:y_cf_from_q_cf}
\begin{aligned}
	&\langle y(t_1) y(t_2) \dots y(t_n) \rangle \\
	&= \int_{-\infty}^{\infty} \int_{-\infty}^{\infty} \dots \int_{-\infty}^{\infty}
		g(t_1, t'_1) g(t_2, t'_2) \dots g(t_n, t'_n)
		\Gamma_q (t_1, t_2, \dots, t_n)
	\, dt_1 dt_2 \dots dt_n
\end{aligned}
\end{equation}
for
\begin{equation}\label{eqn:def_q_correlations}
	\Gamma_q (t_1, t_2, \dots, t_n)
	= \langle q(t_1) q(t_2) \dots q (t_n) \rangle.
\end{equation}
Keeping only the highest order terms in $|a_p|$ in the results in Appendix \ref{sec:full_td_calc_homo}, we find that in the strong-pump regime we can approximate
\begin{align}
	&\langle q(t_1) \rangle = 0
	\label{eqn:q_cf1} \\
	&\langle q(t_1) q(t_2) \rangle = \Gamma_q (t_1, t_2)
	\label{eqn:q_cf2} \\
	&\langle q(t_1) q(t_2) q(t_3) \rangle = 0
	\label{eqn:q_cf3} \\
	&\langle q(t_1) q(t_2) q(t_3) q(t_4) \rangle =
	\Gamma_q (t_1, t_2) \Gamma_q (t_3, t_4)
	+ \Gamma_q (t_1, t_3) \Gamma_q (t_2, t_4)
	+ \Gamma_q (t_1, t_4) \Gamma_q (t_2, t_3)
	\label{eqn:q_cf4}
\end{align}
where
\begin{equation}\label{eqn:def_gamma_q}
	\Gamma_q (t_1, t_2)
	= \bigl\{
		h \nu_0 \delta (t_1 - t_2)
		+ \Re \bigl[ \Gamma (t_1, t_2) e^{2 \pi i \vp (t_1 - t_2)} \bigr]
	\bigr\} P_p
	+ \Gamma_{n_1} (t_1, t_2)
	+ \Gamma_{n_2} (t_1, t_2)
\end{equation}
and
\begin{equation}\label{eqn:def_pump_power}
	P_p = \frac{1}{2} |a_p|^2
\end{equation}
is the total pump power.
The terms that scale as $P_p$ in (\ref{eqn:def_gamma_q}) are those associated with $d(t)$, while the other two terms are due to the detector noise.

(\ref{eqn:q_cf1})--(\ref{eqn:q_cf4}) show that $q(t)$ obeys the Gaussian-moment theorem up to fourth order in the strong-pump regime.
We can therefore treat $q(t)$, and by extension $y(t)$, as a zero-mean Gaussian process.
This will significantly simplify later analyses.

(\ref{eqn:q_cf1})--(\ref{eqn:q_cf4}) can also be obtained by the following argument.
It follows from (\ref{eqn:def_q}) that
\begin{equation}\label{eqn:mq_as_product}
	\mathcal{M}_q [s(t)]
	= \mathcal{M}_d [s(t)] \mathcal{M}_4 [s(t)] \mathcal{M}_5 [s(t)]
\end{equation}
for $\mathcal{M}_d$ given by (\ref{eqn:mgf_d_as_a_product}), $\mathcal{M}_4$ given by (\ref{eqn:def_m4}) and $\mathcal{M}_5$ given by (\ref{eqn:def_m5}).
If we are only interested in the terms of highest order in $|a_p|$ resulting from the differentiation of $\mathcal{M}_d$, it is sufficient to approximate
\begin{equation}\label{eqn:md_approx_v1}
\begin{aligned}
	&\mathcal{M}_d [s(t)]
	= \exp \biggl( 
		\frac{1}{2 h \nu_0} |a_p|^2 \int_{-\infty}^{\infty}
		\{ \cosh(h \nu_0 s(t))  - 1 \} \, dt
	\biggr) \\
	&\times
	\exp \biggl( \frac{|a_p|^2}{(2 h \nu_0)^2}
		\iiint_{-\infty}^{\infty} k^{-1} (t_1, t_2) \Gamma (t_2, t_3)
		\sinh (h \nu_0 s(t_1)) \sinh (h \nu_0 s(t_3))
		e^{- 2 \pi i \vp (t_1 - t_3)}
		\, dt_1 dt_2 dt_3
	\biggr).
\end{aligned}
\end{equation}
We are also only interested in the value of $\mathcal{M}_d$ and its derivatives around $s(t) = 0$.
Around $s(t) = 0$ we can make the expansions
\begin{equation}
\begin{aligned}
	&\cosh(h \nu_0 s(t))  - 1 = \frac{1}{2} \{ h \nu_0 s(t) \}^2 + \smallO \bigl( s(t)^4 \bigr) \\
	&\sin(h \nu_0 s(t)) = h \nu_0 s(t) + \smallO \bigl( s(t)^3 \bigr) \\
	&k^{-1} (t_1, t_2) = \delta (t_1 - t_2) + \smallO \bigl( s(t)^2 \bigr)
\end{aligned}
\end{equation}
so (\ref{eqn:md_approx_v1}) can be written as
\begin{equation}\label{eqn:md_approx_v2}
\begin{aligned}
	&\mathcal{M}_d [s(t)]
	= 
	\exp \biggl( \frac{1}{2} P_p \iint_{-\infty}^{\infty} \Bigr\{
		h \nu_0 \delta (t_1 - t_2)
		+\Gamma (t_1, t_2) e^{- 2 \pi i \vp (t_1 - t_2)}
	\Bigl\} s(t_1) s(t_2) \, dt_1 dt_2 \biggr) \\
	&\times \exp \biggl( P_p \iiint \smallO(s(t)^4) \, dt \biggr)
\end{aligned}
\end{equation}
At this point it is also useful to note that
\begin{equation}\label{eqn:integral_is_real}
	\iint_{-\infty}^{\infty}
		\Gamma (t_1, t_2) e^{- 2 \pi i \vp (t_1 - t_2)}
		s(t_1) s(t_2)
	\, dt_1 dt_2
	= \iint_{-\infty}^{\infty}
		\Re \bigl[ \Gamma (t_1, t_2) e^{- 2 \pi i \vp (t_1 - t_2)} \bigl]
		s(t_1) s(t_2)
	\, dt_1 dt_2,
\end{equation}
as we can show the integral is its own conjugate by switching labels $1 \leftrightarrow 2$ and then making use of the fact $\Gamma(t_1, t_2) = \Gamma^* (t_2, t_1)$.
The second term in (\ref{eqn:md_approx_v2}) only produces non-zero derivatives at $s(t) = 0$ for when the order of the derivative is a multiple of four.
Further, if the order is $4n$ then the derivative scales as $P_p^n$ as versus $P_p^{2n}$ for the derivatives of the first term.
This means the terms generated by the second term will, in general, be negligible compared to those generated and as such can be ignored; this is achieved by setting the second term equal to unity.
Doing so, substituting the result into (\ref{eqn:mq_as_product}) and making use of (\ref{eqn:integral_is_real}), (\ref{eqn:def_m4}) and (\ref{eqn:def_m5}), we obtain the final approximation
\begin{equation}\label{eqn:mq_as_product_gaussian}
	\mathcal{M}_q [s(t)]
	\approx \exp \biggl( \frac{1}{2} \int_{-\infty}^\infty
		\Gamma_{q} (t_1, t_2) s(t_1) s(t_2),
	\, dt_1 dt_2 \biggr)
\end{equation}
where $\Gamma_q (t_1, t_2)$ is as defined by (\ref{eqn:def_gamma_q}).
(\ref{eqn:mq_as_product}) indicates $q(t)$ is a zero-mean Gaussian process with correlation function $\Gamma_q (t_1, t_2)$, consistent with the earlier results.

\subsection{Output noise power spectral density}\label{sec:mgf_fd}

It will be helpful for the sensitivity calculations to convert the correlation functions just derived into the spectral domain.
To do this we will need the spectral domain representations of the correlation functions of the input signal and detector noise.
The former is
\begin{equation}\label{eqn:acf_signal_from_psd}
	\Gamma(t_1, t_2) = 2 \int_0^\infty p(\nu) e^{2 \pi i \nu (t_1 - t_2)} \, d\nu
\end{equation}
where $p(\nu)$ is the single sided power spectral density of the input signal.
Using our definition of NEP, as discussed in Appendix \ref{sec:nep}, we can express the latter as
\begin{equation}\label{eqn:acf_det_noise_from_nep}
	\Gamma_n (t_1, t_2) = \Re \biggl[
		\int_0^\infty \text{NEP}_i^2 (f) e^{2 \pi i f (t_1 - t_2)} \, df
	\biggr]
\end{equation}
where $\text{NEP}_i (f)$ is the internal NEP of the detector system.

In addition, we will make the simplifying assumption that the detector response is time independent as is usually the case in practice.
By this we mean that if the output is $y(t)$ for an input signal $p(t)$, then the output for an input signal $p(t - s)$ is $y(t - s)$.
Mathematically, this implies
\begin{equation}\label{eqn:time_shifted_output}
	y(t) = \int_{-\infty}^{\infty} g(t, t') p(t' - s) \, dt'
	= \int_{-\infty}^{\infty} g(t - s, t') p(t') \, dt',
\end{equation}
for all shifts, which requires
\begin{equation}\label{eqn:time_shifted_output_constraint_on_g}
	g(t_1, t_2 + s) = g(t_1 - s, t_2).
\end{equation}
(\ref{eqn:time_shifted_output_constraint_on_g}) is satisfied if
\begin{equation}\label{eqn:g_function_from_1d_function}
	g(t_1, t_2) = g (t_1 - t_2),
\end{equation}
in which case
\begin{equation}\label{eqn:g_function_spectral_decomposition}
	g(t_1, t_2) = \int_{-\infty}^{\infty}
		\tilde{g} (f) e^{2 \pi i f (t_1 - t_2)}
	\, df
\end{equation}
where
\begin{equation}\label{eqn:def_ft}
	\tilde{g} (\nu) = \int_{-\infty}^{\infty} g(t) e^{-2 \pi i \nu t} \, dt.
\end{equation}
Since $g(t_1 - t_2)$ is real, it can be shown that $\tilde{g}(\nu) = \tilde{g}^*(-\nu)$.
We will use the notation $G(\nu) = |\tilde{g}(\nu)|^2$ to denote the gain function of the filter in the spectral domain.

\subsubsection{Output noise power spectral density for a single filter bank channel}\label{sec:psd_fb}

First we rewrite the expected output signal, (\ref{eqn:expect_y_fb}), in terms of the power spectral density of the input signal.
Substituting (\ref{eqn:g_function_spectral_decomposition}) and (\ref{eqn:acf_signal_from_psd}) into (\ref{eqn:expect_y_fb}) and reordering the integrals, we obtain
\begin{equation}\label{eqn:expect_y_fb_v1}
	\langle y(t) \rangle =
	\int_{f_1=-\infty}^\infty \int_{f_2=0}^\infty
		\tilde{g} (f_1) p(f_2)
		\biggr\{ \int_{t'=-\infty}^\infty e^{2 \pi i f_1 (t - t')} \, dt' \biggr\}
	\, df_1 df_2.
\end{equation}
The integral in parentheses evaluates to $\delta (f_1)$ and we can also then evaluate the integral with respect to $f_2$.
The final result is 
\begin{equation}\label{eqn:expect_y_fb_v2}
	\langle y(t) \rangle =
	\tilde{g} (0) \int_{0}^\infty p(f) df.
\end{equation}
The expected output is therefore simply the total incident power scaled by the steady state responsivity, as expected.

To calculate the spectral decomposition of $\langle \Delta y(t_1) \Delta y(t_2) \rangle$ it is easiest to first calculate the spectral representations of each of the terms in parentheses in (\ref{eqn:fb_dy_correlation}).
Doing so for the first and third term is straightforward and it follows from (\ref{eqn:acf_signal_from_psd}) and (\ref{eqn:acf_det_noise_from_nep}) that
\begin{equation}\label{eqn:psf_dy_term_1}
	\frac{1}{2} h \nu_0 \Gamma (t'_1, t'_2) \delta (t'_1 - t'_2)
	= h \nu_0 \int_0^\infty p(\nu) \, d\nu
	\int_{-\infty}^\infty e^{2 \pi i f (t_1 - t_2)} \, df
\end{equation}
and
\begin{equation}\label{eqn:psf_dy_term_3}
	\Gamma_n (t_1, t_2) =
	\frac{1}{2} \int_{-\infty}^\infty \text{NEP}_i^2 (|f|) e^{2 \pi i f (t_1 - t_2)} \, df.
\end{equation}
The second term is more challenging and we start by using (\ref{eqn:acf_signal_from_psd}) to write
\begin{equation}\label{eqn:psf_dy_term_2_step_1}
	\frac{1}{4} | \Gamma (t'_1, t'_2) |^2
	= \int_0^\infty \int_0^\infty
		p(\nu_1) p(\nu_2) e^{2 \pi i (\nu_1 - \nu_2) (t_1 - t_2)}
	\, d\nu_1 d\nu_2 .
\end{equation}
We then make a change of variables $\nu_1, \nu_2 \rightarrow \nu, \nu + f$ to give
\begin{equation}\label{eqn:psf_dy_term_2_step_2}
	\frac{1}{4} | \Gamma (t'_1, t'_2) |^2
	= \int_{-\infty}^\infty
		\biggl\{ \int_0^\infty p(\nu) p(\nu + f) \, d\nu \biggr\}
		e^{2 \pi i f (t_1 - t_2)}
	\, df .
\end{equation}
which puts the decomposition in the same form as (\ref{eqn:psf_dy_term_1}) and (\ref{eqn:psf_dy_term_3}).

We can now substitute (\ref{eqn:psf_dy_term_1}), (\ref{eqn:psf_dy_term_2_step_2}) and (\ref{eqn:psf_dy_term_3}) into (\ref{eqn:fb_dy_correlation}), then use (\ref{eqn:def_ft}) to evaluate the time integrals.
The final result can be written as
\begin{equation}
	\langle \Delta y(t_1) \Delta y(t_2) \rangle
	= 2 \int_0^\infty G(f) S(f) \, df
\end{equation}
where
\begin{equation}\label{eqn:fb_equivalent_input_psd}
	S(f) = \underbrace{\frac{1}{2} \text{NEP}_i^2 (f)}_{(i)}
	+ \underbrace{h \nu_0 \int_0^\infty p(\nu) \, d\nu}_{(ii)}
	+ \underbrace{\int_0^\infty p(\nu) p(\nu + f) \, d\nu}_{(iii)} \quad (f \geq 0)
\end{equation}
the is the input referenced noise power spectral density of the fluctuations.

It is worth briefly commenting on the different terms in (\ref{eqn:fb_equivalent_input_psd_a_and_b}).
Term (i) is obviously due to the internal noise of the detector.
Terms (ii) and (iii) are due to the input signal.
Term (ii) disappears in the classical limit ($h \nu_0 \rightarrow 0$) and is therefore associated with the quantum nature of the input signal; this term is the photon shot noise discussed in Section \ref{sec:bo_homodyne_scheme}.
It can be seen to correspond to white noise (constant spectral density), as expected for a shot noise process.
Term (iii) persists in the classical noise limit and is the wave noise discussed in Section \ref{sec:bo_homodyne_scheme}.
We see that the wave noise is non-white and has bandwidth of order of that of the input signal.

Had we carried out the analysis accounting for emission noise using (\ref{eqn:mgf_p_zero_mean_gaussian_noise_a_and_b}), the result would instead be
\begin{equation}\label{eqn:fb_equivalent_input_psd_a_and_b}
\begin{aligned}
	S(f) &= \frac{1}{2} \text{NEP}_i^2 (f)
	+ h \nu_0 \int_0^\infty p_a (\nu) \, d\nu
	+ \int_0^\infty p_a (\nu) p_a (\nu + f) \, d\nu \\
	& + h \nu_0 \int_0^\infty p_b (\nu) \, d\nu
	+ \int_0^\infty p_b (\nu) p_b (\nu + f) \, d\nu, 
\end{aligned}
\end{equation}
where $p_a (\nu)$ and $p_b (\nu)$ are the noise power spectral density of the input and output signals respectively.
Since the input and output signals are uncorrelated, the noise resulting from them simply adds incoherently.
Use of (\ref{eqn:fb_equivalent_input_psd_a_and_b}) is only necessary in instances where the emission from the detector output is similar to the input signal.

\subsubsection{Output noise power spectral density in the homodyne scheme}\label{sec:psd_homo}

Again we limit ourselves to the strong-pump regime.
We begin by calculating the single-sided noise power spectral density $S_q (f)$ of $q(t)$ as defined by
(\ref{eqn:def_q}).
To do so, we first note that (\ref{eqn:acf_signal_from_psd}) implies
\begin{equation}
	\Re \bigl[ \Gamma (t_1, t_2) e^{- 2 \pi i \vp (t_1 - t_2)} \bigl]
	= 2 \int_0^{\infty} \Re \bigl[ p(\nu) e^{2 \pi i (\nu - \vp) (t_1 - t_2)} \bigl] \, d\nu.
\end{equation}
Making a first change of variable $\nu = \vp + f$ yields
\begin{equation}
	\Re \bigl[ \Gamma (t_1, t_2) e^{- 2 \pi i \vp (t_1 - t_2)} \bigl]
	= 2 \int_0^{\infty} \Re \bigl[ p(\nu_p + f) e^{2 \pi i f (t_1 - t_2)} \bigl] \, df
	+ 2 \int_{-\nu_p}^{0} \Re \bigl[ p(\nu_p + f) e^{2 \pi i f (t_1 - t_2)} \bigl] \, df.
\end{equation}
where we have split the integral on the right-hand-side into two parts based on the sign of $f$.
Making a second change of variables $f \rightarrow -f$ in the second term and exploiting the fact $\Re[z]=\Re[z^*]$, we then obtain
\begin{equation}\label{eqn:fourier_decomp_mixed_down_term}
	\Re \bigl[ \Gamma (t_1, t_2) e^{- 2 \pi i \vp (t_1 - t_2)} \bigl]
	= 2 \int_0^{\infty} \Re \bigl[ p(\nu_p + f) e^{2 \pi i f (t_1 - t_2)} \bigl] \, df
	+ 2 \int_0^{\nu_p} \Re \bigl[ p(\nu_p - f) e^{2 \pi i f (t_1 - t_2)} \bigl] \, df,
\end{equation}
where both integrals are now with respect to positive frequencies.
Using (\ref{eqn:fourier_decomp_mixed_down_term}) and (\ref{eqn:acf_det_noise_from_nep}) to substitute for the relevant terms in (\ref{eqn:def_gamma_q}) and replacing the $\delta$-function with its Fourier decomposition, we obtain:
\begin{equation}
\begin{aligned}
	\Gamma_q (t_1, t_2)
	&= \int_0^{\infty} \Re \biggl[ \biggl\{
		2 \{ h \nu_0 + p(\nu_p + f) \} P_p
		+ \text{NEP}_{i,1}^2 (f) + \text{NEP}_{i,2}^2 (f) 
	\biggr\} e^{2 \pi i f (t_1 - t_2)} \biggr] \, d\nu \\
	& + 2 P_p \int_0^{\nu_p} \Re \biggl[ 
		p(\nu_p - f) e^{2 \pi i f (t_1 - t_2)} \biggr] \, d\nu,
\end{aligned}
\end{equation}
and from this we can identify
\begin{equation}\label{eqn:psd_q}
	S_q (f) = 2 \{ h \nu_0 + p(\nu_p + f) \} P_p
	+ \text{NEP}_{i,1}^2 (f) + \text{NEP}_{i,2}^2 (f) 
	+ 2 P_p \begin{cases}
		p(\nu_p - f) & f < \nu_p \\
		0 & \text{otherwise}.
	\end{cases}
\end{equation}

It follows from (\ref{eqn:y_cf_from_q_cf}) that $S_y (f) = G(f) S_q (f)$.
Hence (\ref{eqn:psd_q}) implies that the single-sided power spectral density of the noise signal at the output of the homodyne scheme is
\begin{equation}\label{eqn:psd_homo_final}
	S_y (f) = \biggl[
		2 \{ \underbrace{h \nu_0}_{(i)}
		+ \underbrace{p(\vp + f)}_{(ii)}
		+ \underbrace{p(\vp - f)}_{(iii)} \} P_p
		+ \underbrace{2 \text{NEP}_{i}^2 (f)}_{(iv)}
	\biggr]  G(f)
\end{equation}
under the assumption the internal NEP of the two detectors is the same and $f \ll \nu_p$.
As expected from the mixing action described in Section \ref{sec:bo_homodyne_scheme}, the output contains two terms, (ii) and (iii), arising from the downconversion of the input signal.
The first term (ii) corresponds to the upper sideband response (frequencies above $\vp$) and the (iii) to the lower sideband response (frequencies below $\vp$).
Term (iii) is produced by the noise internal to the detector and term (iv) is a term of quantum origin.

(\ref{eqn:psd_homo_final}) can also be argued on the basis of simple mixer theory.
We start by considering the mixing action in more detail.
Assume we can treat the input to each detector as a resistance $Z_0$ and that the detector responds to the power dissipated in that resistor; this is the case for bolometric detectors.
In addition, assume the input signal comprises to tones at frequencies $\nu_p + f$ and $\nu_p - f$.
The input voltages at the two detectors are then
\begin{equation}
	V_1 (t) = \frac{1}{\sqrt{2}} \Re \bigl[
		V_p e^{2 \pi i \nu_p t}
		+ V_u e^{2 \pi i (\nu_p + f) t}
		+ V_l e^{2 \pi i (\nu_p - f) t}
	\bigr]
\end{equation}
and
\begin{equation}
	V_2 (t) = \frac{1}{\sqrt{2}} \Re \bigl[
		V_p e^{2 \pi i \nu_p t} 
		- V_u e^{2 \pi i (\nu_p + f) t}
		- V_l e^{2 \pi i (\nu_p - f) t}
	\bigr]
\end{equation}
and the difference $d(t)$ in the power dissipated in the two resistors simplifies to
\begin{equation}\label{eqn:simplified_power_difference_v1}
\begin{aligned}
	d(t) &= \frac{1}{Z_0} \{ V_1 (t)^2 - V_2 (t)^2 \} \\
	&= \frac{1}{Z_0} \Re \bigl[ \bigl\{
		V^*_p V^{\vphantom *}_u
		+ V^{\vphantom *}_p V^*_l
	\bigr\} e^{2 \pi i f t} \bigl] 
	+ \dots
\end{aligned}
\end{equation}
where $\dots$ indicates higher frequency terms that will be filtered out by the detector response.
(\ref{eqn:simplified_power_difference_v1}) shows the homodyne scheme essentially behaves as a double-sideband mixer.

We can derive an effective `conversion-gain' for the mixer, although we have to be careful as the dimensions of the signal change in down-conversion and it is perhaps safer to refer to a \emph{power-conversion factor}.
Letting $V_n = |V_n| e^{i \phi_n}$ and identifying $|V_n|^2 / 2 Z_0$ as the power $P_n$ incident in signal $n$, (\ref{eqn:simplified_power_difference_v1}) can be rewritten as
\begin{equation}\label{eqn:simplified_power_difference_v2}
\begin{aligned}
	d(t) &= 2 \sqrt{P_p} \, \Re \bigl[ \bigl\{
		\sqrt{P_u} \, e^{i (\phi_u - \phi_p)}
		+ \sqrt{P_l} \, e^{i (\phi_p - \phi_l)}
	\bigr\} e^{2 \pi i f t} \bigl],
\end{aligned}
\end{equation}
keeping only the low frequency terms.
The corresponding `power' in this difference signal is
\begin{equation}\label{eqn:d_power_conversion}
	d(t)^2 = 2 (P_u + P_l) P_p + \dots,
\end{equation}
where again $\dots$ indicates high-frequency terms that can be ignored when considering steady-state powers.
(\ref{eqn:d_power_conversion}) implies the power conversion factor between the input sidebands and $d(t)$ is $2 P_p$.
Accounting for the response functions of the detectors then gives a total power conversion factor of $2 G(f) P_p$ between each sideband and the final output $y(t)$.

Now we are in a position to calculate the noise using standard results.
Following Kerr\,\cite{kerr1997receiver}, the minimum noise power spectral density added by a double-sideband mixer is $h \nu_0$.
This adds to the power spectral density in each sideband and the total is weighted by the power conversion factor to give the contribution to $y(t)$, yielding the signal terms in (\ref{eqn:psd_homo_final}).
The remaining terms in (\ref{eqn:psd_homo_final}) are then found by weighting the single-sided power spectral density of the detector noise, as found by summing the squares of the NEPs, by the detector response power gain.
Although this simple argument reproduces (\ref{eqn:psd_homo_final}), the advantage of the full Poisson-mixture model is that it can also be used to calculate the behaviour for all pump levels, not just when the pump is strong.

\begin{figure}
\centering
\includegraphics{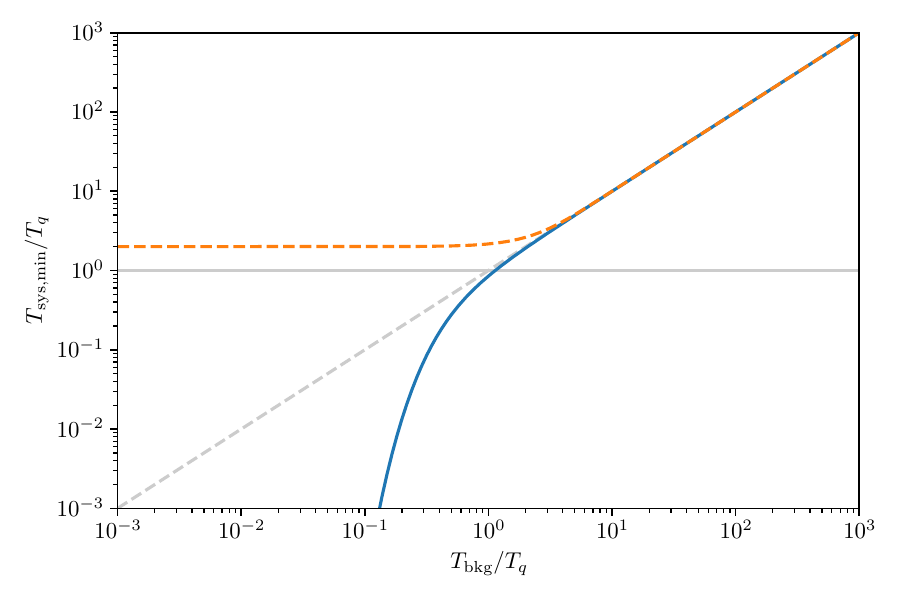}
\caption{\label{fig:tsys_min_plot}
Minimum effective system noise temperature as a function of background temperature in the limit of small fractional bandwidth ($\Delta \nu / \nu_0 \ll 1$).
This value is achieved in the limit where the contribution from the intrinsic noise in the detector is made negligible.
The dashed orange line shows the value for the homodyne scheme and the blue line that for the filter-bank spectrometer channel.
The solid grey line is the quantum noise limit for a radiometer, $\Tbkg$.
This plot illustrates the potential sensitivity advantage of a power detector over a coherent radiometer when the background temperature is below $\Tq$ at the frequency $\nu$ of interest (or equivalently, when the frequency is above $2 k_b \Tbkg / h$).
For reference, $h \nu / (2 k_b)$=23\,mK at 1\,GHz.
The dashed grey line shows $T_\text{sys,min}=\Tbkg$.
}
\end{figure}

\begin{figure}
\centering
\includegraphics{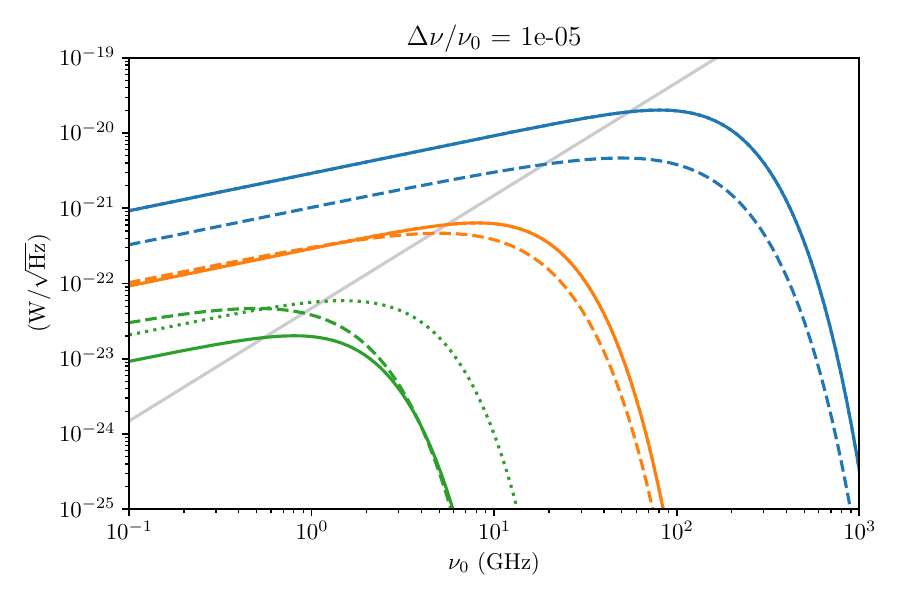}
\caption{\label{fig:nep_plot}
Fundamental contributions to the NEP of an ideal TES filter-bank spectrometer channel as a function of channel frequency $\nu_0$ for different background temperatures.
It is assumed the fractional bandwidth of the channel is $10^{-5}$ (representative of a cavity haloscope), the bath temperature is 15\,mK and that the TES has been designed to saturate at a total input power equal to twice the background loading from the source ($m=1$).
Line colour indicates source temperature: blue=1.5\,K, orange=150\,mK and green=15\,mK.
The dashed lines show the intrinsic NEP of the TES as arising from phonon and Johnson noise, as calculated using (\ref{eqn:tes_nep_approx}). 
The solid lines show the photon contribution to the NEP from the background alone, while the dotted lines show the combined photon noise due to the background and emission from the RF termination on the TES.
In the latter case the temperature of the termination is assumed to be twice that of the bath.
For source temperatures above 15\,mK the solid and dotted lines are nearly coincident.
The grey line shows the equivalent NEP of a quantum noise limited radiometer with the same fractional bandwidth.
The dotted/solid lines are useful generally, as they indicate an upper limit for the intrinsic NEP of any detector type that might be used in place of a TES.
Where the dotted/solid lines are below the grey line, a power detector can in principle be made more sensitive than a coherent radiometer.
}
\end{figure}

\section{Sensitivity}\label{sec:ultimate_sensitivity}

\subsection{Scenario considered}\label{sec:sensitivity}

In this section we will evaluate the sensitivity of the two schemes in an illustrative radiometric application.
We consider an experiment to measure the effective noise temperature of a signal of interest and in the presence of a much stronger background signal over bandwidth $\Delta \nu$ centred on frequency $\nu_0$.
For simplicity, we will assume $\Delta \nu \ll \nu_0$ and that the background radiation is from a black-body source at a physical temperature of $\Tbkg$.
To model an axion search experiment, we set $\Tbkg$ equal to the cavity temperature and $\Delta \nu$ to the bandwidth of the cavity mode.

The appropriate figure of merit for sensitivity in this scenario is the system noise temperature $T_\text{sys}$ as defined by (\ref{eqn:rms_temp_noise_coherent_receiver}).
In Sections \ref{sec:sensitivity_filterbank_spectrometer} and \ref{sec:sensitivity_homodyne_receiver} we derive $T_\text{sys}$ for the filter-bank spectrometer and homodyne receiver respectively.
In both cases we will show $T_\text{sys}$ can be decomposed into a contributions due to noise in the detector and the background loading.
In Section \ref{sec:sensitivity_tes_performance} we consider how the detector contribution scales with detector temperature in the case of an ideal TES.

\subsection{Filter-bank spectrometer performance}\label{sec:sensitivity_filterbank_spectrometer}

To model the performance of the filter-bank spectrometer we make two additional assumptions.
The first is that the detector input behaves like a black-body in emission and that its effective radiometric temperature is equal to the detector's operating temperature $T_0$.
This is the case for bolometric detectors like a TES, where the input is simply a matched resistive load.
The second is that we consider the limit where the modulation frequency tends to zero.
The behaviour of the photon noise for a chopped signal is more complicated and we defer its discussion to a future report.

Given these assumptions, it follows from (\ref{eqn:dp_from_nep}) and (\ref{eqn:fb_equivalent_input_psd}) that the total NEP of single channel is given by
\begin{equation}\label{eqn:total_detector_nep}
	\text{NEP} = \sqrt{
		\text{NEP}_\text{i}^2 (0_+)
		+ \text{NEP}_\text{bb}^2 (\Tbkg)
		+ \text{NEP}_\text{bb}^2 (T_0)
	}
\end{equation}
where
\begin{equation}\label{eqn:bb_photon_nep}
	\text{NEP}_\text{bb}^2 (T)
	= 2 \int_{\nu_0 - \Delta \nu /2}^{\nu_0 + \Delta \nu / 2}
		\bigl\{ h \nu_0 + p (\nu, T) \bigr\} p(\nu, T)
	\, d\nu
\end{equation}
and
\begin{equation}\label{eqn:bb_psd}
	p(\nu, T) = \frac{h \nu}{e^{h \nu / (k_b T)} - 1}.
\end{equation}
$p(\nu, T)$ is the noise power spectral density $p(\nu)$ of a black body at temperature $T$ and $\text{NEP}_\text{bb} (T)$ is the corresponding contribution to the photon NEP.
For calculations it is convenient the quantum noise limit
\begin{equation}\label{eqn:def_tq}
	\Tq = \frac{h \nu_0}{2 k_b}
	\sim 24 [\text{mK/GHz}] \times \nu_0 
\end{equation}
as a scale temperature and rewrite (\ref{eqn:bb_photon_nep}) as
\begin{equation}\label{eqn:bb_photon_nep_rewritten}
	\text{NEP}_\text{bb}^2 (T)
	= 2 k^2_b T \bigr\{ 2 \Tq K_1 (T / \Tq)  + T K_2 (T / \Tq) \bigl\} \Delta \nu
\end{equation}
where
\begin{equation}\label{eqn:bb_def_big_k_n}
	 K_n (t) = \frac{2^n}{t^n \Delta x}
	 \int_{1 - \Delta x /2}^{1 + \Delta x / 2}
	 	\frac{x^n}{(e^{2 x / t} - 1)^n}
	 \, dx,
\end{equation}
and $\Delta x = \Delta \nu / \nu_0$ is the fractional bandwidth.

(\ref{eqn:sys_temp_power_detector}) can be used to convert the total NEP into an equivalent system noise temperature. 
The result is
\begin{equation}\label{eqn:filterbank_tsys}
	T_\text{sys} = \sqrt{
		T_\text{r,d}^2
		+ \{ T_\text{bb} (\Tbkg) \}^2
		+ \{ T_\text{bb} (T_0) \}^2
	} 
\end{equation}
where
\begin{equation}\label{eqn:tr_filterbank_det}
	T_\text{r,d} = \frac{\text{NEP}_\text{i} (0_+)}{k_b \sqrt{2 \Delta \nu}}
\end{equation}
and
\begin{equation}\label{eqn:tsys_filterbank_bb}
	T_\text{bb} (T) = \sqrt{\bigr\{ 2 \Tq K_1 (T / \Tq)  + T K_2 (T / \Tq) \bigl\} T}
\end{equation}
$T_{r,d}$ can be interpreted as the `receiver' noise temperature on the filter-bank spectrometer, as it is present in the absence of any input signal.
However, this analogy should not be taken too far as it does not combine with the other noise contributions in the same way as classical microwave radiometer (which, as we will see, add).
$T_\text{bb} (T)$ is the effective system noise contribution due to blackbody emission and we end up with two contributions of this kind; one due the background and the other due to emission from the detector.
There would also have been a third contribution from the signal were not assuming it was negligible.

The sensitivity is maximized when $T_\text{r,d}$ and $T_\text{bb}(T_0)$ are much smaller than $T_\text{bb}(\Tbkg)$:
\begin{equation}\label{eqn:background_limited_tsys_filterbank}
	T_\text{sys,min} = T_\text{bb}(\Tbkg)
\end{equation}
This is normally referred to as the case where the detector is \emph{background noise limited}. 
The blue line in Figure \ref{fig:tsys_min_plot} shows $T_\text{sys,min}$ as a function of $\Tbkg$ in this regime.
The horizontal axis is scaled in units of $\Tq$, so the trend in $T_\text{sys,min}$ moving left to right is that seen on increasing $\Tbkg$ at fixed $\nu_0$, or on decreasing the frequency $\nu_0$ at fixed $\Tbkg$.
The solid grey line shows the quantum noise limit for the system noise temperature of a coherent microwave radiometer, which is simply $\Tq$.
As can be seen, the filter-bank spectrometer can in principle beat the quantum noise limit of a coherent system provided the background temperature is smaller than this quantum noise limit.

Although it possible to achieve better sensitivity than a coherent system in principle, in practice a number of challenging requirements must be met to do so.
The first and most obvious is that the background source (e.g. the cavity in an axion experiment) must be cooled below the quantum noise limit at the frequency of interest.
Modern dilution fridges can achieve a base temperature of 15\,mK, giving a lower frequency limit of $\sim$1\,GHz.
The second challenge is that it must be possible to engineer a detector with lower intrinsic NEP than the photon NEP of the background; we will consider this problem in more detail shortly.
The third challenge, which is related to the second, is receiver instability. 
We have assumed the modulation frequency can be made zero without consequence, whereas in reality, as discussed in Section \ref{sec:bo_microwave_power_detector_theory}, it is likely $\text{NEP}_i$ will increase as $1/f$.
This means modulation will be necessary, which at the very least requires the corresponding development of a low-noise, low-temperature, switch to enable chopping.

It is instructive to consider what sort of intrinsic NEP levels would be necessary in order to beat the quantum noise limit in a cavity haloscope experiment.
This is illustrated in Figure \ref{fig:nep_plot}, which shows the different NEP contributions in (\ref{eqn:total_detector_nep}) as function of frequency, assuming the fractional bandwidth is fixed at 10$^{-5}$ by the cavity and the detector operating temperature is $T_0$=30\,mK.
The solid lines show $\text{NEP}_\text{bb} (\Tbkg)$ and the dotted lines $\sqrt{\text{NEP}_\text{bb}^2 (\Tbkg) + \text{NEP}_\text{bb}^2 (T_0)}$.
Line colour indicates assumed background temperature: green = 15\,mK (dilution fridge temperature), orange = 150\,mK (ADR temperature) and blue = 1.5\,K (cryocooler temperature).
With the exception of $\Tbkg$ = 15\,mK the dotted lines lies on top of the solid lines, i.e. the contribution from the photon noise from detector emission is negligible.
The grey line shows the quantum noise limit expressed as an equivalent NEP.
The intrinsic detector NEP must be made smaller than the values shown by the solid and dashed lines to achieve background limited performance, and smaller than the grey line to beat the quantum noise limit.
Current TES detector technology can achieve NEPs of order of 10$^{-21}$--10$^{-20}$\,W/$\sqrt{\text{Hz}}$ at $T_0$=100\,mK and it can be seen that this is compatible with performance better than the quantum noise limit at frequencies above 100\,GHz.

\subsection{Homodyne scheme performance}\label{sec:sensitivity_homodyne_receiver}

In Sections \ref{sec:mgf_td_homo} and \ref{sec:psd_homo} we saw that the output of the homodyne system in the strong-pump regime is a Gaussian noise signal.
The noise power spectral density this signal contains, amongst other terms, a copy of the noise power spectral density $p(\nu)$ of the input signal that has been shifted down in frequency by the pump frequency.
Hence we can recover the input signal by processing of the homodyne system to recover its power spectral density and removing the background terms (i) and (iv) in (\ref{eqn:psd_homo_final}).

One way of measuring the power spectral density of the output is to carry out a procedure similar to microwave radiometry, which is reviewed in Appendix \ref{sec:radiometer_theory}.
At a fundamental level, a microwave radiometer measures the power spectral density of an input noise signal at a desired frequency in units of equivalent noise temperature.
Hence we can mimic the process of radiometry (filtering, squaring and integrating the signal in sequence) at low frequencies to measure the power spectral density of the homodyne receiver output.
We can either do this using a filtering circuit, or by logging the data and using digital signal processing techniques.
The latter is particular attractive, as by using an FFT or polyphase filterbank as a channelizer it is possible to measure several frequencies simultaneously.

In the case a radiometric approach is used, the equivalent receiver noise temperature at a particular frequency corresponds to the power spectral density of the output signal, in equivalent input noise temperature units, at the corresponding down-converted frequency.
Hence, if are we interested in the input power spectral density at $\nu$ given pump $\vp$, we to need measure the power spectral density of the output at either $f = \nu - \vp$ ($\nu > \vp$) or $f = \vp - \nu$ ($\nu < \vp$).
It follows then follows from (\ref{eqn:psd_homo_final}) and the results in Appendix \ref{sec:radiometer_theory} that the equivalent system noise temperature is
\begin{equation}\label{eqn:tsys_homodyne_full}
	T_\text{sys} = \frac{S_y (f)}{2 k_b P_p G(f)}
	= 2 \Tq + \frac{p(\nu)}{k_b} + \frac{p(2 \vp - \nu)}{k_b}
		+ \frac{\text{NEP}_{i}^2 (\nu - \vp)}{k_b P_p}
\end{equation}
for $\nu > \vp$.

Substituting in (\ref{eqn:bb_psd}) and assuming a regime where $p(\nu) \approx p(2 \vp - \nu)$, we may simplify (\ref{eqn:tsys_homodyne_full}) to
\begin{equation}\label{eqn:tsys_homodyne_full_2}
	T_\text{sys} = T_\text{r,h} + T_\text{f,h} (\Tbkg)
\end{equation}
where 
\begin{equation}\label{eqn:tsys_receiver_homodyne}
	T_\text{r,h} = \frac{\text{NEP}_{i}^2 (\nu - \vp)}{k_b P_p}
\end{equation}
and
\begin{equation}\label{eqn:tsys_fundamental_homodyne}
	T_\text{f,h} (T) = 2 \Tq + \frac{2 h \nu_0}{k_b} \frac{1}{e^{2 \Tq / T} - 1}
	= 2 \Tq \coth \left( \frac{\Tq}{T} \right)
\end{equation}
for $\Tq$ is evaluated at the pump frequency.
$T_\text{r,h}$ is the equivalent receiver noise due to noise in the power detectors and $T_\text{f,h}$ combines the limiting noise temperature of the receiver and the noise contribution from the background radiation.

The best sensitivity is achieved in the limit $T_\text{r,h} \gg T_{f,h}$, so
\begin{equation}\label{eqn:tsys_min_homodyne}
	T_\text{sys,min} = 2 \Tq \coth \left( \frac{\Tq}{\Tbkg} \right),
\end{equation}
where $\Tq$ is evaluated at the pump frequency.
The orange line in Figure \ref{fig:tsys_min_plot} shows $T_\text{sys,min}$ plotted as a function of $\Tbkg / \Tq$.
We remind the reader that $\Tbkg / \Tq$ is not only proportional to $\Tbkg$, but inversely proportional to $\nu_0$.
We see that when the limiting noise temperature of the homodyne scheme is equivalent to that of the filter-bank spectrometer for $\Tbkg / \Tq \gg 1$, so neither has a sensitivity advantage in this regime.
However, the sensitivity of the homodyne scheme tends to a lower limit as $\Tbkg / \Tq$ decreases below unity, while the that of the filterbank-scheme continues to decrease.
We can approximate (\ref{eqn:tsys_min_homodyne}) as
\begin{equation}\label{eqn:tsys_min_homodyne_approx}
	T_\text{sys,min} \approx
	\begin{cases}
		2 \Tq & \Tbkg \ll \Tq \\
		2 \Tbkg & \Tbkg \gg \Tq
	\end{cases}
\end{equation}
in the two regimes, giving the ultimate lower limit on the system noise temperature as $2\Tq$. 

(\ref{eqn:tsys_min_homodyne_approx}) is consistent with Kerr's results for the system noise temperature of a double sideband mixer receiver when it is used to measure a signal present in only one side band\,\cite{kerr1997receiver}.
This further supports our interpretation of the action of the homodyne receiver as a direct downconversion receiver.
When the signal of interest appears in both sidebands, e.g. a broadband radiometric signal, then the signal is doubled and this halves the system noise temperature; in this case the homodyne scheme can achieve the quantum noise limit $\Tq$.

As with the filter-bank spectrometer, a number of requirements must be met for the homodyne scheme to achieve its ultimate sensitivity.
The first, shared with the filter-bank spectrometer, is that the background temperature must be below the quantum noise temperature at the frequency of interest.
The second is that we need $T_{r, h} \ll 2 \Tq$.
Using (\ref{eqn:tsys_homodyne_full}) The latter condition can more conveniently rewritten as
\begin{equation}\label{eqn:nep_constraint_homo}
	\text{NEP}_{i} (\nu - \vp) \ll \sqrt{h \nu_0 P_p}.
\end{equation}

(\ref{eqn:nep_constraint_homo}) highlights a number of attractive features of the homodyne scheme.
Firstly, it is possible to increase the range of $\text{NEP}_i$ at which the ultimate sensitivity can be achieved by increasing the pump power.
In practice other factors, such as saturation power, will limit the pump that can be applied, however this feature still adds flexibility.
The second is that the homodyne scheme is inherently able to mitigate the effects of detector instability.
This is seen in (\ref{eqn:nep_constraint_homo}) in the fact we are free to choose the pump frequency to place $\nu - \vp$ at the minimum in $\text{NEP}_i$.
However, it follows more generally from the fact we are able to choose the pump frequency to place the down-converted signal anywhere in the output bandwidth of the detector.
This avoids the need for a separate modulation technology.

\subsection{Achieving ultimate sensitivity with a Transition-Edge-Sensor (TES)}\label{sec:sensitivity_tes_performance}

As of yet we have said nothing on what performance is possible with current or near-term detector technology.
This is the subject of the companion report\,\cite{goldie2020first}, which considers a range of superconducting microwave detector types and their different implementations in detail.
The report identifies TES as particularly promising technology and so we will use it here as illustrative technology to consider requirements on detector cooling in the filterbank and homodyne-schemes.
The results hear apply to the ideal case and the companion note explores what can be achieved with current and near-term technology.

An overview of TES technology is given in Appendix \ref{sec:tes_physics}.
The key result we will use in this section is that the limiting intrinsic NEP of a TES is given by
\begin{equation}\label{eqn:tes_nep_final_main}
	\frac{\text{NEP}_{\text{i}}^2}{4 k_b T_b P_0}
	= \frac{n^2 \bigr( m + 1 \bigl) \bigr( t^{2 n + 1} - 1 \bigl)}
	{\bigr( 2 n + 1 \bigl) \bigr( t^n - 1 \bigr)^2}
	+ m t.
\end{equation}
where $\Tb$ is the bath temperature the device operates from; $P_0$ is the steady-state signal power loading; $m P_0$ is the additional input power required to saturate the device; $t = T_0 / \Tb$ where $T_0$ is the operating temperature of the device; and $n$ is a design dependent parameter.
For the types of designs of interest, $m \sim 1$ and $n \sim 1$.
It can the be shown that the minimum value of the intrinsic NEP is 
\begin{equation}\label{eqn:tes_nep_approx}
	\text{NEP}_{\text{i}} \sim \sqrt{25 k_b \Tb P_0}
\end{equation}
and that this achieved for $T_0 \sim 2 \Tb$.
This analysis assumes the thermal conductance between the TES and the bath is the dominant source of heat loss in the device and that the phonon noise is described by the expression given by Mather\,\cite{mather1982bolometer}.
An enhanced analysis would also take into account the radiation conductance of the input signal coupling and modifications to the phonon noise spectrum at low temperatures, but we avoid these additional complexities for now.

In the measurement scenario considered the steady-state loading in the filter-bank spectrometer is determined by the background blackbody radiation, so
\begin{equation}\label{eqn:bb_power_loading}
	P_0 = k_b \Tbkg K_1 (\Tbkg / \Tq) \Delta \nu
\end{equation}
for $K_1$ is as defined by (\ref{eqn:bb_def_big_k_n}).
Inserting this result into (\ref{eqn:tes_nep_approx}) yields
\begin{equation}\label{eqn:tes_nep_approx_fb}
	\text{NEP}_{\text{i,fb}} \sim 5 k_b \sqrt{\Tbkg \Tb K_1 (\Tbkg / \Tq) \Delta \nu}.
\end{equation}
Hence we see that the only free parameter in determining the TES sensitivity in this instance is the bath temperature.

Current refrigeration technology puts a lower limit on the bath temperature of 15\,mK.
The dashed lines in Figure \ref{fig:nep_plot} show $\text{NEP}_{\text{i,fb}}$ as calculated using (\ref{eqn:tes_nep_approx}) assuming the same parameters as for the photon noise calculations, i.e. $\Delta x$=10$^{-5}$ and $\Tbkg$=15\,mK (green line), 150\,mK (orange line) and 1.5\,K (blue line).
The plot indicates a TES can in principle achieve background-limited or near-background-limited performance at all frequencies considered for bath temperatures above 15\,mK and above 1\,GHz for a background at 15\,mK.

In the case of the homodyne scheme, the background is set by the pump power.
Each detector receives half the pump power, so $P_0 = P_p / 2$ and
\begin{equation}\label{eqn:tes_nep_approx_homo}
	\text{NEP}_{\text{i,h}} \sim \sqrt{25 k_b \Tb P_p / 2}.
\end{equation}
For the scheme to realise its ultimate sensitivity requires $T_\text{r,h} \ll 2 \Tq$ and so using (\ref{eqn:tsys_receiver_homodyne}) and (\ref{eqn:tes_nep_approx_homo}) we find the corresponding requirement on the TES bath temperature is
\begin{equation}\label{eqn:tb_constraint_homo}
	\Tb \ll \frac{2 h \nu_0}{25 k_b} .
\end{equation}
Assuming a minimum bath temperature of 15\,mK, we find the homodyne receiver can be made quantum noise limited for $\nu_0 \gg 4$\,GHz.

It is also useful to put a quantitive limit on the value of $\text{NEP}_\text{i,h}$ at which quantum noise limited performance can be achieved for a TES-based homodyne receiver.
One such limit is imposed by the fact the steady-state power dissipated in the two TESs cannot exceed the cooling power $P_c$ of the fridge.
In the case considered, the combined steady-state power dissipation is twice the pump power.
This comprises the power dissipated by the pump itself and the reserve of Joule power that is exchanged for signal power; the latter is equal to the pump power as we are assuming $m=1$.
Requiring (\ref{eqn:tsys_receiver_homodyne}) is much less that $2 T_q$ and setting $P_p = P_c / 2$, we find the requirement
\begin{equation}\label{eqn:nep_constraint_homo_cooling}
	\text{NEP}_\text{i,h} \ll\sqrt{\frac{h \nu_0 P_c}{2}}.
\end{equation}
Modern dilution fridges can achieve a cooling power of 15\,$\mu$W at 20\,mK\,\cite{blufors}, yielding $\text{NEP}_\text{i,h} \ll$ 10$^{-15}$\,W/$\sqrt{\text{Hz}}$ at $\nu_0$=1\,GHz for a receiver operating in such a system.

\section{Conclusions}\label{sec:conclusions}

In this report we have derived expressions for the sensitivity of a filter bank spectrometer and direct conversion homodyne receiver both implemented using superconducting microwave power detectors.
We then used these results to explore the potential sensitivity of the two schemes for a radiometric measurement similar to the case of cavity haloscope readout.
In the case of the homodyne scheme we have concentrated on the strong-pump regime, however our method is general and the full results can be used to model any combination of signal levels.
Our analysis points to a general strategy for superconducting microwave detector development for cavity haloscope readout, with near- and long-term phases.

Near-term, efforts should focus on homodyne receiver development.
The technology to implement such systems is straightforward to develop and the potential operational advantages are significant.
For example, the analysis of this report has shown that a TES-based receiver cooled to at least 15\,mK can achieve an effective receiver noise temperature less than twice the quantum limit at frequencies above 5\,GHz.
We believe such a receiver will be easier to use then competing paramp technologies at these frequencies, having simpler tuning, wider tuning-range, `cleaner' optics and requiring less cold electronics.
The homodyne scheme also the advantage that it is intrinsically resilient to detector instabilities, which circumvents what can be a significant operational issue for superconducting microwave detectors.

Longer-term, the focus should move to filter-bank spectrometer development.
This scheme can circumvent the quantum noise limit of a coherent radiometer and as such offers the ultimate sensitivity performance.
However, the technical development challenges in doing so are significant.
For example, in the case of a cavity-haloscope experiment comprising a 1\,GHz cavity cooled to 10\,mK and having a fractional bandwidth of $10^{-5}$, our calculations show the detector would need to have an internal NEP less than 10$^{-23}$\,W/$\sqrt{\text{Hz}}$.
This increases to 10$^{-20}$\,W/$\sqrt{\text{Hz}}$ for a 100\,GHz cavity, but this figure is still at the limits of current technology.
In addition, methods to modulate the cavity signal would need to be developed to mitigate the effect of detector instabilities.
Despite these challenges the potential rewards are significant and more study is needed to identify suitable technological routes forward.

\newpage

\bibliographystyle{unsrtnat}
\newcommand*{\doi}[1]{\href{http://dx.doi.org/#1}{doi: #1}}
\bibliography{references}

\pagebreak
\appendixpage
\appendix

\section{Useful mathematical results}\label{sec:useful_mathematical_results}

\subsection{Results for Gaussian integrals}\label{sec:gaussian_integrals}

Consider the scalar function
\begin{equation}\label{eqn:gi_def_f}
	f(\mathbf{z}) = \exp \left( -\mathbf{z}^\dagger \cdot H \cdot \mathbf{z} \right)
\end{equation}
where $\mathbf{z}$ is an element complex vector and $H$ is a positive definite Hermitian matrix.
If $\mathbf{z} = \mathbf{x} + i \mathbf{y}$ for real vectors $\mathbf{x}$ and $\mathbf{y}$, then we can integrate over all values of $\mathbf{z}$ by integrating over all values of $\mathbf{x}$ and $\mathbf{y}$.
With that in mind, let
\begin{equation}\label{eqn:gi_chi_wrt_to_x_and_y}
	\chi  = \int\limits_{x_n} \int\limits_{y_n} 
		f(\mathbf{x} + i \mathbf{y})
	\, \prod_n dx_n dy_n.
\end{equation}
Since
\begin{equation}\label{eqn:gi_def_z_measure}
	dz dz^* = -2i dx_n dy_n,
\end{equation}
we could equivalently write
\begin{equation}\label{eqn:gi_chi_wrt_z}
	\chi  = \int\limits_{z^{\vphantom *}_n} \int\limits_{z^*_n} 
		f(\mathbf{x} + i \mathbf{y})
	\, \prod_n \frac{dz^{\vphantom *}_n dz^*_n}{2 i}.
\end{equation}
where the measure is now expressed only in terms of the $z_n$.

To evaluate the integral, we first note that since $H$ is a positive definite Hermitian matrix it admits a decomposition
\begin{equation}\label{eqn:gi_h_decomp}
	H = U \cdot S \cdot U^\dagger
\end{equation}
where
\begin{equation}\label{eqn:gi_def_s}
	S_{mn} = \lambda_m \delta_{mn}
\end{equation}
where the $\{\lambda_m\}$ are real and positive and $U$ is a unitary matrix satisfying
\begin{equation}\label{eqn:gi_u_unitary}
	U^\dagger \cdot U = U \cdot U^\dagger = I.
\end{equation}
We can therefore make a change of variables $\mathbf{k} = U^\dagger \mathbf{z}$ in (\ref{eqn:gi_chi_wrt_z}) to obtain
\begin{equation}\label{eqn:gi_chi_wrt_k}
	\chi  = \int\limits_{z^{\vphantom *}_n} \int\limits_{z^*_n} 
		\exp \left( -\mathbf{k}^\dagger \cdot S \cdot \mathbf{k} \right)
	\, \prod_n \frac{i dk^{\vphantom *}_n dk^*_n}{2}.
\end{equation}
Changing the integration variables again to the real and imaginary parts $\mathbf{u}$ and $\mathbf{v}$ of $\mathbf{k}$, we can then evaluate the integral using the normal rules for real Gaussian integrals:
\begin{equation}\label{eqn:gi_chi_wrt_u_and_v}
	\chi  = \prod_n \int\limits_{u_n} \int\limits_{v_n}
		e^{-\lambda_n (u_n^2 + v_n^2)}
	\, du_ndv_n
	= \prod_{n=1}^N \frac{\pi}{\lambda_n}.
\end{equation}
Making the identification $\det(H) = \prod_{n=1}^N \lambda_n$, we hence obtain the final result
\begin{equation}\label{eqn:gi_int1}
	\frac{1}{\det(H)} = \int
		\exp \left( -\mathbf{z}^\dagger \cdot H \cdot \mathbf{z} \right)
	\, \prod_n \frac{dz^{\vphantom *}_n dz^*_n}{2 \pi i}.
\end{equation}

(\ref{eqn:gi_int1}) can be used to obtain an additional result that is of use in this work.
Consider the integral
\begin{equation}\label{eqn:gi_with_mean}
	f(A, \mathbf{b}) = \int
		\exp \left(
			\mathbf{z}^\dagger \cdot A \cdot \mathbf{z}
			- \mathbf{b}^\dagger \cdot \mathbf{z}
			- \mathbf{z}^\dagger \cdot \mathbf{b}
			- (\mathbf{z} - \boldsymbol{\mu})^\dagger
			\cdot \Sigma^{-1} \cdot (\mathbf{z} - \boldsymbol{\mu})
		\right)
	\, \prod_n \frac{dz^{\vphantom *}_n dz^*_n}{2 \pi i}.
\end{equation}
where $\Sigma$ and $A$ are invertible positive-definite Hermitian matrices.
The exponent can be refactored as
\begin{equation}
\begin{aligned}
	& \mathbf{z}^\dagger \cdot \boldsymbol{A} \cdot \mathbf{z}
		+ \mathbf{b}^\dagger \cdot \mathbf{z}
		+ \mathbf{z}^\dagger \cdot \mathbf{b}
		- (\mathbf{z} - \boldsymbol{\mu})^\dagger
			\cdot \boldsymbol{\Sigma}^{-1} \cdot (\mathbf{z} - \boldsymbol{\mu}) \\
	&= -(\mathbf{z} - \mathbf{p})^\dagger 
		\cdot (\boldsymbol{\Sigma}^{-1} - \boldsymbol{A}) \cdot (\mathbf{z} - \mathbf{p})
		- \boldsymbol{\mu}^\dagger \cdot \boldsymbol{\Sigma}^{-1} \cdot \boldsymbol{\mu} \\
	&+ (\boldsymbol{\mu} - \boldsymbol{\Sigma} \cdot \mathbf{b})^\dagger
		\cdot \boldsymbol{\Sigma}^{-1}
		\cdot (\boldsymbol{I} - \boldsymbol{\Sigma} \cdot \boldsymbol{A})^{-1}
		\cdot (\boldsymbol{\mu} - \boldsymbol{\Sigma} \cdot \mathbf{b})
\end{aligned}
\end{equation}
where
\begin{equation}
	\mathbf{p} =  (\boldsymbol{I} - \boldsymbol{\Sigma} \cdot \boldsymbol{A})^{-1}
		\cdot (\boldsymbol{\mu} + \boldsymbol{\Sigma} \mathbf{b}).
\end{equation}
Therefore, if we make the change of variables $\mathbf{z} - \mathbf{p} \rightarrow \mathbf{z}$ in (\ref{eqn:gi_with_mean}), we can apply (\ref{eqn:gi_int1}) to obtain
\begin{equation}
\begin{aligned}
	&\frac{1}{\det(\boldsymbol{\Sigma})} \int
		\exp \left(
			\mathbf{z}^\dagger \cdot \boldsymbol{A} \cdot \mathbf{z}
			- \mathbf{b}^\dagger \cdot \mathbf{z}
			- \mathbf{z}^\dagger \cdot \mathbf{b}
			- (\mathbf{z} - \boldsymbol{\mu})^\dagger
			\cdot \boldsymbol{\Sigma}^{-1} \cdot (\mathbf{z} - \boldsymbol{\mu})
		\right)
	\, \prod_n \frac{dz^{\vphantom *}_n dz^*_n}{2 \pi i} \\
	&= \frac{1}{\det(\boldsymbol{I} - \boldsymbol{\Sigma} \cdot \boldsymbol{A})}
	\exp \left(
		- \boldsymbol{\mu}^\dagger \cdot \boldsymbol{\Sigma}^{-1} \cdot \boldsymbol{\mu}
		+ (\boldsymbol{\mu} - \boldsymbol{\Sigma} \cdot \mathbf{b})^\dagger
		\cdot \boldsymbol{\Sigma}^{-1}
		\cdot (\boldsymbol{I} - \boldsymbol{\Sigma} \cdot \boldsymbol{A})^{-1}
		\cdot (\boldsymbol{\mu} - \boldsymbol{\Sigma} \cdot \mathbf{b})
	\right).
\end{aligned}
\end{equation}

\subsection{Useful functional derivatives}\label{sec:useful_functional_derivatives}

In this section we will derive a general expressions for the functional derivatives of the determinant and inverses of operator functions of the form $f(t_1, t_2)$.
In both cases we will start with a matrix prototype and take the relevant continuum limit.

In the case of the inverse, we start with the matrix identity
\begin{equation}
	\boldsymbol{M}^{-1} \cdot \boldsymbol{M} = \boldsymbol{I}.
\end{equation}
for invertible matrix $\boldsymbol{M}$.
Differentiating this expression with respect to a parameter $s$ yields
\begin{equation}
	\frac{d\boldsymbol{M}^{-1}}{ds} \cdot \boldsymbol{M}
	+ \boldsymbol{M}^{-1} \cdot \frac{d\boldsymbol{M}}{ds}
	= \boldsymbol{0},
\end{equation}
which may be rearranged to
\begin{equation}
	\frac{d\boldsymbol{M}^{-1}}{ds} =
	- \boldsymbol{M}^{-1}
	\cdot \frac{d\boldsymbol{M}}{ds}
	\cdot \boldsymbol{M}^{-1}.
\end{equation}
The continuum version of this expression is
\begin{equation}\label{eqn:fd_inv}
	\frac{\delta g (t_1, t_2)}{\delta s(t_3)}
	= - \iint g(t_1, t_4) \frac{\delta f(t_4, t_5)}{\delta s(t_3)}  g(t_5, t_2) \, dt_4 dt_5
\end{equation}
where $g(t_1, t_2)$ is defined by
\begin{equation}\label{eqn:def_inv_f}
	\iint g(t_1, t_2) f(t_2, t_3) h(t_3) \, dt_2 dt_4 = h(t_1).
\end{equation}

In the case of the determinant, we can start with Jacobi's formula:
\begin{equation}
	\frac{d}{ds} \det(\boldsymbol{M})
	= \det(\boldsymbol{M}) \Tr \left[
		\boldsymbol{M}^{-1} \cdot \frac{d\boldsymbol{M}}{ds}
	\right].
\end{equation}
This has a straightforward continuum limit of the form
\begin{equation}\label{eqn:fd_det}
	\frac{\delta \det[f]}{\delta s(t_n)}
	= \det[f] \iint g(t_1, t_2) \frac{\delta f(t_2, t_1)}{\delta s(t_n)} \, dt_1 dt_2
\end{equation}
where $g(t_1, t_2)$ is as defined by (\ref{eqn:def_inv_f}).

\subsection{Sinc functions as delta functions}\label{sec:sinc_functions_as_delta_functions}

Sinc functions can be treated like delta functions under certain conditions.
Consider the integral
\begin{equation}\label{eqn:sinc_int_demo_integral}
	\int_{-\infty}^{\infty} \sinc (\{\nu' - \nu \} \tau) S(\nu') \, d\nu'
\end{equation}
where we use the definition
\begin{equation}\label{eqn:sinc_int_def_sinc}
	\sinc x = \frac{\sin (\pi x)}{\pi x}
\end{equation}
and $\tau$ is a positive constant.
As $\tau$ is increased, the $\sinc$ function in (\ref{eqn:sinc_int_demo_integral}) becomes increasingly sharply peaked in the region $|\nu - \nu'| \tau < \pi$.
For sufficiently large $\tau$, will $S(\nu)$ no longer vary significantly over this region and we may approximate
\begin{equation}\label{eqn:sinc_int_demo_integral_approx}
	\int \sinc (\{\nu' - \nu \} \tau) S(\nu') \, d\nu'
	\approx S(\nu) \int_{-\infty}^{\infty} \sinc (\{\nu' - \nu \} \tau) \, d\nu'
	= \frac{1}{\tau} S(\nu).
\end{equation}
Hence we see that as $\tau \rightarrow \infty$, we have
\begin{equation}\label{eqn:sinc_int_sinc_to_delta}
	\sinc (\{\nu' - \nu \} \tau) \rightarrow \delta(\nu - \nu') / \tau.
\end{equation}

The same reasoning can be applied to the square of a $\sinc$ function.
Repeating the preceding analysis, we have
\begin{equation}\label{eqn:sinc_int_sinc_squared_int}
	\int_{-\infty}^{\infty} \sinc^2 (\{\nu' - \nu \} \tau) S(\nu') \, d\nu'
	\approx S(\nu) \int_{-\infty}^{\infty} \sinc^2 (\{\nu' - \nu \} \tau) \, d\nu'.
\end{equation}
Since
\begin{equation}\label{eqn:sinc_int_sinc_squared_int_eval}
	\int_{-\infty}^{\infty} \sinc^2 (\{\nu' - \nu \} \tau) \, d\nu' = 1 / \tau,
\end{equation}
we therefore see
\begin{equation}\label{eqn:sinc_int_sinc_squared_approx}
	\sinc^2 (\{\nu' - \nu \} \tau) \rightarrow \delta(\nu - \nu') / \tau
\end{equation}
as $\tau \rightarrow \infty$.

\newpage

\section{Power waves}\label{sec:tl_signals}

In this section we provide more information about the representation of the input signal that is used throughout the analysis.
It is assumed this signal is supplied to the input ports along an ideal, lossless, transmission line with characteristic impedance $\eta$.

Consider a single input line.
Let $z$ denote position along the line as measured from the input port, so the port itself is at $z=0$.
It is well known that we can write the current $I$ in the \emph{negative} $z$-direction and voltage $V$ at any point and time on the line in the form
\begin{equation}\label{eqn:tl_v}
	V(z, t) = \int_0^\infty \Re \biggl[ \biggl\{
		V_+ (\nu) e^{i k z} + V_- (\nu) e^{-i k z}
	\biggr\} e^{2 \pi i \nu t} \biggr] \, d\nu
\end{equation}
and
\begin{equation}\label{eqn:tl_i}
	I(z, t) = \int_0^\infty \Re \biggl[ \biggl\{
		I_+ (\nu) e^{i k z} + I_- (\nu) e^{-i k z}
	\biggr\} e^{2 \pi i \nu t} \biggr] \, d\nu,
\end{equation}
where
\begin{equation}\label{eqn:tl_i_from_v}
	\eta V_{\pm} (\nu) = \pm I_{\pm} (\nu)
\end{equation}
and
\begin{equation}\label{eqn:k_from_nu}
	k = 2 \pi c / \nu.
\end{equation}
$V_+ (\nu)$ and $V_- (\nu)$ can be interpreted as the complex amplitudes of voltage waves of frequency $\nu$ that travelling in the negative and positive directions respectively.
$I_+ (\nu)$ and $I_- (\nu)$ can be interpreted similarly.

Now let us define new signals
\begin{align}
	V_+ (t) &= \frac{1}{2} \bigl\{ V(t) + \eta I(t) \bigr\} \label{eqn:tl_v_plus} \\
	V_- (t) &= \frac{1}{2} \bigl\{ V(t) - \eta I(t) \bigr\}.  \label{eqn:tl_v_minus}
\end{align}
Using (\ref{eqn:tl_v})--(\ref{eqn:tl_i_from_v}) to substitute, we can alternatively express (\ref{eqn:tl_v_plus}) and (\ref{eqn:tl_v_minus}) in terms of voltage wave amplitudes as
\begin{align}
	V_+ (z, t) &= \int_0^\infty \Re \Bigl[ 
		V_+ (\nu) e^{i k z} e^{2 \pi i \nu t}
	\Bigr] \, d\nu \label{eqn:tl_v_plus_decomp} \\
	V_- (z, t) &= \int_0^\infty \Re \Bigl[ 
		V_- (\nu) e^{-i k z} e^{2 \pi i \nu t}
	\Bigr] \, d\nu. \label{eqn:tl_v_minus_decomp}
\end{align}
In this form it is clear that $V_+ (t)$ corresponds to the component of the signal moving towards the port and $V_- (t)$ to that leaving the port.

(\ref{eqn:tl_v_plus_decomp}) and (\ref{eqn:tl_v_minus_decomp}) make it possible to define a representation of the input signal at the port.
In the absence of a reflection $V_- (\nu)$ is zero and $V_+ (\nu)$, the input wave, entirely characterises the signal.
This suggests the input signal can be entirely represented in terms of the complex analytic signal
\begin{equation}\label{eqn:tl_input_signal_prototype}
	a(t) = \frac{1}{\kappa} \int_0^\infty
		V_+ (\nu) e^{2 \pi i \nu t} \, d\nu
\end{equation}
doe arbitrary normalisation $\kappa$, where the port voltage and current are
\begin{equation}\label{eqn:tl_port_voltage}
	V(0, t) = \kappa \Re \bigl[ a(t) \bigr]
\end{equation}
and
\begin{equation}\label{eqn:tl_port_current}
	\eta I(0, t) = \kappa \Re \bigl[ a(t) \bigr].
\end{equation}

All that is left is to specify $\kappa$.
We are interested in the input power supplied by a quasi-monochromatic input.
The latter requires $V_+ (\nu)$ is non-zero only over frequencies $2 |\nu - \nu_0| < \Delta \nu$, where $\Delta \nu \ll \nu_0$.
In this case, we can use (\ref{eqn:tl_port_voltage}), (\ref{eqn:tl_port_voltage}) and  (\ref{eqn:tl_input_signal_prototype}) to write the instantaneous input power as
\begin{equation}
\begin{aligned}
	P(t) = V(0, t) I(0, t) = \frac{\kappa^2}{2 \eta} \Bigl\{
		|a(t)|^2+ \Re \bigl[ \{ a(t) \}^2 \bigr]
	\Bigr\}
\end{aligned}
\end{equation}
where
\begin{equation}\label{eqn:tl_mod_a_squared}
\begin{aligned}
	&|a(t)|^2
	= \frac{1}{\kappa^2}
		\int_{\nu_0 - \Delta \nu/2}^{\nu_0 + \Delta \nu/2}
		\int_{\nu_0 - \Delta \nu/2}^{\nu_0 + \Delta \nu/2}
		V_+ (\nu_1) V_+ (\nu_2) e^{2 \pi i (\nu_1 - \nu_2) t}
	\, d\nu_1 d\nu_2 \\
	&=  \frac{1}{\kappa^2} \int_{-\Delta \nu}^{\Delta \nu} \biggl\{
		\int_{\nu_0 - \{\Delta \nu - |f|\} / 2}^{\nu_0 + \{\Delta \nu - |f|\} / 2}
		V_+ (\nu + f / 2) V_+ (\nu - f / 2) \, d\nu
	\biggr\} e^{2 \pi i f t} \, df
\end{aligned}
\end{equation}
and
\begin{equation}\label{eqn:tl_a_squared}
\begin{aligned}
	&\Re \bigl[ \{ a(t) \}^2 \bigr]
	= \frac{1}{\kappa^2} \Re \biggl[
		\int_{\nu_0 - \Delta \nu/2}^{\nu_0 + \Delta \nu/2}
		\int_{\nu_0 - \Delta \nu/2}^{\nu_0 + \Delta \nu/2}
		V_+ (\nu_1) V_+ (\nu_2) e^{2 \pi i (\nu_1 + \nu_2) t}
	\, d\nu_1 d\nu_2 \biggr] \\
	&= \frac{1}{2 \kappa^2}
		\int_{2 \nu_0 -\Delta \nu}^{2 \nu_0 + \Delta \nu} \biggl\{
		\int^{\Delta \nu - |2 \nu_0 - f|}_{|2 \nu_0 - f| - \Delta \nu}
		V_+ ((f + \nu) / 2) V_+ ((f - \nu) / 2) \, d\nu
	\biggr\} e^{2 \pi i f t} \, df
\end{aligned}
\end{equation}
(\ref{eqn:tl_mod_a_squared}) shows $|a(t)|^2$ contains only terms with frequencies in the range $0 < \nu < \Delta \nu$ and is therefore slowly varying.
(\ref{eqn:tl_a_squared}), by contrast, shows $\Re \bigl[ \{ a(t) \}^2 \bigr]$ contains only terms near $2 \nu_0$ in frequency and is rapidly varying.
Consequently, we can associate $|a(t)|^2$ with the slowly-varying power that would be retained on time averaging over multiple cycles.
We would also normally associate $|V_+ (\nu)|^2 / (2 \eta)$ with this average power flow.
Hence if we choose $1 / \kappa = \sqrt{\eta}$, i.e.
\begin{equation}\label{eqn:input_signal_prototype}
\begin{aligned}
	a(t) &= \frac{1}{\sqrt{\eta}} \int_0^\infty
		V_+ (\nu) e^{2 \pi i \nu t} \, d\nu \\ 
	V_+ (\nu) &= \frac{1}{2} \int_{-\infty}^\infty
		\bigl\{ V (0,t) + \eta I(0, t) \bigr\} e^{-2 \pi i \nu t} \, dt,
\end{aligned}
\end{equation}
then the cycle averaged power flow is
\begin{equation}
	P_\text{av} (t) = \frac{1}{2} |a(t)|^2
\end{equation}
as required in the main note.

\newpage

\section{Noise equivalent power}\label{sec:nep}

\subsection{Canonical derivation}\label{sec:nep_canonical}

The noise equivalent power of a detector system is defined as the input power level that gives a signal to noise ratio of unity in an output bandwidth of 1\,Hz.
In this section we will relate the NEP to the output noise, responsivity and modulation strategy employed by the system.

Consider a system where the input power level is modulated into a sine wave and the power detector output is then filtered to define the output bandwidth.
In this case we can write the real-valued output of the detector as
\begin{equation}\label{eqn:nep_detector_output}
\begin{aligned}
	d(t) &= p(t) + n_d (t) \\
	s(t) &= \begin{cases}
		\sqrt{2} \eta_m r(f_0) p_i \cos (2 \pi f_0 t) & f_0 \neq 0 \\
		r(0) p_i & f_0 = 0
	\end{cases}
\end{aligned}
\end{equation}
Here $s(t)$ is the signal due to the input power and $n(t)$ is the noise at the detector output.
In the case of the signal, $p_i$ is the input power level of interest, $f_0$ is the modulation frequency, $r(f_0)$ is the responsivity of the detector for signals modulated at $f_0 \neq 0$ and $0 < \eta_m < 1$ is an assumed modulation efficiency when $f_0 \neq 0$.
For example, if input is chopped and the detector only responds to the fundamental frequency, $\eta_m = \sqrt{2} / \pi$.
We will assume the noise is a zero-mean, stationary, signal, in which case can be characterised by the correlation function
\begin{equation}\label{eqn:nep_output_noise_acf}
	\Gamma_{n_d} (t) = \langle n(t) n(t + \tau) \rangle
	= 2 \Re \biggl[ \int_0^\infty S(f) e^{2 \pi i f \tau} \, df \biggr],
\end{equation}
where $S(f)$ is the output noise power spectral density.
The resulting output at the filter is
\begin{equation}\label{eqn:nep_filter_output}
\begin{aligned}
	y(t) &= d(t) + n (t) \\
	s(t) &= \eta_m p_i \sqrt{G(f_0)} R(f_0) \cos (2 \pi f_0 t + \phi)
\end{aligned}
\end{equation}
where $G(f)$ is the frequency domain filter gain, $\phi$ is the phase shift due to the filter and
\begin{equation}\label{eqn:nep_filter_noise_acf}
	\Gamma_{n} (t) = \langle n(t) n(t + \tau) \rangle
	= 2 \Re \biggl[ \int_0^\infty G(f) S(f) e^{2 \pi i f \tau} \, df \biggr].
\end{equation}

The signal to noise ratio filter output is the ratio of the time-averaged power (i.e. square) of the useful signal, $P_\text{sig}$, to the time-averaged average power in the noise, $P_\text{noise}$.
Here, the former is
\begin{equation}\label{eqn:nep_p_sig}
	 P_\text{sig} = \begin{cases}
	 	\eta_m^2 p_i^2 G(f_0) \{ r(f_0) \}^2 & f_0 \neq 0 \\
	 	\{ r(f_0) \}^2 G(0) p_i & f_0 = 0
	\end{cases},
\end{equation}
when $f_0$ is non-zero.
The latter can be calculated using the fact a sufficiently long time average of a random variable is equivalent to the ensemble average, so
\begin{equation}\label{eqn:nep_p_noise}
	 P_\text{noise} = \langle \{n(t)\}^2 \rangle = \Gamma_n (0)
	 = 2 \int_0^\infty G(f) S(f) \, df.
\end{equation}
Combining these results, we find
\begin{equation}\label{eqn:nep_first}
	\text{SNR} =
		\frac{\{r (f_0) \}^2 G(f_0) p_i^2}{2 \int_0^\infty G(f) S(f) \, df} \times
	\begin{cases}
		1 & f_0 = 0 \\
		\eta_m^2 & f_0 \neq 0
	\end{cases}.
\end{equation}

In order to calculate the NEP, we need to define the output bandwidth for the scenario considered.
In practical system, the filter bandwidth is normally centred on and symmetric about the modulation frequency, as well as being sufficiently narrow that $S(f)$ does not vary appreciably.
We may therefore approximate
\begin{equation}\label{eqn:noise_approx}
	P_\text{noise} = 2 \int_0^\infty S(f) G(f) \, df \approx
	2 S(f_0) \Delta f
\end{equation}
where
\begin{equation}\label{eqn:bandwidth}
	\Delta \nu = \frac{1}{G(f_0)} \int_{0}^{\infty} G(f) \, df
\end{equation}
is the effective bandwidth.
Now we can write
\begin{equation}\label{eqn:nep_approx_nep}
	\text{SNR} \approx \frac{\{r(f_0)\}^2 p^2_i}{2 S(f_0)\Delta f}
	\times \begin{cases}
		1 & f = f_0 \\
		\eta_m^2 & f \neq f_0
	\end{cases}.
\end{equation}

We can now find the NEP by setting $\text{SNR} = 1$, $p_i = \text{NEP}$ and $\Delta \nu =$ 1\,Hz in (\ref{eqn:nep_approx_nep}) and rearranging, in line with the definition.
The result is
\begin{equation}\label{eqn:nep_final_result}
	\text{NEP} (f_0) = \sqrt{\frac{2 S(f_0)}{\{r (f_0) \}^2}}
	\times \begin{cases}
		1 & f = f_0 \\
		\frac{1}{\eta_m^2} & f \neq f_0
	\end{cases}.
\end{equation}
We see that the NEP is the effective single-sided amplitude power spectral density, $\sqrt{2 S(f_0)}$, divided by the amplitude responsivity, $r(f_0)$ as we might expect.
Alternatively, we can view the NEP as the root of the effective input referenced noise power spectral density, as given by $2 S(f_0) / \{ r(f_0) \}^2$.

\subsection{Cram\'er Rao bound}

An alternate way of thinking about the problem is using the Cram\'er-Rao, which removes the need to specifically consider a filter.
Assume the detector output is of the form
\begin{equation}\label{eqn:crb_d}
	d(t) = p_i g(t) + n(t)
\end{equation}
where $g(t)$ is the modulation function and $n(t)$ is a zero-mean Gaussian noise process.
Assume we have access to the signal over an interval $|t| < \tau / 2$.
The corresponding Cram\'er-Rao bound on the RMS error of any estimator of $p_i$ is
\begin{equation}\label{eqn:crb_rms_err_full}
	\Delta p_i^2 \geq \frac{1}{
		\iint_{-\tau/2}^{+\tau/2} \Gamma^{-1} (t_1, t_2) g(t_1) g(t_2) \, dt_1 dt_2
	} 
\end{equation}
where $\Gamma^{-1} (t_1, t_2)$ is defined by
\begin{equation}\label{eqn:crb_inv_gamma}
	\iint_{-\tau/2}^{\tau/2} \Gamma^{-1} (t_1, t_2) \Gamma(t_2, t_3) f(t_3) \, dt_2 dt_3
	= f(t_1).
\end{equation}
In this section we will show how (\ref{eqn:crb_rms_err_full}) can be used as a more general basis for the definition of NEP.

The first step in calculating (\ref{eqn:crb_inv_gamma}) is evaluating the inverse of the correlation function.
To do, it is useful to introduce a set of basis functions of the form
\begin{equation}\label{eqn:cr_bound_basis}
	f_n (t) = \frac{1}{\sqrt{\tau}} e^{2 \pi i n t / \tau}
\end{equation}
defined for all positive and negative values of integer $n$.
These basis functions satisfy the orthonormality relationship
\begin{equation}\label{eqn:cr_bound_basis_orthonormality}
	\int_{-\tau/2}^{\tau/2} f^*_m (t) f_n(t) \, dt = \delta_{mn}.
\end{equation}
We can expand $\Gamma (t_1, t_2)$ and $\Gamma^{-1} (t_1, t_2)$ in terms of these basis functions as
\begin{equation}\label{eqn:cr_bound_gamma_expansion}
	\Gamma (t_1, t_2) = \sum_{m,n} \Gamma^{\vphantom 1}_{mn}
		f^{\vphantom *}_m (t) f^*_n (t)
\end{equation}
and
\begin{equation}\label{eqn:cr_bound_gamma_inv_expansion}
	\Gamma^{-1} (t_1, t_2) = \sum_{m,n} \Gamma^{-1}_{mn}
		f^{\vphantom *}_m (t) f^*_n (t)
\end{equation}
where
\begin{equation}
	\Gamma_{mn} = \int_{-\tau/2}^{\tau/2} \int_{-\tau/2}^{\tau/2}
		\Gamma (t_1, t_2) e^{-2 \pi i (m t_1 - n t_2) / \tau}
	\, dt_1 dt_2
\end{equation}
and $\boldsymbol{\Sigma}^{-1} \cdot \boldsymbol{\Sigma}^{\vphantom 1} = \boldsymbol{I}$.
Using the Wiener-Khinchin theorem we have
\begin{equation}
\begin{aligned}
	\Gamma_{mn}
	&= \tau \int_{-\infty}^{\infty}
		S(f) \sinc ((f - m / \tau) \tau) \sinc ((f - n / \tau) \tau)
	\, df \\
	&= \int_{-\infty}^{\infty}
		S((x + n) / \tau) \sinc (x - m + n) \sinc (x)
	\, dx.
\end{aligned}
\end{equation}
The product of sincs in this integral is sharply peaked at a frequency of $(m - n) / \tau$ and provided $f \tau$ is large enough and $S(f)$ varies sufficiently slowly, we can approximate
\begin{equation}
\begin{aligned}
	\Gamma_{mn}
	&\approx S(m / \tau) \int_{-\infty}^{\infty}
		\sinc (x - m + n) \sinc (x)
	\, dx.
\end{aligned}
\end{equation}
The integral evaluates to
\begin{equation}
	\int_{-\infty}^{\infty}
		\sinc (x - m + n) \sinc (x)
	\, dx
	= \sinc(m - n) = \delta_{mn}.
\end{equation}
Hence we can approximate $\Sigma_{mn} \approx S(m / \tau) \delta_{mn}$ and $\Sigma^{-1}_{mn} = \delta_{mn} / S(m / \tau)$, so
\begin{equation}\label{eqn:crb_inv_gamma_sum}
	\Gamma^{-1} (t_1, t_2)
	= \sum_m \frac{1}{\tau S(m / \tau)} e^{2 \pi i m (t_1 - t_2) / \tau}.
\end{equation}

Now we can calculate the Cram\'er-Rao bound for the modulation scheme considered.
In that case
\begin{equation}\label{eqn:crb_g}
	g(t) = \sqrt{2} \eta_m r(f_0) \Re \bigl[ e^{2 \pi k t / \tau} \bigr]
\end{equation}
for integer $k$.
Using (\ref{eqn:crb_inv_gamma_sum}) and (\ref{eqn:crb_g}), we have
\begin{equation}
\begin{aligned}
	&\iint_{-\tau/2}^{+\tau/2} \Gamma^{-1} (t_1, t_2) g(t_1) g(t_2) \, dt_1 dt_2 \\
	 &= \frac{\tau \eta^2_m \{ r(k/\tau) \}^2}{2} \sum_m
		\frac{1}{S(m / \tau)} \Bigl\{ \sinc^2 (m - k) + \sinc^2 (m + k) \Bigr\} \\
	&+ \tau \eta^2_m \{ r(k/\tau) \}^2 \sum_m
		\frac{1}{S(m / \tau)} \sinc (m + k) \sinc (m - k)  \\
	&= \frac{\tau \eta^2_m \{ r(k/\tau) \}^2}{S(k/\tau)}.
\end{aligned}
\end{equation}
Hence the Cram\'er-Rao bound yields
\begin{equation}\label{eqn:crb_rms_err_final}
	\Delta p_i^2 \geq 
	\frac{1}{\eta_m^2}
	\times \underbrace{\frac{2 S(f_0)}{\{ r(f_0) \}^2}}_{\{\text{NEP}(f_0)\}^2}
	\times \underbrace{\frac{1}{2 \tau}}_{\Delta f}.
\end{equation}
Here we have identified $\Delta \nu = 1 / \tau$ as the effective output bandwidth had we averaged over the record length $\tau$ and the NEP as previously defined.
(\ref{eqn:crb_rms_err_final}) gives the result that is typically used for bolometer sensitivity calculations,
\begin{equation}
	\Delta p_\text{\text{rms}} \approx \frac{\text{NEP} (f_0)}{\eta_m \sqrt{2 \tau}}
\end{equation}
for integration time $\tau$ and provides a statistical motivation for the NEP as defined as a measure of sensitivity.

\newpage

\section{Full time domain correlation functions in the homodyne scheme}\label{sec:full_td_calc_homo}

\subsection{Strategy}\label{sec:full_td_calc_homo_strategy}

The calculation of the correlation functions of the homodyne receiver output is algebraically involved as the number of terms increases rapidly with the order of the derivative.
It is first necessary to evaluate the derivatives of the individual $\mathcal{M}_n [s(t)]$, then these can be combined to give the derivatives of the full function.

The results for the functional derivatives of the individual $\mathcal{M}_n [s(t)]$ are listed in Section \ref{sec:deriv_individual_homo_mgf}.
These were calculated both by hand and using a computer algebra system and the method by which the latter was carried out is described in Section \ref{sec:ft_in_sympy}.
Section \ref{sec:deriv_individual_homo_mgf} gives both the full results and the appropriate approximate results in the strong-pump regime described in the main report.

In this section we will work in terms of $q(t)$ as defined by (\ref{eqn:def_q}), as the results for $y(t)$ follow by simple linear transformation and it avoids the extra algebra of dealing with $g(t_1, t_2)$.
The expected value of $q(t)$ can be written in terms of the derivatives of the $\mathcal{M}_n$ as
\begin{equation}\label{eqn:exp_y_homo_from_m}
	\langle q(t_1) \rangle
	= \sum_{n=1}^5 \left( \frac{\delta \mathcal{M}_n [s(t)]}{\delta s(t_1)} \right)_{s(t)=0}.
\end{equation}
Using the results from Section \ref{sec:deriv_individual_homo_mgf}, we then have
\begin{equation}\label{eqn:exp_d}
	\langle q (t_1) \rangle = 0
\end{equation}
as all the first order derivatives are zero.
For a perfectly balanced system the expected value of $q (t)$ is zero, as would be expected physically.

The correlation function of $q (t)$ involves taking the second order derivative.
Because the first order derivatives of the $\mathcal{M}_n$ are all zero for $s(t) = 0$, we only need consider terms involving the second order derivatives of the constituent functionals.
Hence we may simplify
\begin{equation}\label{eqn:exp_d2_from_m}
	\langle q(t_1) q(t_2) \rangle
	= \sum_{n=1}^5 \left( \frac{\delta^2 \mathcal{M}_n [s(t)]}
	{\delta s(t_1) \delta s(t_2)} \right)_{s(t)=0}.
\end{equation}
Using the results from Section \ref{sec:deriv_individual_homo_mgf} in the strong pump regime, we find
\begin{equation}\label{eqn:exp_d2}
\begin{aligned}
	\langle q(t_1) q(t_2) \rangle
	& \sim \underbrace{ \,
		\frac{1}{2} h \nu_0 \left| a_p \right|^2 \delta(t_1 - t_2)
	}_{(i)}
	+ \underbrace{ \frac{1}{2} |a_p|^2 \Re \Bigl[
		\Gamma (t_1, t_2) e^{-2 \pi i \vp (t_1 - t_2)}
	\Bigr]
	}_{(ii)} \\
	&+ \underbrace{
		\Gamma_{n_1} (t_1, t_2) + \Gamma_{n_2} (t_1, t_2) 
	}_{(iii)}.
\end{aligned}
\end{equation}
The correlation function consists of three terms.
Term (i) corresponds to the shot noise associated with the detection of the pump signal; the shot noise is uncorrelated between the detectors and does not cancel, unlike the wave noise.
Term (ii) is the down-converted signal of interest, as described by $\Gamma_s$, weighted by the pump power.
Term (iii) is the internal noise in the detectors.

The third order correlation function can be calculated in the same manner.
Ignoring terms involving first order derivatives, we may again simplify
\begin{equation}\label{eqn:exp_d3_from_m}
\begin{aligned}
	&\langle q(t_1) q(t_2) q(t_3) \rangle \\
	&= \sum_{n=1}^5 \left( \frac{\delta^3 \mathcal{M}_n [s(t)]}
	{\delta s(t_1) \delta s(t_2) \delta s(t_3)} \right)_{s(t)=0}.
\end{aligned}
\end{equation}
However, all of these derivatives evaluate to zero and we obtain
\begin{equation}\label{eqn:exp_d3}
\begin{aligned}
	\langle q(t_1) q(t_2) q(t_3) \rangle = 0.
\end{aligned}
\end{equation}

Finally, we will need the fourth order correlation function to calculate the radiometric sensitivity.
Using the same approach as previously, we have
\begin{equation}\label{eqn:exp_d4_from_m}
\begin{aligned}
	&\langle q(t_1) q(t_2) q(t_3) q(t_4) \rangle \\
	&= \sum_{n=1}^5 \biggr( \frac{\delta^4 \mathcal{M}_n}
	{\delta s(t_1) \delta s(t_2) \delta s(t_3) \delta s(t_4)} \biggr)_{s(t)=0} \\
	& + \sum_{(r, s)} \sum_{(m, n)} \biggr( \frac{\delta^4 \mathcal{M}_r}{\delta s(t_m) \delta s(t_n)} \biggr)_{s(t)=0}
	\biggr( \frac{\delta^4 \mathcal{M}_s}{\delta s(t_p) \delta s(t_q)} \biggr)_{s(t)=0},
\end{aligned}
\end{equation}
Here the first sum in the second term is taken over all combinations $(r, s)$ drawn from $\{1, 2, 3, 4, 5\}$ without replacement and without order.
The second sum is taken over all combinations $(m, n)$ drawn from $\{1, 2, 3, 4\}$ without replacement and order and $p$ and $q$ are the leftover elements in each choice.
This simplifies to
\begin{equation}\label{eqn:exp_d4_factored}
\begin{aligned}
	&\langle q(t_1) q(t_2) q(t_3) q(t_4) \rangle \\
	&= \biggl\{
		\frac{1}{2} h \nu_0 \delta(t_1 - t_3)
		+ \Re \Bigl[ \Gamma (t_1, t_3) e^{-2 \pi i \nu_0 (t_1 - t_3)} \Bigr]
		+ \sum_{m=1}^2 \Gamma_{n_m} (t_1, t_3)
	\biggr\} \\
	& \times \biggl\{
		\frac{1}{2} h \nu_0 \delta(t_2 - t_4)
		+ \frac{1}{2} \Re \Bigl[ \Gamma (t_2, t_4) e^{-2 \pi i \nu_0 (t_2 - t_4)} \Bigr]
		+ \sum_{m=1}^2 \Gamma_{n_m} (t_2, t_4)
	\biggr\} \left| a_p \right|^4 \\
	&+ \biggl\{
		\frac{1}{2} h \nu_0 \delta(t_1 - t_2)
		+ \frac{1}{2} \Re \Bigl[ \Gamma (t_1, t_2) e^{-2 \pi i \nu_0 (t_1 - t_2)} \Bigr]
		+ \sum_{m=1}^2 \Gamma_{n_m} (t_1, t_2)
	\biggr\} \\
	& \times \biggl\{
		\frac{1}{2} h \nu_0 \delta(t_3 - t_4)
		+ \frac{1}{2} \Re \Bigl[ \Gamma (t_3, t_4) e^{-2 \pi i \nu_0 (t_3 - t_4)} \Bigr]
		+ \sum_{m=1}^2 \Gamma_{n_m} (t_3, t_4)
	\biggr\} \left| a_p \right|^4 \\
	&+ \biggl\{
		\frac{1}{2} h \nu_0 \delta(t_1 - t_4)
		+ \frac{1}{2} \Re \Bigl[ \Gamma (t_1, t_4) e^{-2 \pi i \nu_0 (t_1 - t_4)} \Bigr]
		+ \sum_{m=1}^2 \Gamma_{n_m} (t_1, t_4)
		\biggr\} \\
	& \times \biggl\{
		\frac{1}{2} h \nu_0 \delta(t_2 - t_3)
		+ \frac{1}{2} \Re \Bigl[ \Gamma (t_2, t_3) e^{-2 \pi i \nu_0 (t_2 - t_3)} \Bigr]
		+ \sum_{m=1}^2 \Gamma_{n_m} (t_2, t_3)
	\biggr\} \left| a_p \right|^4.
\end{aligned}
\end{equation}
in the strong pump limit.

(\ref{eqn:exp_d}), (\ref{eqn:exp_d2}), (\ref{eqn:exp_d3}) and (\ref{eqn:exp_d4_factored}) taken together imply that in the strong pump regime the difference signal behaves (at least to fourth-order) as a zero mean Gaussian random signal with characteristic correlation function
\begin{equation}\label{eqn:d_correlation_function_with_quantum_noise}
	\Gamma_q (t_1, t_2) =\biggl\{
		\frac{1}{2} h \nu_0 \delta(t_1 - t_2)
		+ \frac{1}{2} \Re \Bigl[ \Gamma (t_1, t_2) e^{-2 \pi i \nu_0 (t_1 - t_2)} \Bigr]
		+ \sum_{m=1}^2 \Gamma_{n_m} (t_1, t_2)
	\biggr\} \left| a_p \right|^2.
\end{equation}

\subsection{Evaluation of Fourier transforms using sympy}\label{sec:ft_in_sympy}

The software used to evaluate the higher order moments, \emph{sympy}, does not handle Fourier transforms well.
Instead the following procedure was used, which we will illustrate using the term
\begin{equation}\label{eqn:sympy_demo_term_1}
\begin{aligned}
	& h(\nu_1, \nu_2, \nu_3, \nu_4) = \\
	&\iiiint
		\Gamma(t_1, t_1) \Gamma^* (t_3, t_4)
		a(t_3) a^* (t_4) \delta (t_2 - t_1)
		e^{-2 \pi i (\nu_1 t_1 + \nu_2 t_2 + \nu_3 t_3 + \nu_4 t_4)}
		\, dt_1 dt_2 dt_3 dt_4,
\end{aligned}
\end{equation}
where the integrals with respect to $\{t_n\}$ are all taken from $-\infty$ to $\infty$.

Firstly instances of $\Gamma$, $a$ and $\delta$ are replaced with their Fourier decompositions:
\begin{equation}\label{eqn:sympy_fd_a}
	a (t) = a_p e^{2 \pi i \nu_0 t}
\end{equation}
\begin{equation}\label{eqn:sympy_fd_gamma}
	\Gamma (t_1, t_2) = \int p(f) e^{2 \pi i f (t_1 - t_2)} \, df
\end{equation}
\begin{equation}\label{eqn:sympy_fd_delta}
	\delta (t) = \int e^{2 \pi i f t} \, df.
\end{equation}
The integrals are taken from $-\infty$ to $\infty$ and $p(f) = 0$ for $f < 0$.
A consistent set of dummy variables of the form $f_n$ are used for these decompositions and the integrals are restricted to positive values.
Replacement order is $a$, $\Gamma$ and then $\delta$, which ensures the $f_n$ with the lowest values of $n$ are used for the arguments of the noise power spectral density $p$.
In the case of the test term, this yields
\begin{equation}\label{eqn:sympy_demo_term_2}
\begin{aligned}
	& h(\nu_1, \nu_2, \nu_3, \nu_4) = |a_p|^2  \iiiint \iiint \\
	& p(f_1) p(f_2)
		e^{2 \pi i (-\nu_1 - f_3) t_1}
		e^{2 \pi i (-\nu_2 + f_3) t_2}
		e^{2 \pi i (-\nu_3 - f_2 + \nu_0) t_3}
		e^{2 \pi i (-\nu_4 + f_2 - \nu_0) t_4} \\
	&\, df_1 df_2 df_3 dt_1 dt_2 dt_3 dt_4.
\end{aligned}
\end{equation}

Next the integrals with respect to time are evaluated.
In practice this is most simply achieved by dropping the integrals with respect to time and making the replacements
\begin{equation}
	e^{2 \pi i x t_n} \rightarrow \delta (x).
\end{equation}
In the case of the test term, we obtain
\begin{equation}\label{eqn:sympy_demo_term_3}
\begin{aligned}
	& h(\nu_1, \nu_2, \nu_3, \nu_4) = \\
	&|a_p|^2 \iiint p(f_1) p(f_2)
		\delta(f_3 + \nu_1) \delta (f_3 - \nu_2)
		\delta (-\nu_3 - f_2 + \nu_0) \delta (f_2 - \nu_4 - \nu_0)
		\, df_1 df_2 df_3.
\end{aligned}
\end{equation}

Finally, we evaluate the integrals over $f_n$ where possible and \emph{sympy} handles this automatically.
In the case of the test term this yields
\begin{equation}\label{eqn:sympy_demo_term}
\begin{aligned}
	& h(\nu_1, \nu_2, \nu_3, \nu_4) = 
	|a_p|^2  p(\nu_4 + \nu_0) \delta(\nu_1 + \nu_2) \delta (\nu_3 + \nu_4)
	\int p(f_1) \, df_1.
\end{aligned}
\end{equation}

\subsection{Derivatives of the terms in the difference signal moment generating functionals}\label{sec:deriv_individual_homo_mgf}

The functional derivatives of the $\mathcal{M}_n [s(t)]$ are as follows:

\begin{equation}
	\mathcal{M}_n [0] = 1 \quad \forall  \quad n
\end{equation}
\begin{equation}
	\frac{\delta \mathcal{M}_n [0]}{\delta s(t_1)} = 0
	\quad \forall \quad n
\end{equation}

\begin{align}
	\frac{\delta^2 \mathcal{M}_1 [0]}{\delta s(t_1) \delta s(t_2)}
	&= \frac{1}{2} h \nu_0 \Gamma(t_1, t_1) \delta(t_1 - t_2) \\
	\frac{\delta^2 \mathcal{M}_2 [0]}{\delta s(t_1) \delta s(t_2)}
	&= \frac{1}{2} h \nu_0 |a_p|^2 \delta(t_1 - t_2) \\
	\frac{\delta^2 \mathcal{M}_3 [0]}{\delta s(t_1) \delta s(t_2)}
	&= \frac{1}{2} |a_p|^2 \Re \Bigl[
		\Gamma (t_1, t_2) e^{-2 \pi i \vp (t_1 - t_2)}
	\Bigr] \\
	\frac{\delta^2 \mathcal{M}_4 [0]}{\delta s(t_1) \delta s(t_2)}
	&= \Gamma_{n_1} (t_1, t_2) \\
	\frac{\delta^2 \mathcal{M}_5 [0]}{\delta s(t_1) \delta s(t_2)}
	&= \Gamma_{n_2} (t_1, t_2)
\end{align}

\begin{equation}
	\frac{\delta^3 \mathcal{M}_n [0]}{\delta s(t_1) \delta s(t_2) \delta s(t_3)}
	= 0 \quad \forall \quad n
\end{equation}
\begin{equation}
\begin{aligned}
\frac{\delta^4 \mathcal{M}_{1} \bigl[ 0 \bigr]}
	{\delta s(t_1) \delta s(t_2) \delta s(t_3) \delta s(t_4)}
	&= \frac{1}{2} (h \nu_0)^3 \Gamma(t_1, t_1)
		\delta(t_1 - t_2) \delta(t_1 - t_3) \delta(t_1 - t_4) \\
	&+ \frac{1}{4} (h \nu_0)^2 \Gamma(t_1, t_1) \Gamma(t_3, t_3)
		\delta(t_1 - t_2) \delta(t_3 - t_4) \\
	&+ \frac{1}{4} (h \nu_0)^2 \Gamma(t_1, t_1) \Gamma(t_2, t_2)
		\delta(t_1 - t_3) \delta(t_2 - t_4) \\
	&+ \frac{1}{4} (h \nu_0)^2 \Gamma(t_1, t_1) \Gamma(t_2, t_2)
		\delta(t_2 - t_3) \delta(t_1 - t_4) \\
	&+ \frac{1}{4} (h \nu_0)^2 |\Gamma(t_1, t_3)|^2
		\delta(t_1 - t_2) \delta(t_3 - t_4) \\
	&+ \frac{1}{4} (h \nu_0)^2 |\Gamma(t_1, t_2)|^2
		\delta(t_1 - t_3) \delta(t_2 - t_4) \\
	&+ \frac{1}{4} (h \nu_0)^2 |\Gamma(t_1, t_2)|^2
		\delta(t_1 - t_4) \delta(t_2 - t_3)
\end{aligned}
\end{equation}
\begin{equation}
\begin{aligned}
\frac{\delta^4 \mathcal{M}_{2} \bigl[ 0 \bigr]}
	{\delta s(t_1) \delta s(t_2) \delta s(t_3) \delta s(t_4)}
	&= \frac{1}{4} (h \nu_0)^2 |a_p|^4
		\delta(t_1 - t_3) \delta(t_2 - t_4) \\
	&+ \frac{1}{4} (h \nu_0)^2 |a_p|^4
		\delta(t_1 - t_4) \delta(t_2 - t_3) \\
	&+ \frac{1}{4} (h \nu_0)^2 |a_p|^4
		\delta (t_1 - t_2) \delta (t_3 - t_4) \\
	&+ \frac{1}{2} (h \nu_0)^3 |a_p|^2
		\delta (t_1 - t_2) \delta (t_1 - t_3) \delta(t_1 - t_4)
\end{aligned}
\end{equation}
\begin{equation}
\begin{aligned}
\frac{\delta^4 \mathcal{M}_{3} \bigl[ 0 \bigr]}
{\delta s(t_1) \delta s(t_2) \delta s(t_3) \delta s(t_4)}
&= \frac{1}{2} (h \nu_0)^2 |a_p|^2
	\Re \Bigl[\Gamma (t_1, t_2) e^{-2 \pi i \vp (t_1 - t_2)} \Bigr]
	\delta (t_1 - t_3) \delta (t_1 - t_4) \\
&+ \frac{1}{2} (h \nu_0)^2 |a_p|^2
	\Re \Bigl[ \Gamma (t_1, t_2) e^{-2 \pi i \vp (t_1 - t_2)} \Bigr]
	\delta (t_2 - t_3) \delta (t_2 - t_4) \\
&+ \frac{1}{2} (h \nu_0)^2 |a_p|^2
	\Re \Bigl[ \Gamma (t_1, t_3) e^{-2 \pi i \vp (t_1 - t_3)} \Bigr]
	\delta (t_1 - t_2) \delta (t_1 - t_4) \\ 
&+ \frac{1}{4} (h \nu_0)^2 |a_p|^2
	\Re \Bigl[ \Gamma (t_1, t_4) e^{-2 \pi i \vp (t_1 - t_4)} \Bigr]
	\delta (t_1 - t_2) \delta (t_1 - t_3) \\
&+ \frac{1}{4} h \nu_0 |a_p|^2
	\Re \Bigl[\Gamma (t_1, t_3) e^{-2 \pi i \vp (t_4 - t_3)} \Bigr]
	\delta (t_1 - t_2) \\
&+ \frac{1}{4} h \nu_0 |a_p|^2
	\Re \Bigl[ \Gamma (t_1, t_2) e^{-2 \pi i \vp (t_4 - t_2)} \Bigr]
	\delta (t_1 - t_3) \\
&+ \frac{1}{4} h \nu_0 |a_p|^2
	\Re \Bigl[ \Gamma (t_1, t_2) e^{-2 \pi i \vp (t_3 - t_2)} \Bigr]
	\delta (t_1 - t_4) \\
&+ \frac{1}{4} h \nu_0 |a_p|^2
	\Re \Bigl[ \Gamma (t_2, t_1) e^{-2 \pi i \vp (t_4 - t_1)} \Bigr]
	\delta (t_2 - t_3) \\
&+ \frac{1}{4} h \nu_0 |a_p|^2
	\Re \Bigl[ \Gamma (t_2, t_1) e^{-2 \pi i \vp (t_3 - t_1)} \Bigr]
	\delta (t_2 - t_4) \\
&+ \frac{1}{4} h \nu_0 |a_p|^2
	\Re \Bigl[ \Gamma (t_3, t_1) e^{-2 \pi i \vp (t_2 - t_1)} \Bigr]
	\delta (t_3 - t_4) \\
&+ \frac{1}{4} |a_p|^4
	\Re \Bigl[ \Gamma (t_3, t_1) e^{-2 \pi i \vp (t_3 - t_1)} \Bigr]
	\Re \Bigl[ \Gamma (t_4, t_2) e^{-2 \pi i \vp (t_4 - t_2)} \Bigr] \\
&+ \frac{1}{4} |a_p|^4
	\Re \Bigl[ \Gamma (t_3, t_2) e^{-2 \pi i \vp (t_3 - t_2)} \Bigr]
	\Re \Bigl[ \Gamma (t_4, t_1) e^{-2 \pi i \vp (t_4 - t_1)} \Bigr] \\
&+ \frac{1}{4} |a_p|^4
	\Re \Bigl[ \Gamma (t_2, t_1) e^{-2 \pi i \vp (t_2 - t_1)} \Bigr]
	\Re \Bigl[ \Gamma (t_4, t_3) e^{-2 \pi i \vp (t_4 - t_3)} \Bigr]
\end{aligned}
\end{equation}
\begin{equation}
\begin{aligned}
\frac{\delta^4 \mathcal{M}_{4} \bigl[ 0 \bigr]}
	{\delta s(t_1) \delta s(t_2) \delta s(t_3) \delta s(t_4)}
	&= \Gamma_{n_1} (t_1, t_2) \Gamma_{n_1} (t_3, t_4)
	+ \Gamma_{n_1} (t_1, t_3) \Gamma_{n_1} (t_3, t_2)
	+ \Gamma_{n_1} (t_1, t_4) \Gamma_{n_1} (t_2, t_3)
\end{aligned}
\end{equation}
\begin{equation}
\begin{aligned}
\frac{\delta^4 \mathcal{M}_{5} \bigl[ 0 \bigr]}
	{\delta s(t_1) \delta s(t_2) \delta s(t_3) \delta s(t_4)}
	&= \Gamma_{n_2} (t_1, t_2) \Gamma_{n_2} (t_3, t_4)
	+ \Gamma_{n_2} (t_1, t_3) \Gamma_{n_2} (t_3, t_2)
	+ \Gamma_{n_2} (t_1, t_4) \Gamma_{n_2} (t_2, t_3)
\end{aligned}
\end{equation}

In the strong-pump regime we keep only terms of order $|a_p|^n$ for the $n^\text{th}$ correlation function in the photon noise terms, yielding the approximations:

\begin{equation}
	\mathcal{M}_n [0] = 1 \quad \forall  \quad n
\end{equation}
\begin{equation}
	\frac{\delta \mathcal{M}_n [0]}{\delta s(t_1)} = 0
	\quad \forall \quad n
\end{equation}

\begin{align}
	\frac{\delta^2 \mathcal{M}_1 [0]}{\delta s(t_1) \delta s(t_2)}
	&\approx 0 \\
	\frac{\delta^2 \mathcal{M}_2 [0]}{\delta s(t_1) \delta s(t_2)}
	&= \frac{1}{2} h \nu_0 |a_p|^2 \delta(t_1 - t_2) \\
	\frac{\delta^2 \mathcal{M}_3 [0]}{\delta s(t_1) \delta s(t_2)}
	&= \frac{1}{2} |a_p|^2 \Re \Bigl[
		\Gamma (t_1, t_2) e^{-2 \pi i \vp (t_1 - t_2)}
	\Bigr] \\
	\frac{\delta^2 \mathcal{M}_4 [0]}{\delta s(t_1) \delta s(t_2)}
	&= \Gamma_{n_1} (t_1, t_2) \\
	\frac{\delta^2 \mathcal{M}_5 [0]}{\delta s(t_1) \delta s(t_2)}
	&= \Gamma_{n_2} (t_1, t_2)
\end{align}
\begin{equation}
	\frac{\delta^3 \mathcal{M}_n [0]}{\delta s(t_1) \delta s(t_2) \delta s(t_3)}
	= 0 \quad \forall \quad n
\end{equation}

\begin{equation}
\begin{aligned}
\frac{\delta^4 \mathcal{M}_{1} \bigl[ 0 \bigr]}
	{\delta s(t_1) \delta s(t_2) \delta s(t_3) \delta s(t_4)}
	&\approx 0
\end{aligned}
\end{equation}
\begin{equation}
\begin{aligned}
\frac{\delta^4 \mathcal{M}_{2} \bigl[ 0 \bigr]}
	{\delta s(t_1) \delta s(t_2) \delta s(t_3) \delta s(t_4)}
	&\approx \frac{1}{4} (h \nu_0)^2 |a_p|^4
		\delta(t_1 - t_3) \delta(t_2 - t_4) \\
	&+ \frac{1}{4} (h \nu_0)^2 |a_p|^4
		\delta(t_1 - t_4) \delta(t_2 - t_3) \\
	&+ \frac{1}{4} (h \nu_0)^2 |a_p|^4
		\delta (t_1 - t_2) \delta (t_3 - t_4)
\end{aligned}
\end{equation}
\begin{equation}
\begin{aligned}
\frac{\delta^4 \mathcal{M}_{3} \bigl[ 0 \bigr]}
{\delta s(t_1) \delta s(t_2) \delta s(t_3) \delta s(t_4)}
&\approx \frac{1}{4} |a_p|^4
	\Re \Bigl[ \Gamma (t_3, t_1) e^{-2 \pi i \vp (t_3 - t_1)} \Bigr]
	\Re \Bigl[ \Gamma (t_4, t_2) e^{-2 \pi i \vp (t_4 - t_2)} \Bigr] \\
&+ \frac{1}{4} |a_p|^4
	\Re \Bigl[ \Gamma (t_3, t_2) e^{-2 \pi i \vp (t_3 - t_2)} \Bigr]
	\Re \Bigl[ \Gamma (t_4, t_1) e^{-2 \pi i \vp (t_4 - t_1)} \Bigr] \\
&+ \frac{1}{4} |a_p|^4
	\Re \Bigl[ \Gamma (t_2, t_1) e^{-2 \pi i \vp (t_2 - t_1)} \Bigr]
	\Re \Bigl[ \Gamma (t_4, t_3) e^{-2 \pi i \vp (t_4 - t_3)} \Bigr]
\end{aligned}
\end{equation}
\begin{equation}
\begin{aligned}
\frac{\delta^4 \mathcal{M}_{4} \bigl[ 0 \bigr]}
	{\delta s(t_1) \delta s(t_2) \delta s(t_3) \delta s(t_4)}
	&= \Gamma_{n_1} (t_1, t_2) \Gamma_{n_1} (t_3, t_4)
	+ \Gamma_{n_1} (t_1, t_3) \Gamma_{n_1} (t_3, t_2)
	+ \Gamma_{n_1} (t_1, t_4) \Gamma_{n_1} (t_2, t_3)
\end{aligned}
\end{equation}
\begin{equation}
\begin{aligned}
\frac{\delta^4 \mathcal{M}_{5} \bigl[ 0 \bigr]}
	{\delta s(t_1) \delta s(t_2) \delta s(t_3) \delta s(t_4)}
	&= \Gamma_{n_2} (t_1, t_2) \Gamma_{n_2} (t_3, t_4)
	+ \Gamma_{n_2} (t_1, t_3) \Gamma_{n_2} (t_3, t_2)
	+ \Gamma_{n_2} (t_1, t_4) \Gamma_{n_2} (t_2, t_3)
\end{aligned}
\end{equation}

\newpage

\section{Microwave radiometer theory}\label{sec:radiometer_theory}

In this section we present a derivation of the radiometer equation.

\subsection{Basic operation}\label{sec:rt_asic_operation}

The basic function of a microwave radiometer is to measure the power spectral density (PSD) of a stationary random signal at a desired frequency $\nu_0$.
This measurement can be treated as comprising three steps: 1) pre-detection filtering, 2) detection and 3) post-detection filtering.
In step 1 the signal is bandpass-filtered to select the desired frequency component and possibly amplified.
In step 2, the output step 1 is processed by a square-law detector.
Finally, in step 3 the output of the detector is filtered to give the final measurement.
In heterodyne radiometers step 1 will also include frequency down-conversion steps, however a single frequency analysis can still be used by defining an equivalent input PSD that includes image responses.

Denoting the radiometer output as $y(t)$, we can express the operations described mathematically as
\begin{equation}\label{eqn:rt_radiometer_output_eqn}
\begin{aligned}
	y(t) 
	&= \int_{-\infty}^{\infty} h(t - t_1)
		\left\{ \int_{-\infty}^{\infty} g(t_1 - t_2) x(t_2) \, dt_2 \right\}^2
	\, dt_1 \\
	&= \int_{-\infty}^{\infty} \int_{-\infty}^{\infty} \int_{-\infty}^{\infty}
		h(t - t_1) g(t_1 - t_2) g(t_1 - t_3) x(t_2) x(t_3)
	\, dt_1 dt_2 dt_3
\end{aligned}
\end{equation}
where $x(t)$ is the effective input signal and $h(t)$ and $g(t)$ are the time-domain response functions associated with the pre- and post-detection filtering respectively.
$x(t)$ is the sum of the true input signal and an equivalent input signal needed to produce any additional noise added in the pre-detection stages.

We will assume $x(t)$ is stationary and, without any loss of generality, that it is real, in which case it follows from Wiener-Khinchin theorem that
\begin{equation}\label{eqn:rt_wk_x}
	\langle x(t) x(t + \tau) \rangle = \frac{1}{2} \int_{-\infty}^{\infty}
		\{ p_s (\nu) + p_n (\nu) \} e^{2 \pi i \nu t}
	\, d\nu
\end{equation}
where the angle brackets denote statistically expected value, $p_s (\nu)$ is the single-sided PSD of the input signal and $p_d (\nu)$ is the input-referenced single-sided PSD of any noise added in the receiver pre-detection.
Both PSDs are even, real-valued, functions of $\nu$.
Taking the expectation value of (\ref{eqn:rt_radiometer_output_eqn}), using (\ref{eqn:rt_wk_x}) to substitute and evaluating the resulting Fourier transforms, we find the expected value of the output signal under this assumption is
\begin{equation}\label{eqn:exp_rt}
\begin{aligned}
	\langle y(t) \rangle 
	&= \tilde{h}(0) \int_{0}^{\infty}
		G(\nu) \{ p_s (\nu) + p_n (\nu) \}
	\, d\nu.
\end{aligned}
\end{equation}
Here we use the same notation for Fourier transforms as in the main note, where $\tilde{f}(\nu)$ is the Fourier transform of $f(t)$ and $F(\nu) = |\tilde{f}(\nu)|^2$.
To obtain (\ref{eqn:exp_rt}) we have also exploited the fact $p_s (\nu)$ , $p_n (\nu)$ and $G(\nu)$ are all even functions to restrict the integral to positive frequencies.

In a practical radiometer the pre-detection filter function will be sharply peaked at the frequency of interest.
We may then further approximate (\ref{eqn:exp_rt}) by
\begin{equation}\label{eqn:exp_rt_approx}
\begin{aligned}
	\langle y(t) \rangle 
	&\approx \tilde{h}(0) \{ p_s (\nu_0) + p_n (\nu_0) \}
	\int_0^\infty G(\nu) \, d\nu.
\end{aligned}
\end{equation}
From (\ref{eqn:exp_rt_approx}) we see that we expect to be able to estimate the signal PSD at the desired frequency from the radiometer output provided that we have calibrated the filter parameters and receiver noise PSD.

\subsection{Sensitivity}\label{sec:rt_sensitivity}

The expected variance 
\begin{equation}
	\sigma_y^2 = \langle (y (t) - \langle y(t) \rangle)^2 \rangle
\end{equation}
of the output provides a measure of the smallest detectable value of $p_s (\nu_0)$ and therefore of the sensitivity of the radiometer.

We will make the additional assumption $x(t)$ is a Gaussian random process, in which case it is possible to obtain an approximate analytic for $\sigma_y$.
If $x(t)$ is Gaussian, it follows from the Gaussian moment theorem that
\begin{equation}\label{eqn:gaussian_moment_theorem}
\begin{aligned}
	&\langle x(t_1) x(t_2) x(t_3) x(t_4) \rangle \\
	&= \langle x(t_1) x(t_2) \rangle \langle x(t_3) x(t_4) \rangle
	+ \langle x(t_1) x(t_3) \rangle \langle x(t_2) x(t_4) \rangle
	+ \langle x(t_1) x(t_4) \rangle \langle x(t_2) x(t_3) \rangle.
\end{aligned}
\end{equation}
Using (\ref{eqn:gaussian_moment_theorem}) and (\ref{eqn:rt_radiometer_output_eqn}), it follows after some algebra that
\begin{equation}\label{eqn:rt_var_y_step_1}
\begin{aligned}
	\sigma_y^2 &=
		2 \int_{-\infty}^{\infty} \int_{-\infty}^{\infty} \int_{-\infty}^{\infty}
		\int_{-\infty}^{\infty} \int_{-\infty}^{\infty} \int_{-\infty}^{\infty} \\
	& \times h(t - t_1) g(t_1 - t_2) g(t_1 - t_3) h(t - t_4) g(t_4 - t_5) g(t_4 - t_6) \\
	& \times \langle x(t_2) x(t_5) \rangle \langle x(t_3) x(t_6) \rangle
	\, dt_1 dt_2 dt_3 dt_4 dt_5 dt_6.
\end{aligned}
\end{equation}
(\ref{eqn:rt_wk_x}) can then be used to substitute for the correlation functions and if we then evaluate the resulting Fourier transforms, we find we can express the variance in terms of the PSDs as
\begin{equation}\label{eqn:rt_var_y_step_2}
\begin{aligned}
	\sigma_y^2
	&= \frac{1}{2} \int_{-\infty}^{\infty} \int_{-\infty}^{\infty}
		H (\nu_1 + \nu_2) G(\nu_1) G(\nu_2)
		\{p_s(\nu_1) + p_n (\nu_1) \} \{p_s(\nu_2) + p_n (\nu_2) \}
	\, d\nu_1 d\nu_2
\end{aligned}
\end{equation}
In a practical radiometer the post-detection filtering should be sharply peaked around zero-frequencies to remove noise; an example would be averaging the detector output over a long time period.
We may then approximate
\begin{equation}
	H(\nu) \approx \delta (\nu) \int_{-\infty}^\infty H(\nu) \, d\nu
\end{equation}
and then, remembering that $G(\nu)$ is also assumed to be sharply peaked, (\ref{eqn:rt_var_y_step_2}) simplifies to
\begin{equation}\label{eqn:rt_var_y_step_3}
\begin{aligned}
	\sigma_y^2
	&\approx \int_{0}^\infty H(\nu) \, d\nu
	\times 2 \int_{0}^{\infty} G(\nu)^2 \, d\nu
	\times \{ p_s(\nu_0) + p_n (\nu_0) \}^2.
\end{aligned}
\end{equation}

Our primary interest is in $p_0 = p_s (\nu_0) + p_n(\nu_0)$.
We can divide (\ref{eqn:rt_var_y_step_3}) by the square of (\ref{eqn:exp_rt_approx}) to give an equivalent variance in the measurement of $p_0$ and we find
\begin{equation}\label{eqn:rt_var_p}
	\sigma_{p_0}^2 \approx \frac{2 p_0^2 \Delta f}{\Delta \nu_n},
\end{equation}
where $\Delta f$ is the baseband bandwidth of the post detection filter as defined by
\begin{equation}\label{eqn:rt_post_det_bandwidth}
	\Delta f = \frac{\int_0^\infty H(\nu) \, d\nu}{H(0)}
\end{equation}
and
\begin{equation}\label{eqn:rt_pre_det_bandwidth}
	\Delta \nu_n = \frac{(\int_0^\infty G(\nu) \, d\nu)^2}
	{\int_0^\infty G(\nu)^2 \, d\nu}
\end{equation}
is the effective noise bandwidth of the pre-detection filter.
In the case $G(\nu)$ approximates a top-hat function, $\Delta \nu_n$ is approximately equal to the bandwidth of the filter.

\subsection{Radiometer equation}\label{sec:rt_radiometer_eqn}

We are now in a position to derive the radiometer equation.
The output of a microwave radiometer is normally scaled in units of equivalent noise temperature $T (\nu)$, where $p(\nu) = k_b T (\nu)$.
In addition, the post-detection filtering is usually assumed to be time-integration, in which case the appropriate filtering function is
\begin{equation}\label{eqn:rt_h_time_int}
	h(t) = \begin{cases}
		1 / \tau & |t| < \tau / 2 \\
		0 & \text{otherwise}
	\end{cases}
\end{equation}
for integration time $\tau$.
The corresponding baseband bandwidth can be calculated efficiently using Parseval's theorem and we obtain
\begin{equation}\label{eqn:rt_f_time_int}
	\Delta f = \frac{\int_{-\infty}^\infty h(t)^2 \, dt}{2\int_{-\infty}^\infty h(t) \, dt}
	= \frac{1}{2 \tau}.
\end{equation}
Rescaling (\ref{eqn:rt_var_p}) and substituting in (\ref{eqn:rt_f_time_int}) and then taking the square root to obtain the root-mean-square error, we obtain the radiometer equation
\begin{equation}\label{eqn:rt_radiometer_eqn}
	\Delta T_\text{rms} = \frac{T_\text{sys}}{\sqrt{ \tau \Delta \nu_n}}.
\end{equation}
for $T_\text{sys} = (p_s (\nu_0) + p_n (\nu_0)) / k_b$.

\newpage

\section{Transition-Edge Sensors}\label{sec:tes_physics}

\subsection{Introduction}\label{sec:tes_introduction}

Transition-Edge Sensors (TESs) are a type of bolometric detector.
As such, they detect the power absorbed from an input signal by the change in temperature it produces in a heat capacity that has been thermally isolated from its surroundings.
In a TES, these changes in temperature are sensed using the change in resistance of a superconducting film held at the point of its superconducting transition (the `transition-edge').
In modern designs, the absorber and film are usually suspended together on a micro-machined island.
Thermal isolation is then achieved by patterning the connecting film between the island and its surroundings into a number of narrow `legs' to restrict heat flow.

The superconducting film must be voltage-biased into the superconducting transition for stable operation.
The bias voltage $V_b$ results in power dissipation $V_b^2 / R$ in the film, where $R$ is the film resistance.
In the absence of a signal, this is balanced by heat flow down the legs to the surrounding thermal bath.
When a signal is applied, the additional power input acts to increase the island temperature.
However, this causes the resistance $R$ of the superconducting film to increase, which decreases the Ohmic power dissipation and so partially negate the temperature increase.
This negative electrothermal feedback mechanism acts to keep the island temperature within the superconducting transition over wide input power ranges.
Since Joule power is being exchanged for signal power, the TES will saturate at the latter exceeds the value of the former before a signal is applied.

The ultimate sensitivity of a TES is limited by two intrinsic noise sources that cannot be eliminated.
The first is the intrinsic thermodynamic fluctuations of the energy of the island.
This is usually attributed to fluctuations in the power flow down the legs and called phonon noise.
The second second is Johnson-Nyquist associated with the resistance of the superconducting film.
This fundamental sensitivity can only be achieved if other sources of noise, such as from the readout electronics, can be made sufficiently small.

In this section we will derive the relationship between the limiting sensitivity of the TES (as characterised by NEP), saturation power and the bath temperature used in the main text.
To do so, we will first derive expressions for the power flow and phonon noise in the leg structures.
These results will then be used as inputs to coupled large and small-signals models of a TES, from which we will derive an expression for the NEP as a function of device operating temperature when the saturation power is fixed.
We will then determine the minimal value of the NEP and the operating temperature at which it is achieved.

\subsection{Leg model}\label{sec:leg_model}

Simple models of bolometers usually assume a lumped thermal conductance $G$ between the heat capacity $C$ and the bath at temperature $T_b$.
The difference in temperature $\Delta T$ between the heat capacity and bath is also usually assumed to be small ($\Delta T \ll T_b$).
In this case the heat flow $q$ from island to bath is simply $G \Delta T$ and the phonon-limited NEP is $\sqrt{4 k_b G T_b^2}$, with the latter corresponding to an effective noise heat flow in the conductance satisfying $\langle q(t_1) q(t_2) \rangle = 2 k_b G T_b^2 \delta (t_1 - t_2)$.

The situation is much more complicated in a TES, as both assumptions are usually violated; the leg is a distributed conductance with a temperature profile and, for reasons we will see, the operating temperature is typically around twice the bath temperature.
In addition, the conductivity of the leg material is also usually temperature dependent.
This significantly alters both the heat flow and phonon NEP and the aim of this section will be to derive expressions for both.

We start by considering heat flow.
Consider a leg of uniform cross-sectional area $A$ made out of a material with temperature dependent thermal conductivity
\begin{equation}\label{eqn:tes_leg_conductivity}
	k(T) = n \kappa T^{n -1}
\end{equation}
where $n$ and $\kappa$ are model parameters.
Let $x$ denote position along the leg and assume we hold the temperature at point $x=x_1$ at $T_1$ and $x = x_2$ ($x_2 > x_1$) at $T_2$.
Conservation of energy requires that 
\begin{equation}\label{eqn:tes_heat_flow_constraint}
	A k(T) \frac{dT}{dt} = - q_{21} (T_2, T_1)
\end{equation}
at any point $x_1 < x < x_2$, where $q_{12}(T_2, T_1)$ is the equilibrium heat flow along the section of material in the positive $x$-direction.
Integrating (\ref{eqn:tes_heat_flow_constraint}) over $x_1 < x < x_2$, we find the heat flow can be written in terms of the temperatures of the ends of the section as
\begin{equation}\label{eqn:heat_flow_12}
	q_{21} (T_2, T_1) = -\frac{A}{x_2 - x_1} \int_{T_1}^{T_2} k(T) \, dT
	= \frac{\kappa A}{l} \bigl\{ T_1^n - T_2^n \bigr\}
\end{equation}
Hence the total heat flow from bath to the TES island will be
\begin{equation}\label{eqn:leg_heat_flow}
	Q(T) = \frac{\kappa N A}{l} \bigl\{ T_b^n - T^n \bigr\}
\end{equation}
if the island temperature is $T$ and the device has $N$ legs of length $l$.

Now we consider the phonon noise.
Consider dividing the leg up into three sections: 1) $0 < x < x'$, 2) $x' < x < x' + \delta x$ and 3) $x' + \delta x < x < l$.
We start by considering the case where a small, zero-mean, noise phonon-noise heat flow $\Delta q_N$ is introduced across section 2.
This will result in changes in temperature $\Delta T_1$ at $x=x'$ and $\Delta T_2$ at $x=x' + \delta x$ about the equilibrium values.
Similarly, it will result in a change of heat-flow $\Delta q_m$ in section $m$, with conservation of energy requiring
\begin{equation}\label{eqn:tes_q_constraint_1}
	\Delta q_1 = \Delta q_2 + \Delta q_N
\end{equation}
and
\begin{equation}\label{eqn:tes_q_constraint_2}
	\Delta q_2 + \Delta q_N = \Delta q_3.
\end{equation}
If follows from (\ref{eqn:heat_flow_12}) that we can express the change in the heat flow in a section of leg when the end temperatures change as
\begin{equation}\label{eqn:tes_perturbative_power_flows}
\begin{aligned}
	&q_{21} (T_2 + \Delta T_2, T_1 + \Delta T_1)
	= q_{12} (T_2, T_1) + G_{21} \Delta T_1 - G_{22} \Delta T_2 \\
	&G_{2m} = \frac{A k(T_m)}{x_2 - x_1}
\end{aligned}
\end{equation}
to first order.
Hence if the temperatures of the ends of the leg are kept constant, we expect
\begin{equation}\label{eqn:tes_section_power_flows}
\begin{aligned}
	&\Delta p_1 = -G_{11} \Delta T_1
	&&\Delta p_2 = G_{21} \Delta T_1 - G_{22} \Delta T_2
	&&\Delta p_3 = G_{i2} \Delta T_2
	&& \\
	& G_{21}= \frac{\kappa A k (T(x'))}{\delta x}
	&& G_{22}= \frac{\kappa A k (T(x' + \delta x))}{\delta x}
	&& G_{11} = \frac{\delta x}{x'} G_{21}
	&& G_{i2} = \frac{\delta x}{l - x'} G_{22}.
\end{aligned}
\end{equation}
To determine the resultant noise heat flow into the island, we need to calculate $\Delta q_3$ given $\Delta q_N$.
Solving (\ref{eqn:tes_q_constraint_1}) and (\ref{eqn:tes_q_constraint_2}) given (\ref{eqn:tes_section_power_flows}), we find
\begin{equation}
	\Delta p_3 = \frac{\delta x}{(l + \delta x)} \Delta q_N.
\end{equation}
Given the results for a lumped heat capacity, we therefore expect $\langle \Delta p_3 (t) \rangle = 0$ and 
\begin{equation}
\begin{aligned}
	\langle \Delta p_3 (t_1) \Delta p_3 (t_2)  \rangle
	&\sim \frac{\delta x^2}{(l + \delta x)^2}
		\langle \Delta q_N (t_1) \Delta q_N (t_2)  \rangle \\
	&\sim \frac{2 A k_b}{l^2}
	 k(T(x')) \bigl\{ T(x') \bigr\}^2 \delta x \, \delta(t_1 - t_2)
\end{aligned}
\end{equation}
where the final result applies in the limit $\delta x \rightarrow 0$.

Now consider dividing each leg up into $M$ such segments.
Each section behaves statistically independently, so the total noise power flow $\Delta Q (t)$ into the island will then satisfy
\begin{equation}\label{eqn:tes_total_noise_as_sum}
	\langle \Delta Q(t_1) \Delta Q(t_2) \rangle
	= \frac{2 A k_b N}{l^2} \delta(t_1 - t_2) \sum_{m=0}^{M-1}
	k(T(m \delta x)) \bigl\{ T(m \delta x) \bigr\}^2 \delta x	.
\end{equation}
Taking the limit $\delta x \rightarrow 0$, we then obtain
\begin{equation}\label{eqn:tes_total_noise_as_integral}
	\langle \Delta Q(t_1) \Delta Q(t_2) \rangle
	= \frac{2 k_b N A}{l^2} \delta(t_1 - t_2) \int_0^l
	k(T(x)) \bigl\{ T(x) \bigr\}^2 d x.
\end{equation}

Ideally we would like to make a change of variables in (\ref{eqn:tes_total_noise_as_integral}) to express the integral with respect to temperature, so as to avoid having to calculate the temperature profile.
We can do so by noting that (\ref{eqn:tes_heat_flow_constraint}) and (\ref{eqn:heat_flow_12}) together imply
\begin{equation}
	k(T) \frac{dT}{dx} = \frac{1}{l} \int_{T_b}^{T} k(T') \, dT'.
\end{equation}
Hence we may write
\begin{equation}\label{eqn:tes_total_noise_as_integral_v2}
\begin{aligned}
	\langle \Delta Q(t_1) \Delta Q(t_2) \rangle
	&= \frac{2 k_b N A}{l}
	\frac{\int_0^l \bigl\{ T(x) k(T(x))\bigr\}^2 \frac{d T}{dx} d x}
	{\int_{T_b}^{T} k(T') \, dT'}
	\delta(t_1 - t_2) \\
	&= \frac{2 k_b N A}{l}
	\frac{\int_{T_b}^T \bigl\{ T'  k(T')\bigr\}^2 \, dT'}
	{\int_{T_b}^{T} k(T') \, dT'}
	\delta(t_1 - t_2).
\end{aligned}
\end{equation}
Finally, using (\ref{eqn:tes_leg_conductivity}) to substitute for $k(T)$ and evaluating the integral, we find
\begin{equation}\label{eqn:tes_total_phonon_noise}
\begin{aligned}
	\langle \Delta Q(t_1) \Delta Q(t_2) \rangle
	&= \underbrace{
		\frac{2 \kappa n k_b N A T_b^{n+1}}{l}
		\frac{\vphantom{T_b^{2n +1}}}{\vphantom{T_b^{n + 1}}}
	}_{(i)}
	\underbrace{
		\frac{(T_b^{2 n + 1} - T^{2n + 1}) n}
		{(2 n + 1) (T_b^n - T^n) T_b^{n + 1}}
	}_{(ii)}
	\delta(t_1 - t_2),
\end{aligned}
\end{equation}
which is the result we will use going forward.
Term (i) can be identified as the result that would be expected if the temperature of the leg was uniform and equal to the bath temperature.
Term (ii) therefore represents a modifying factor that accounts for the non-equilibrium temperature.

\subsection{Device model}\label{sec:device_model}

In this section we will derive a set of equations modelling the behaviour of a TES in response to input power.
We will assume there is some background input power $P_0$ and that the device has come into equilibrium at temperature $T_0$ (the \emph{operating} temperature) and operating resistance $R_0 = R(T_0)$.
If a small amount of additional input power $\Delta P (t)$ is applied, we then expect the island temperature and device current to change by $\Delta T (t)$ and $\Delta R (t)$ respectively.
The aim will be to derive a state equation linking the operating parameters and dynamical equations for the changes in these quantities.

We start by considering the behaviour of the device temperature.
Conservation of energy requires in general that
\begin{equation}\label{eqn:tes_island_heat_flow_equation}
	C \frac{dT}{dt} = Q(T)  + \Delta Q_N (t) + R(T) I(t)^2 + P(t) 
\end{equation}
where $T$ is the island temperature, $C$ is the heat capacity and $I$ is current in the resistive film
The first and second terms on the right are the heat flow down the legs and the noise in the latter as given by (\ref{eqn:leg_heat_flow}) and (\ref{eqn:tes_total_phonon_noise}), respectively.
The second term is the Joule power dissipation in the TES.
The third term is the the input signal power.
When additional signal power is applied, we can expand the conductance of the superconducting film at the new temperature as
\begin{equation}
	R(T_0 + \Delta T)
	= R_0 + \frac{\alpha R_0}{T_0} \Delta T + O(\Delta T^2)
\end{equation}
to first order, where $\alpha = T dR / (R_0 dT)$ is a dimensionless quantity that characterises the gradient of the transition-edge at the operating point.
Expanding the other variables in (\ref{eqn:tes_island_heat_flow_equation}) as the sum of their time-independent value at the operating point and a time dependent fluctuation term, $X = X_0 + \Delta X$, we find the operating point parameters must satisfy
\begin{equation}\label{eqn:tes_temperature_op_point}
	0 = Q(T_0) + P_0 + P_J.
\end{equation}
where $P_J = R_0 I_0^2$ is the quiescent Joule power dissipation.
Similarly, the fluctuations are described, to first order, by the dynamical equation
\begin{equation}\label{eqn:tes_small_signal_temperature}
	C \frac{d \Delta T}{dt}
	= - \left\{ G - \frac{\alpha P_J}{T_0} \right\} \Delta T (t)
	+ 2 I_0 R_0 \Delta I (t)
	+ \Delta P (t) + \Delta Q_N (t),
\end{equation}
where $G = n \kappa N A T_0^{n - 1} / l$.

Now we must consider the readout circuit.
As described in the introduction, changes in the input power in the TES are detected from the corresponding changes in the current $I$ in the superconducting film.
A TES is therefore readout with what is, in effect, an ammeter and the device of choice is usually a SQUID amplifier.
The input impedance of a SQUID is inductive, so the simplest possible model of the corresponding readout circuit is an inductance $L$ in series with the resistance $R$ of the film, across which the bias voltage is applied.
The behaviour of such a circuit is described by the differential equation.
\begin{equation}\label{eqn:tes_readout_circuit_de}
	L \frac{dI}{dt} + R(T) I + \Delta V_N = V_b,
\end{equation}
where $\Delta V_N$ is the Johnson noise voltage across the film resistance.
Carrying out the same small signal expansion as for the island heat flow equation, we find the operating point parameters satisfy
\begin{equation}
	V_b = R_0 I_0
\end{equation}
and the behaviour of the small signal quantities is described by
\begin{equation}\label{eqn:tes_small_signal_current}
	L \frac{d \Delta I}{dt}
	= -R_0 \Delta I - \frac{\alpha I_0 R_0}{T_0} \Delta T - \Delta V_N,
\end{equation}
again to first order.

We are now in a position to calculate the small signal frequency domain behaviour of the device.
(\ref{eqn:tes_small_signal_temperature}) and (\ref{eqn:tes_small_signal_current}) can be combined into a single matrix equation
\begin{equation}\label{eqn:tes_small_signal_equation}
	\frac{d}{dt} \left( \begin{array}{c}
		\Delta I \\
		\Delta T
	\end{array} \right)
	=
	\left( \begin{array}{cc}
		-R_0 / L & - \alpha V_b / (L T_0) \\
		2 V_b / C & -(G - \alpha P_J / T_0) / C 
	\end{array} \right)
	\left( \begin{array}{c}
		\Delta I \\
		\Delta T
	\end{array} \right)
	+\left( \begin{array}{c}
		-\Delta V_N / L \\
		\Delta P / C + \Delta Q_N / C
	\end{array} \right).
\end{equation}
Now assume each quantity is sinusoidally varying: $\Delta X(t) = \Delta \tilde{X} (f) e^{2 \pi i f t}$.
(\ref{eqn:tes_small_signal_equation}) can then be solved for $\Delta \tilde{I}(f)$ in terms of $\Delta P (f)$ using matrix inversion and the final result is
\begin{equation}\label{eqn:tes_small_signal_equation_solution}
\begin{aligned}
	\left( \begin{array}{c}
		\Delta \tilde{I} (f) \\
		\Delta \tilde{T} (f)
	\end{array} \right)
	=
	\frac{1}{\mathcal{R}(f)}
	\left( \begin{array}{cc}
		2 \pi i f + (G - \alpha P_J / T_0) / C & -\alpha V_b / (L T_0) \\
		2 V_b / C & 2 \pi i f + R_0 / L 
	\end{array} \right)
	\left( \begin{array}{c}
		-\Delta \tilde{V}_N (f) / L \\
		\Delta \tilde{P} (f) / C + \Delta \tilde{Q} (f) / C
	\end{array} \right),
\end{aligned}
\end{equation}
where
\begin{equation}
\begin{aligned}
	\mathcal{R} (f)
	& = \{ 2 \pi i f + R_0 / L \} \{ 2 \pi i f + (G - \alpha P_J / T_0) / C \}
	+ 2 \alpha V^2_b / (L C T_0)
\end{aligned}
\end{equation}
This relation determines the scaling of the change in device current for given change in input power.

\subsection{Sensitivity}\label{sec:tes_sensitivity}

(\ref{eqn:tes_small_signal_equation_solution}) implies that we can form an estimator $\hat{P}$ of the amplitude of the change in input power from the change in device current according to
\begin{equation}\label{eqn:tes_power_estimator}
	\hat{P} (f) = \frac{L C \mathcal{R} (f)}{R_0+ 2 \pi i f L} \Delta \tilde{I} (f).
\end{equation}
The expected value of this estimator is
\begin{equation}\label{eqn:tes_exp_power_estimator}
	\langle \hat{P} (f) \rangle
	= \Delta \tilde{P} (f) 
	+ \langle \Delta \tilde{Q}_N (f) \rangle
	+\frac{G T_0 - \alpha P_J + 2 \pi i f C T_0}{\alpha V_b}
	\langle \Delta \tilde{V}_N (f) \rangle
	= \Delta \tilde{P} (f),
\end{equation}
as required.
The expected variance is
\begin{equation}\label{eqn:tes_var_power_estimator}
\begin{aligned}
	&\langle \Delta \hat{P}^2 \rangle (f)
	= \langle \Delta \tilde{Q}_N^2 \rangle (f)
	+\frac{(G T_0 - \alpha P_J)^2 + (2 \pi f C T_0)^2}{\alpha^2 V^2_b}
	\langle \Delta \tilde{V}_N^2 \rangle (f) \\
	&= 
	\frac{2 n k_b G (T_b^{2 n + 1} - T_0^{2n + 1})}{(2 n + 1) (T_b^n - T_0^n) T_0^{n + 1}}
	+\frac{2 k_b \{ (G T_0 - \alpha P_J)^2 + (2 \pi f C T_0)^2 \} T_0}{\alpha^2 P_J}
\end{aligned}
\end{equation}
using (\ref{eqn:tes_total_phonon_noise}) to substitute for the phonon power and the standard result for Johnson noise in a resistor to substitute for the variance of the noise voltage.
In the case of a steady state change in signal power, the corresponding NEP is therefore given by
\begin{equation}\label{eqn:tes_nep_intermediate}
	\frac{\text{NEP}^2}{4 k_b}
	= \frac{n G(T_b^{2 n + 1} - T_0^{2n + 1})}{(2 n + 1) (T_b^n - T_0^n) T_0^{n - 1}}
	+\frac{(G T_0 - \alpha P_J)^2 T_0}{\alpha^2 P_J}
\end{equation}

(\ref{eqn:tes_nep_intermediate}) can be simplified further by making use of the relationships between operating parameters.
Using (\ref{eqn:leg_heat_flow}) and the definition of $G$, (\ref{eqn:tes_temperature_op_point}) can be re-expressed as
\begin{equation}\label{eqn:tes_temperature_op_point_rewritten}
	\frac{G( T_b^n - T_0^n)}{n T_0^{n - 1}}
	+ P_0 + P_J = 0.
\end{equation}
Based on the arguments in Section \ref{sec:tes_introduction}, we can also identify $P_J$ as the approximate saturation power $P_s$ of the device.
Using (\ref{eqn:tes_temperature_op_point_rewritten}) to eliminate $G$ and making the identification $P_s \sim P_J$, (\ref{eqn:tes_nep_intermediate}) becomes
\begin{equation}\label{eqn:tes_nep_intermediate_2}
	\frac{\text{NEP}^2}{4 k_b P_0}
	= \frac{n^2 (T_0^{2 n + 1} - T_b^{2n + 1}) (P_0 + P_s)}{(2 n + 1) (T_0^n - T_b^n)^2 P_0}
	+ \left\{ 1 + \frac{n T_0^n (P_0 + P_s)}{\alpha (T_b^n - T_0^n) P_s} \right\}^2 \frac{P_s T_0}{P_0}.
\end{equation}

(\ref{eqn:tes_nep_intermediate_2}) can be simplified by making two additional assumptions.
Firstly, we assume we are interested in the maximum possible sensitivity as achieved in the limit $\alpha \rightarrow \infty$, i.e. a very sharp transition.
This is the limit of maximum possible sensitivity.
The second is that we are interested in making the saturation power some factor $m$ of the background loading.
We will refer to $m$ as the \emph{saturation margin} of the detector.
In this case the NEP simplifies to
\begin{equation}\label{eqn:tes_nep_final}
	\frac{\text{NEP}^2}{4 k_b T_b P_0}
	= \frac{n^2 \bigr( m + 1 \bigl) \bigr( t^{2 n + 1} - 1 \bigl)}
	{\bigr( 2 n + 1 \bigl) \bigr( t^n - 1 \bigr)^2}
	+ m t.
\end{equation}
where $t = T_0 / T_b$.

(\ref{eqn:tes_nep_final}) is the final result of this section.
It is useful in the context of this report because it shows how the ultimate NEP of a TES scales with the experimentally determined parameters $T_b$, $P_0$ and $m$.
In practice $T_b$ is set by the cooling technology, $P_0$ the background loading (haloscope or pump power) and $m$ by the expected variation in this background.
We can optimize for $t$ via the design of the TES.
Since the device must be biased into the transition, the achievable values of $t$ are set by the range over which the superconducting critical temperature of the film can be varied and what thermal conductances are needed.
(\ref{eqn:tes_nep_final}) does not guarantee this ultimate sensitivity can be achieved and this will likely be determined by other issues such as whether a device with the necessary parameters can be engineered, stability of the readout circuit and whether other noise sources (e.g. the SQUID) can be minimized.
These implementation issues are discussed in more detail in the companion report.

\end{document}